\documentclass[aps,prb,reprint,superscriptaddress]{revtex4-2}
\pdfoutput=1
\usepackage{amsmath}
\usepackage{upgreek}
\usepackage{float}
\usepackage{graphicx}
\usepackage{tabularx}
\usepackage{hyperref}
\usepackage[export]{adjustbox}
\usepackage{subfig}
\usepackage[T2A]{fontenc}
\usepackage[T1]{fontenc}
\usepackage[russian, english]{babel}
\usepackage[version=3]{mhchem}
\usepackage{siunitx}
\usepackage{balance}
\DeclareSIUnit{\molar}{M}
\DeclareSIUnit{\channel}{channel}
\usepackage{xcolor}

\begin{document}

\title{A criticial view on e$_g$ occupancy as a descriptor for oxygen evolution catalytic activity in \ce{LiMn2O4} nanoparticles}

\author{Florian Sch\"onewald}
\email{f.schoenewald@uni-goettingen.de}
\affiliation{Universit\"at G\"ottingen, Institut f\"ur Materialphysik, Friedrich-Hund-Platz 1, 37077 G\"ottingen, Germany.}
\author{Marco Eckhoff}
\affiliation{Universit\"at G\"ottingen, Institut f\"ur Physikalische Chemie, Theoretische Chemie, Tammannstra{\ss}e 6, 37077 G\"ottingen, Germany.}
\author{Max Baumung}
\affiliation{Universit\"at G\"ottingen, Institut f\"ur Materialphysik, Friedrich-Hund-Platz 1, 37077 G\"ottingen, Germany.}
\author{Marcel Risch}
\affiliation{Universit\"at G\"ottingen, Institut f\"ur Materialphysik, Friedrich-Hund-Platz 1, 37077 G\"ottingen, Germany.}
\affiliation{Helmholtz-Zentrum Berlin f\"ur Materialien und Energie, Hahn-Meitner-Platz 1, 14109 Berlin, Germany.}
\author{Peter E. Bl\"ochl}
\affiliation{Technische Universit\"at Clausthal, Institut f\"ur Theoretische Physik, Leibnizstra{\ss}e 10, 38678 Clausthal-Zellerfeld, Germany.}
\affiliation{Universit\"at G\"ottingen, Institut f\"ur Theoretische Physik, Friedrich-Hund-Platz 1, 37077 G\"ottingen, Germany.}
\author{J\"org Behler}
\affiliation{Universit\"at G\"ottingen, Institut f\"ur Physikalische Chemie, Theoretische Chemie, Tammannstra{\ss}e 6, 37077 G\"ottingen, Germany.}
\affiliation{The International Center for Advanced Studies of Energy Conversion (ICASEC), University of G\"ottingen, 37077 G\"ottingen, Germany.}
\author{Cynthia A. Volkert}
\email{cynthia.volkert@phys.uni-goettingen.de}
\affiliation{Universit\"at G\"ottingen, Institut f\"ur Materialphysik, Friedrich-Hund-Platz 1, 37077 G\"ottingen, Germany.}
\affiliation{The International Center for Advanced Studies of Energy Conversion (ICASEC), University of G\"ottingen, 37077 G\"ottingen, Germany.}

\date{\today}

\begin{abstract}
	We investigate the effect of the surface electronic structure and composition of LiMn$_2$O$_4$ nanoparticles on the electrocatalytic oxygen evolution reaction (OER). Scanning transmission electron microscopy (STEM) electron energy loss spectroscopy (EELS) studies combined with  density functional theory (DFT) based simulations of the EEL spectra reveal a \SI{3}{\nm} thick surface layer with reduced average Mn oxidation state and increased Mn concentration in the pristine nanoparticles. This is attributed to Mn$^{2+}$ partially replacing Li$^+$ at the tetrahedral sites of the spinel lattice. During electrocatalytic OER cycling, the near-surface tetrahedral Mn is leached out, thereby increasing the oxidation state of the octahedrally coordinated Mn. Using rotating ring-disc electrode (RRDE) based detection of O and Mn during the OER, we show that the oxygen evolution remains constant while the Mn$^{2+}$ is removed, revealing that near-surface tetrahedrally coordinated Mn has no effect on the OER activity of \ce{LiMn2O4}. This is surprising since the e$_g$ occupancy of Mn in octahedral sites, which is widely used as a descriptor of OER activity, changes significantly in the surface layer during cycling. The fact that e$_g$ occupancy fails to correlate with OER activity here indicates that octahedral cation valence is not a fundamental measure of activity, either because the active surface state is not affected by tetrahedral Mn or because other details of the band structure or metal-oxygen bonding character, more strongly regulate the rate-limiting steps for OER.
\end{abstract}


\maketitle
\section{Introduction}
The splitting of water to \ce{H2} and \ce{O2} is one of the most attractive methods to store solar energy. It is currently rate limited by the transfer of four electrons and four protons during the oxygen evolution reaction (OER), so that searching for optimized OER catalysts has become a central research goal. However, gaining mechanistic information about the active interface between a catalyst and the aqueous electrolyte during catalytic reaction is a huge challenge for both experiment and theory. As a result, an important approach to catalyst development and discovery has been the attempt to define materials "descriptors" that correlate with catalytic activity and thus with the rate-limiting step for the oxygen evolution reaction. A wide range of recent studies have identified the strength of oxygen adsorption on the catalyst's surface as a requirement for OER activity (e.g. \citet{Man2011a, Hong2013}), in agreement with the Sabatier principle. In fact, it is often found that OER catalytic activity on oxides follows a volcano relationship when plotted against the difference between the binding strengths of the O* and HO* intermediates\cite{Man2011a,Hong2013}.\\

Many catalyst studies have focused on manganese oxides due to their abundance, low cost and because Mn is the active site of nature's catalyst for the OER\cite{Wei2017, Man2011a,Cady2015,Hong2016, Suntivich2011a, Robinson2013}. Although it is the active state of the Mn at the catalyst surface that determines activity, various aspects of the electronic and geometric structure of the bulk state have been used successfully as descriptors of catalytic activity for the OER.\cite{Calle-Vallejo2015a} For example, comparative experimental studies have shown that activity of Mn oxides correlates very well with the number of the Jahn-Teller active Mn$^{3+}$ ions in the bulk\cite{Robinson2013, Smith2016, Chan2018}. For the case of spinels, it has been reported that the valence or $3d$ occupancy of the octahedral cation in the bulk material is an accurate descriptor of OER activity\cite{Wei2017, Cady2015,Hong2016}. Volcano plots of OER activity for a variety of oxides, including several Mn oxide spinels, have a maximum at an e$_g$ occupancy of around one, corresponding to a Mn valence of 3.0+ \cite{Wei2017, Cady2015,Hong2016, Man2011a}. It is presumed that these Mn$^{3+}$ ions have the necessary catalyst-reactant bond strength and bond flexibility at the catalyst surface to catalyze the rate limiting steps in the oxygen evolution reaction. Of course, other details of the electronic and geometric structure, such as metal-oxygen covalency and bond angles or lengths\cite{Wei2017, Cady2015,Abrashev2019, Hong2016}, also play a role in determining activity. But what remains surprising is that catalytic activity at the solid/electrolyte interface correlates so well with bulk descriptors. The catalyst surface is fundamentally different from the bulk due to broken crystal symmetry, the contact with an aqueous solution, the electrocatalytic potentials, and the presence of reactants. \citet{Calle-Vallejo2015a} propose that the resolution of this contradiction can be found in the similar dependence of bulk descriptors and the energetics of surface adsorption on the outer electrons of the metal oxide.\\

In this study, we work with the battery electrode spinel material \ce{LiMn2O4}, which provides a model system where the catalytically-relevant octahedral Mn $d$ states can be tailored by various routes without changing the crystal symmetry. It has a fixed octahedral Mn-O framework and the octahedral Mn valence can be selected through the occupancy and oxidation state of the ions on the tetrahedral and octahedral sites. We gain access to the subtleties of the local structure in the nominal \ce{LiMn2O4} nanoparticles by comparing STEM-EEL spectra with simulated spectra and reveal a core-shell structure where the shell is reduced through the presence of anti-site tetrahedral Mn. We make the surprising observation that changing the near-surface octahedral Mn oxidation state through dissolution of tetrahedral Mn has no effect on the OER catalytic activity, while tetrahedral Li occupancy does effect OER activity\cite{Baumung2019}. This makes clear that near-surface octahedral Mn valence, which should be affected by both tetrahedral ions in a similar way, is not a sufficient descriptor of OER catalytic activity. More differentiated measures than Mn valence or e$_g$ occupancy, which include both geometric and electronic structure, will be required to describe the active catalytic surface.
\section{Methods}

\subsection{Sample material}
The material used in this study is nominal \ce{LiMn2O4} powder from Sigma Aldrich with particle sizes \SI{<0.5}{\micro\meter}. The particles were investigated in their pristine state and after being used as the electrocatalyst for the OER. They were characterized with a variety of methods including:
\vspace{0.5cm}
\begin{itemize}%
    \setlength\itemsep{-0.4em}
    \item Scanning electron microscopy  (SEM)
    \item X-ray diffraction (XRD)
    \item X-ray absorption spectroscopy (XAS)
    \item Transmission electron microscopy (TEM) imaging
    \item Scanning transmission electron microscopy electron energy loss spectrometry (STEM-EELS)
\end{itemize}%
All studies were performed on material from a single powder batch.
\subsection{XRD measurements}
A $\theta-2\theta$ XRD scan has been carried out using a Bruker D8 diffractometer with \ce{Cu} $K_\alpha$ source in the angular range \mbox{$15^\circ\leq 2\theta < 12^\circ$} with a total measurement time of \SI{106}{\hour}. The \ce{LiMn2O4} powder was pressed with glue on a plastic sample plate. 
\subsection{Electrochemical measurements and TEM sample preparation}
Oxygen electrocatalysis experiments were conducted at room temperature using a Rotating Ring-Disk Electrode (RRDE) system set to detect either oxygen or manganese. Pristine \ce{LiMn2O4} powder \mbox{(10\,mg)} and acid-treated acetylene carbon black \mbox{(2\,mg)} (Alfa Aesar $\geq99.9\%$) were mixed by ultrasonication for 30 minutes in \mbox{2\,ml} tetrahydrofuran (VWR $\geq99.9\%$ stabilized) to form the catalyst ink. \SI{10}{\micro\l} of the ink, corresponding to \SI{50}{\micro\g} active material, was drop cast onto a \SI{4}{\milli\meter} diameter glassy carbon rotating disk electrode. The RRDE-electrode consists of the glassy carbon disk electrode and a concentric platinum ring electrode. The electrocatalytic experiments were performed in a \SI{0.1}{\molar} \ce{NaOH} electrolyte (\SI{1}{\molar} \ce{NaOH}, Merck TitriPUR, diluted with ultrapure water Milli-Q  $R\geq\SI{18.2}{\mega\ohm}$) with a Pt counter electrode and a saturated calomel reference electrode (ALS Japan Co Ltd., RE-2B). The electrolyte was purged with \ce{Ar} (5.0) gas from AirLiquide Alphagaz. The oxygen evolved at the disk was measured at the ring by potentiostatic reduction at \SI{0.4}{\V} vs. RHE, while the disk was swept at \SI{10}{\mV\per\s} for CV scans between \SIrange{1.25}{1.75}{\V} vs. RHE for ten cycles. The iR drop was corrected in post processing using the resistance obtained by electrochemical impedance spectroscopy measurements at open-circuit \cite{Baumung2019a}. Additional information of the electrochemical methods are presented in the SI\dag\\

TEM samples of pristine material have been prepared by drop casting an isopropanol-particle dispersion on a lacey carbon-copper TEM grid or by immersing the TEM grid in the dispersion. For post-mortem TEM analysis after OER, the particles were rinsed from the electrode with isopropanol and then drop-cast on a TEM grid. For a control experiment, pristine \ce{LiMn2O4} particles were stored for 5 minutes in the NaOH electrolyte, then washed with isopropanol and transferred by drop-casting to a TEM grid.
\subsection{Transmission electron microscopy, EELS, and Mn valence determination}
High resolution TEM imaging and STEM-EELS were performed with an image Cs corrected FEI Titan environmental-TEM at \SI{300}{\kV} acceleration voltage, equipped with a Quantum 965ER Gatan Image Filter and X-FEG by FEI. High resolution TEM images and STEM-EEL spectrum images have been recorded from 12 pristine particles and 8 OER-cycled particles with diameters varying from \SIrange{15}{250}{\nm} and \SI{1}{nm} step sizes. Dual EEL spectra were recorded with \SI{0.1}{\eV\per\channel} dispersion simultaneously from the low-loss region \mbox{\SIrange[range-phrase = --]{0}{180}{\eV}} as well as from the O-$K$ and Mn $L_{2,3}$ edge region (\SIrange[range-phrase = --]{480}{680}{\eV}) for each scanning point. The zero-loss peak (ZLP) exhibits a FWHM of \SI{1.3}{\eV}.\\

The EEL spectra reflect transitions among the Mn $3d$ and O $4s$ and $4p$ levels and thus provide detailed insights into the electronic structure. The O pre-peak observed at \SI{530}{\eV} is sensitive to Mn $3d$ occupancy and the hybridization strength of O $2p$ and Mn $3d$ orbitals, while the O $K$ peaks, centered around \SI{545}{\eV} reflect transitions to the O $1p$ state mixed with the Mn $4sp$ band above the Fermi level \cite{DeGroot1989}. The Mn $L$ edge between \SI{640}{\eV} and \SI{655}{\eV}, which reflects transitions from Mn $2p_{3/2}$ and $2p_{1/2}$ levels to unoccupied Mn $3d$ states, has been reported to be sensitive to the $3d$ occupancy, i.e. Mn oxidation state \cite{Leapman1982, Sparrow1984}.\\
It is often useful to use the Mn valence as a simple proxy for the complex electronic structure. An easily applicable and qualitative methods uses the Mn $L_3/L_2$ intensity ratio which is sensitive to the $3d$ occupancy  \cite{Leapman1982,Sparrow1984, Kurata1993,Varela2009} and has been used to determine Mn valence in a variety of transition metal-based materials \cite{Graetz2004, Varela2009, Tan2012, Zhang2010}. However, it is known to fall short in capturing the oxidation state in some manganese oxides \cite{Botton1995} and can only determine Mn valence to within 0.3 even when optimized \cite{Riedl2006}. Furthermore, available reference data include perovskites \cite{Varela2009, Tan2012} or Mn oxides with different structures \cite{Zhang2010, Graetz2004, Tan2012} but not explicitly transition metal oxide spinels. Thus, applying these calibrations it has to be assumed that crystal structure is not important. However the  reported trends with changing Mn oxidation state are distinct.\\

A second method based on the O $K$ to Mn $L$ peak distance has been calibrated using a wide variety of manganese oxides with oxidation states between $3^+$ and $4^+$, and is claimed to be more sensitive to Mn valence than the Mn $L_3/L_2$ intensity method\cite{Zhang2010}. In fact, in a recent careful EELS study on \ce{LiMn2O4} and Li$_2$Mn$_2$O$_4$ \cite{Erichsen2020}, this energy difference method produced more reproducible results than the peak intensity method. Thus this calibration method has been applied to calculate Mn valence maps from the EEL spectrum images. Since the particles were neither aligned in a specific zone axis nor scanned with atomic resolution during the TEM-EELS measurements, the EEL spectra and Mn valences reflect the average oxygen and manganese states in the unit cell. But differences as large as 0.3 in the valences predicted by the two methods as well as deviations from the expected valences by as much as 0.3 suggest that it is still best used as only a qualitative measure of Mn valence. We note here that there are other methods that can provide a more accurate and precise determination of the Mn valence when there is sufficient energy resolution and calibration in the EEL spectra \cite{Zhang2010, Livi2012}.\\

However, instead of focusing on Mn valence alone, we will also compare our measured EEL spectra to simulated spectra for different \ce{LiMn2O4}-based structures. This so-called "fingerprinting" approach allows us to take advantage of the multifaceted information provided about the electronic structure by the EEL spectra that is not captured by the simple proxy of Mn valence.
\subsection{Simulation methods}
Density functional theory calculations applying the local hybrid density functional PBE0r \cite{Sotoudeh2017, Eckhoff2020} were carried out using the Car-Parrinello Projector Augmented-Wave (CP-PAW) code (version from September 28, 2016) \cite{Bloechl1994, CP-PAW}. The general settings of the calculations were taken from our previous study \cite{Eckhoff2020}. The plane wave cutoff was increased to $35\,E_\mathrm{h}$ for the auxiliary wave functions and to $140\,E_\mathrm{h}$ for the auxiliary densities. The $\mathbf{k}$-point grid was enlarged to $3\times3\times3$ for the \ce{LiMn2O4} unit cell. Comparable $\mathbf{k}$-point densities were chosen for the other compositions studied here. The number of unoccupied Kohn-Sham orbitals was increased to 280. These improved settings yield accurate densities of states even for energies up to 35 eV above the Fermi level.\\

The densities of states were used to calculate the EEL spectra at $0\,$K \cite{Wooten1972, Egerton2007, Seah1979}.
The model for the pristine bulk structure of \ce{LiMn2O4} used in the simulations was taken from Ref. \citenum{Akimoto2004}. The cubic spinel lattice constant was fixed at the experimentally determined value and the atomic positions inside the unit cell were optimized. Details of the Mn$^\mathrm{3+}$/Mn$^\mathrm{4+}$ distribution and their spin states as well as the electronic density of states (DOS) can be found in our previous studies \cite{Eckhoff2020, Eckhoff2020b}. This structure was taken as basis for the oxygen and manganese defect model structures, where the cubic spinel lattice constant was again fixed to the experimental value of the \ce{LiMn2O4} bulk structure while the atomic positions were reoptimized. Note that the absolute energies of the simulated EEL spectra are arbitrary and may be shifted to fit the experimental spectra. The simulations have a higher energy resolution than the experiment due to the energy spread of the TEM electron gun (FWHM$ =  $\SI{1.3}{\eV}). For better comparison, the simulated scattering intensities were broadened by convolution with the ZLP as recorded without a sample in the beam.\\

Further details about the experimental and simulation methods are provided in the SI\dag, including all PBE0r geometry optimized structures used for the calculations in this work.
\section{Results} 
\subsection{Pristine particles}
\begin{figure}[h]
	\centering
	\subfloat[]{%
		\includegraphics[width=0.45\columnwidth]{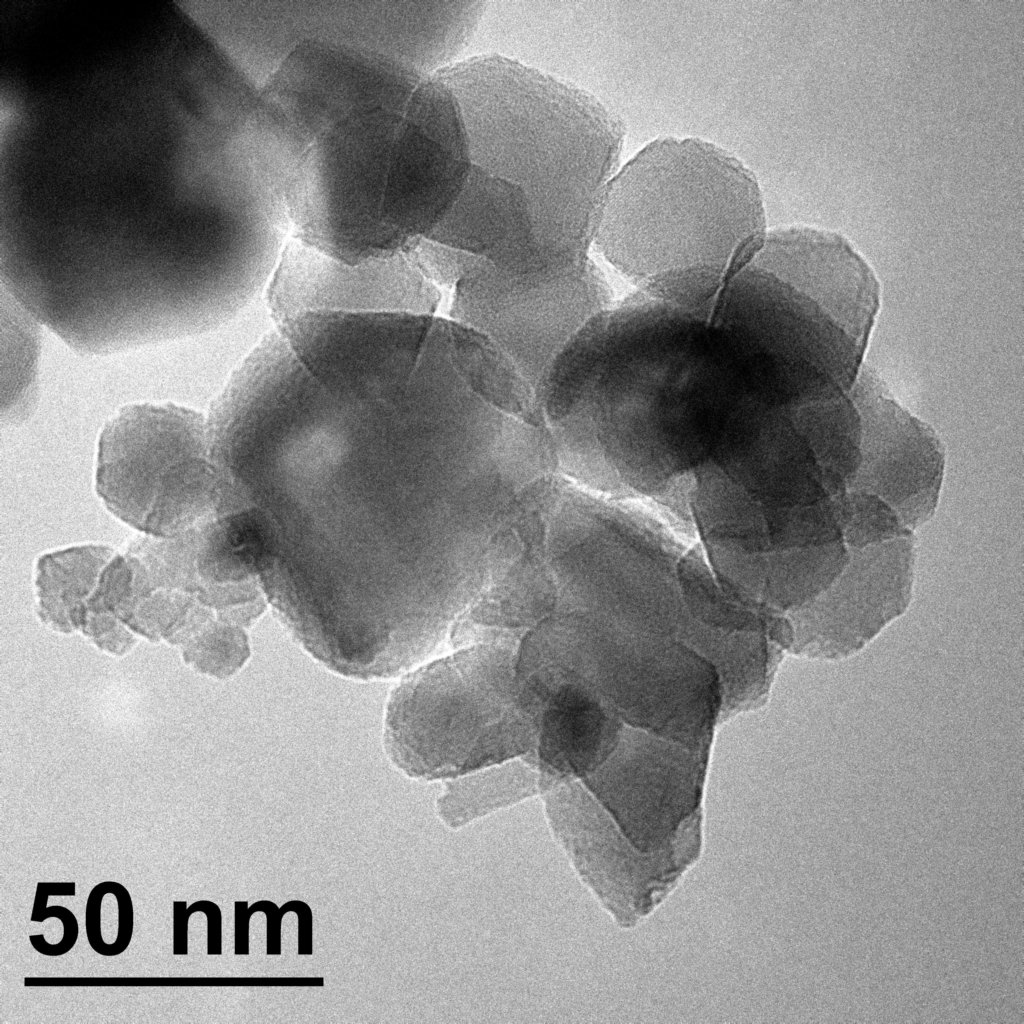}}
	\vspace{0.01\columnwidth}
	\subfloat[]{%
		\includegraphics[width=0.45\columnwidth]{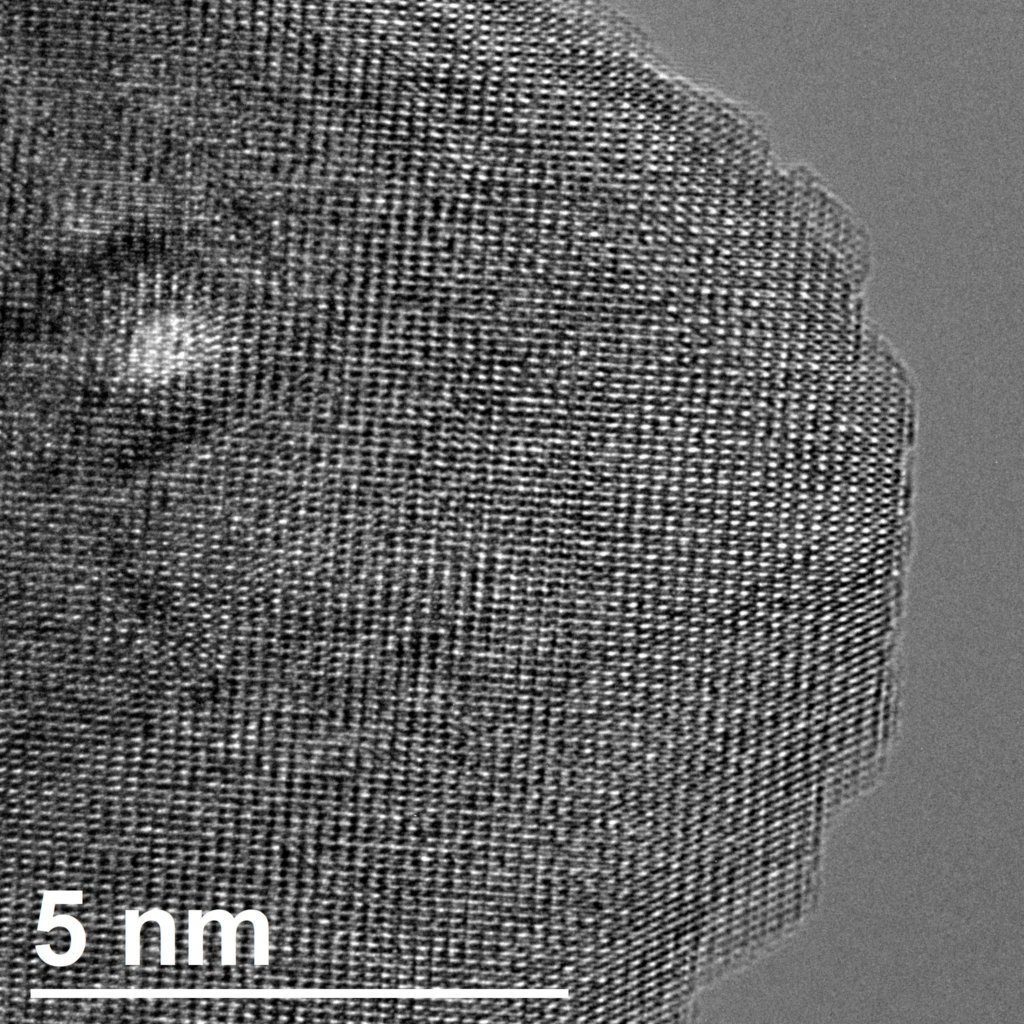}}
	\vspace{-0.5cm}
	\hspace{0.05\textwidth}
	\subfloat[]{%
		\includegraphics[width=0.95\columnwidth]{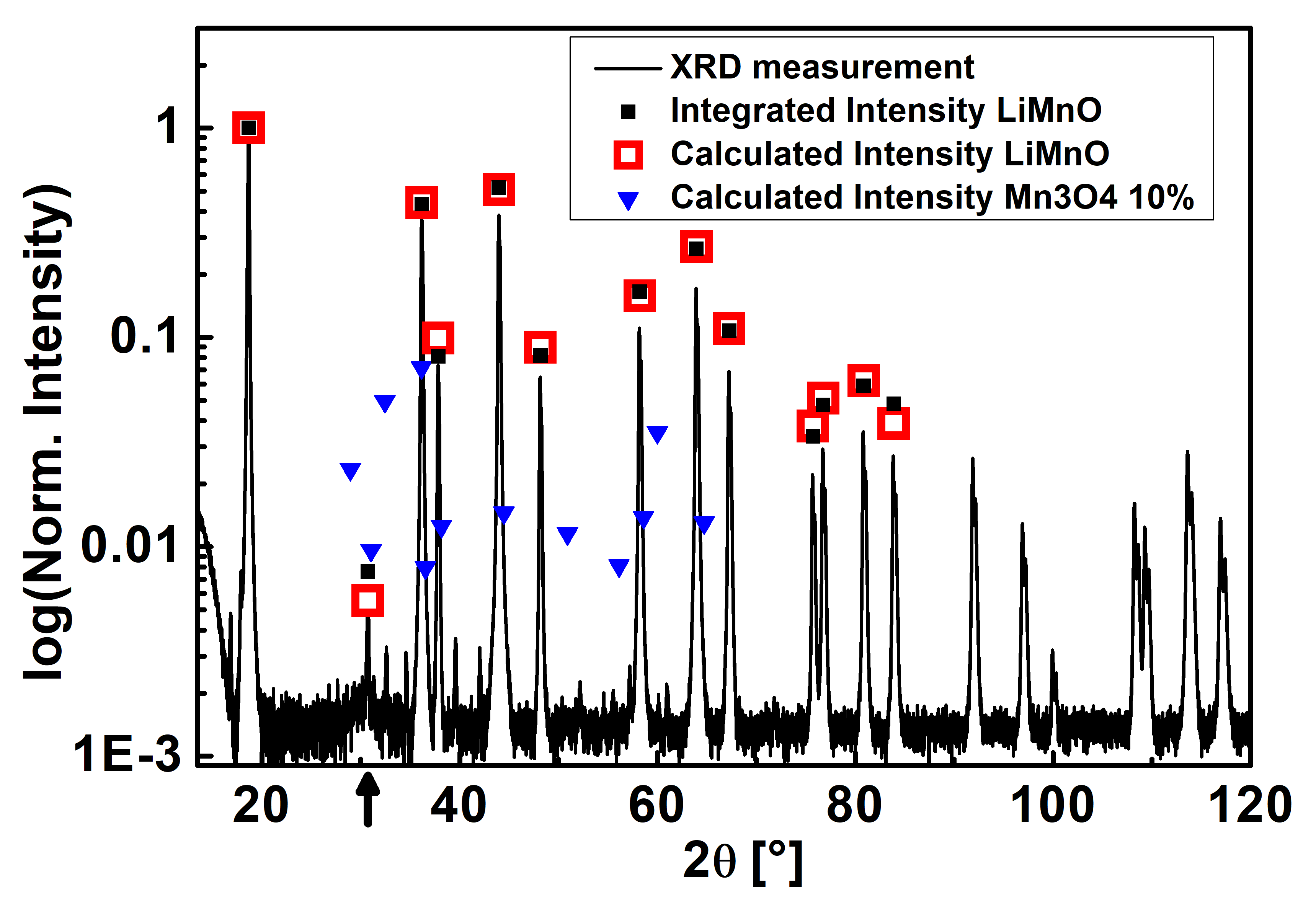}}
	\vspace{-0.3cm}	
	\caption{Pristine \ce{LiMn2O4} powder characterization. (a) TEM overview image showing facetted particle morphology and (b) HR-TEM image showing the single crystal nature of a typical particle. (c) $\theta-2\theta$ XRD pattern from the pristine powder with integrated peak intensities (black dots), calculated peak intensities for \ce{LiMn2O4} (red squares), and expected peak intensities for Mn$_3$O$_4$ present with 10\% volume share (blue triangles).}
	\label{fig:XRDandoverview}
\end{figure}

SEM and TEM studies of the pristine particles show that they are partially facetted single crystals with diameters ranging from \SIrange{10}{400}{\nm} and a mean diameter of \SI{44(14)}{\nm} (Fig.\ref{fig:XRDandoverview}a)\cite{Baumung2019a}. The larger particles show more facetted shapes than the smaller particles and are mainly truncated octahedrons and some truncated rhombic dodecahedrons. High resolution TEM images confirm that the crystalline structure reaches to the surface but also shows atomic scale surface roughness (Fig.\ref{fig:XRDandoverview}b). The main peaks in a $\theta-2\theta$ X-ray pattern of the pristine particles (Fig.\ref{fig:XRDandoverview}c) can be indexed as cubic spinel with $a=\SI{8.234(4)}{\angstrom}$, which is in good agreement with literature values for \ce{LiMn2O4} (e.g. \citet{Thackeray1997, Xia2001}). The peak intensities calculated for the structure using Vesta \cite{Momma2011} (marked with red open squares) compare reasonably well with the integrated experimental peak intensity (black points) after the background had been accounted for, although the measured (220) peak intensity (marked with an arrow in Fig.\ref{fig:XRDandoverview}c) exceeds the calculated one. The effect of replacing Li with Mn on the $8a$ tetrahedral sites as simulated using Vesta \cite{Momma2011} leads to a strong increase in the (220) peak intensity, suggesting the presence of some tetrahedrally coordinated Mn in the nanoparticles (see also \citet{Cady2015}). Applying the Scherrer equation to the \ce{LiMn2O4} pattern, an averaged crystallite size of \SI{61(9)}{\nm} is found. This value is significantly larger than the mean diameter obtained from TEM and SEM micrographs in this study and in our previous publications \cite{Baumung2019a} and can be attributed to the fact that Scherrer broadening is weighted by the particle volume.\\%

Additional unmarked peaks found in the X-ray pattern can be indexed as two cubic phases with $a=\SI{8.603(3)}{\angstrom}$ and $a=\SI{9.123(4)}{\angstrom}$ with a volume fraction \SI{<1}{\%}. It should be explicitly mentioned that the unmarked peaks are not consistent with tetragonal Mn$_3$O$_4$, which was reported to occur on \ce{LiMn2O4} particles and thin film surfaces  \cite{Tang2014,Tang2014b,Amos2016,Gao2019,Liu2019}. The expected peak intensities for a 10\% volume share of \ce{Mn3O4} are indicated by blue triangles. In fact, by requiring that the 4 most intense peaks of the tetragonal \ce{Mn3O4} X-ray pattern should exceed twice the measurement noise level to be detected, we can conclude that the volume fraction of tetragonal \ce{Mn3O4} is less than \SI{1}{\%}. We can also rule out the presence of epitaxial \ce{Mn3O4} on the surface of the \ce{LiMn2O4} nanoparticles due to the large lattice mismatch and since no peaks were found that could be associated with strained \ce{Mn3O4}.\\

\begin{figure}[h]%
	\centering
	
	\subfloat[]{%
		\includegraphics[width=0.9\columnwidth]{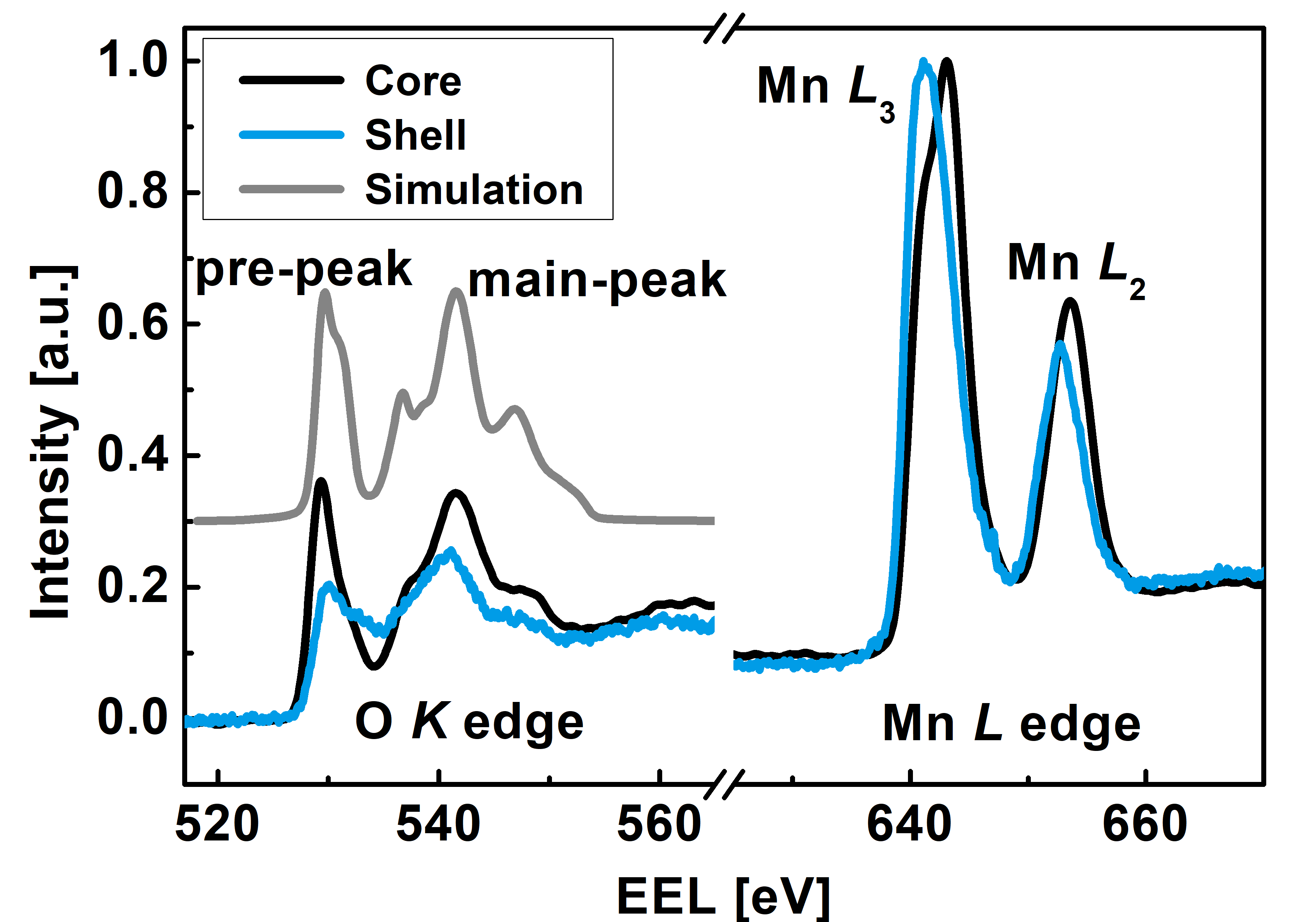}}
	\vspace{-0.45cm}
	\subfloat[]{%
		\includegraphics[width=0.9\columnwidth]{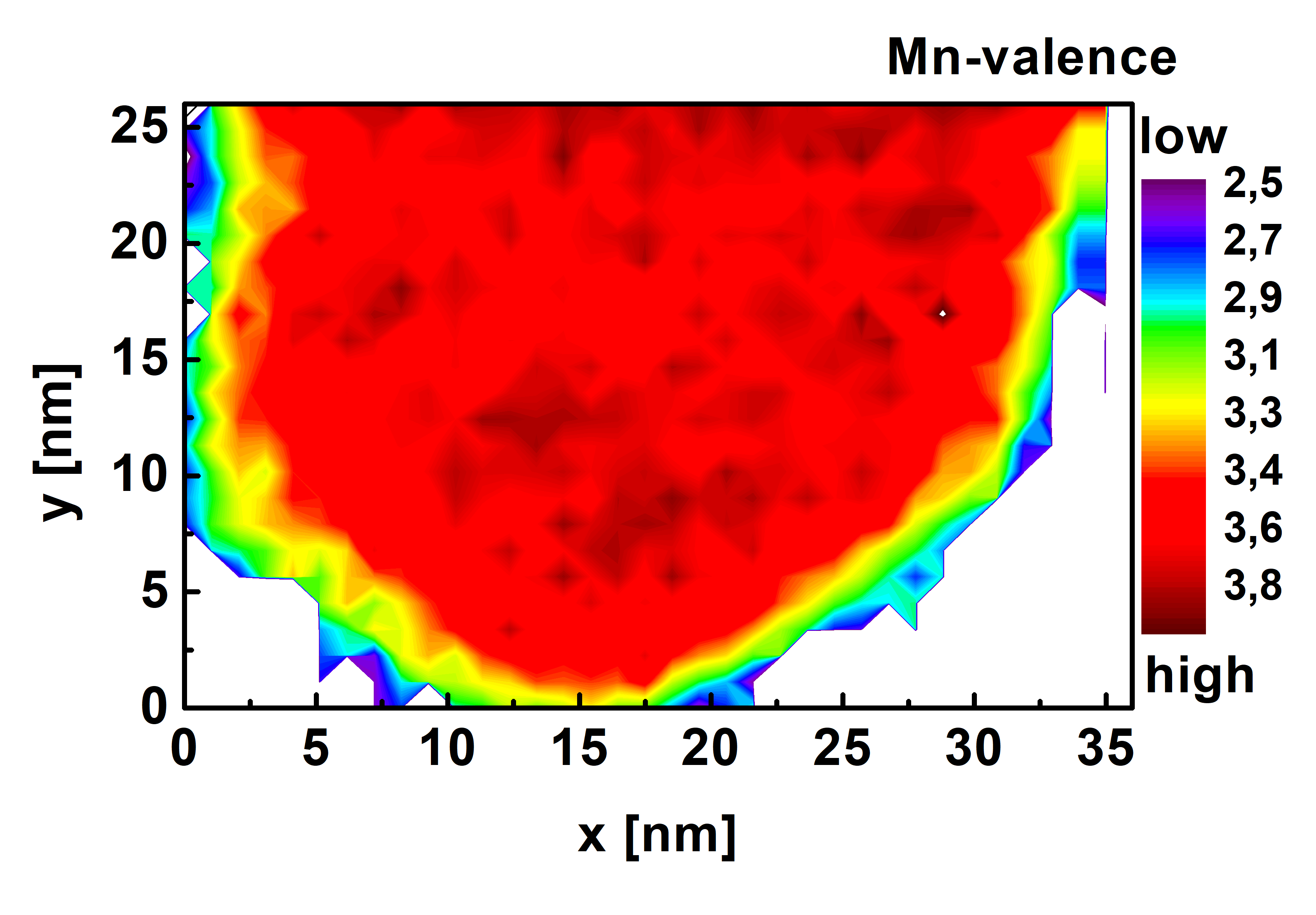}}
	\vspace{-0.25cm}
	\caption{EELS study of pristine \ce{LiMn2O4} particles. (a) Averaged O $K$ and Mn $L$ spectra from the core and shell regions of one particle. The intensities are normalized by the Mn $L_3$ height. A simulated O $K$ spectrum is included for comparison, that was scaled in intensity and shifted in energy to the core spectrum. The Li $K$ edge is marked with an arrow. (b) A Mn valence map from a pristine nanoparticle with maximum relative thickness $t/\lambda=0.36$. The valence has been calculated using the O $K$ pre-peak to Mn $L_3$ peak energy difference.}
	\label{fig:Valencemap}
\end{figure}

STEM-EEL spectra and maps were obtained from 12 different pristine particles that were selected for their TEM transparency and position over holes in the TEM grid carbon layer. The particles have different diameters with relative maximum thicknesses $t/\lambda$ between 0.18 and 0.81. Typical portions of the spectra are shown in Fig. \ref{fig:Valencemap}a. The center and edge spectra in both the O $K$ and Mn $L$ energy ranges are quite different from each other for all 12 particles (Fig. \ref{fig:Valencemap}a), suggesting that the particles have a core-shell structure. The shell spectra show some systematic differences from the core spectra for all 12 particles: decrease of the O pre-peak intensity compared to the second O $4sp$ peak; shift to lower energies of the O $4sp$ peak; decrease of the O $4sp$ peak height compared to the Mn $L_3$ peak at  \SI{640}{\eV}; shift to lower energies of the Mn $L_{2,3}$ peak maxima; increase of the $L_3/L_2$ intensity ratio from $L_3/L_2=2.13(5)$ to $L_3/L_2=2.63(17)$; less pronounced low energy shoulder on the Mn $L_3$ peak; and a slight decrease of the Li $K$ edge at \SI{60}{\eV} (see SI Fig.3\dag). No evidence for impurities could be found in the EEL spectra, meaning that stoichiometry changes must be the cause of the observed core-shell structure. The Li $K$ edge is contained in the low loss spectra next to the Mn $M$ edge at \SI{60}{\eV} \cite{Erichsen2020}. This low loss regime is displayed in Fig.3 of the SI\dag\  for both the shell and core regions and indicate that the Li/Mn ratio is slightly smaller near the surface than in the bulk.\\

A simulated transition spectrum for the O $K$ region of \ce{LiMn2O4} based on DFT calculations is also included in Fig.\ref{fig:Valencemap}(a). It agrees quite well with the experimental core spectrum concerning the energy difference between the main and pre O $K$ peaks, supporting the results of the X-ray diffraction study (Fig.\ref{fig:XRDandoverview}) that the majority component of the particles is \ce{LiMn2O4} and lending validity to the simulation method. However, a comparison of the O $K$ pre-peak to main peak intensity ratio, which is slightly higher for the measured core spectrum than the simulation, as well as a difference in pre-peak to main peak distance suggest that Mn in the core region of the particles is oxidized relative to the simulated defect-free \ce{LiMn2O4} \cite{Varela2009, DeGroot1989}.\\

Both the experimental core spectrum and the simulated spectrum of \ce{LiMn2O4} differ significantly from the shell spectrum. The differences in EEL-spectra between the core and shell regions can be expressed in simplified terms as average Mn valence using either the O $K$ pre-peak to Mn $L_3$ peak energy difference \cite{Zhang2010} or the $L_3/L_2$ intensity ratios  \cite{Varela2009}. Both methods clearly indicate that Mn near the surface is reduced. A Mn valence map from a representative particle is shown in Fig.\ref{fig:Valencemap}b and clearly exhibits a reduced surface layer with a thickness of about \SI{3}{\nm}. Given the map step size of \SI{1}{\nm}, we conclude that the reduced state extends to at least within \SI{1}{\nm} of the surface. In fact, a core-shell model with a \SI{3}{\nm} thick shell is found to describe all 12 particles that were studied giving an average valence of $3.82(6)+$ in the core region and of $2.93(21)+$ in the shell region. No direct effect of the particle relative thickness on the evaluated Mn valence was detected. We note that the measured particle core Mn valence value is significantly larger than the expected effective Mn valence of 3.5+ for \ce{LiMn2O4}, supporting the idea that the particle cores are oxidized \\%

In order to identify the possible structural changes that account for reduction in the shell region and the oxidation in the core region, the O $K$ transition spectra were simulated for \ce{LiMn2O4} containing three different types of defects. These are excess Mn, which was simulated by filling the 8a tetrahedral sites of the $\lambda$-MnO$_2$ structure partially with Mn (and Li), oxygen vacancies, and a change in the Li content. Manganese residing on tetrahedral sites has been reported in previous experimental publications, \cite{Tang2014,Tang2014b,Amos2016,Gao2019,Liu2019}, oxygen vacancies at transition metal oxide surface regions are widely discussed \cite{Ganduglia-Pirovano2007, Huang2011} and the Li content in \ce{LiMn2O4} can be changed easily and directly influences the Mn valence \cite{Thackeray1997}. The defective structures and energy levels were calculated using DFT. Note that the small simulation cell size of one unit cell places strong restrictions on the possible defect concentrations. The EEL scattering intensities were obtained from the calculated energy levels and the scattering cross-sections.\\%

\begin{figure}[!h]%
	\centering%
	\subfloat[]{%
		\includegraphics[width=0.9\columnwidth]{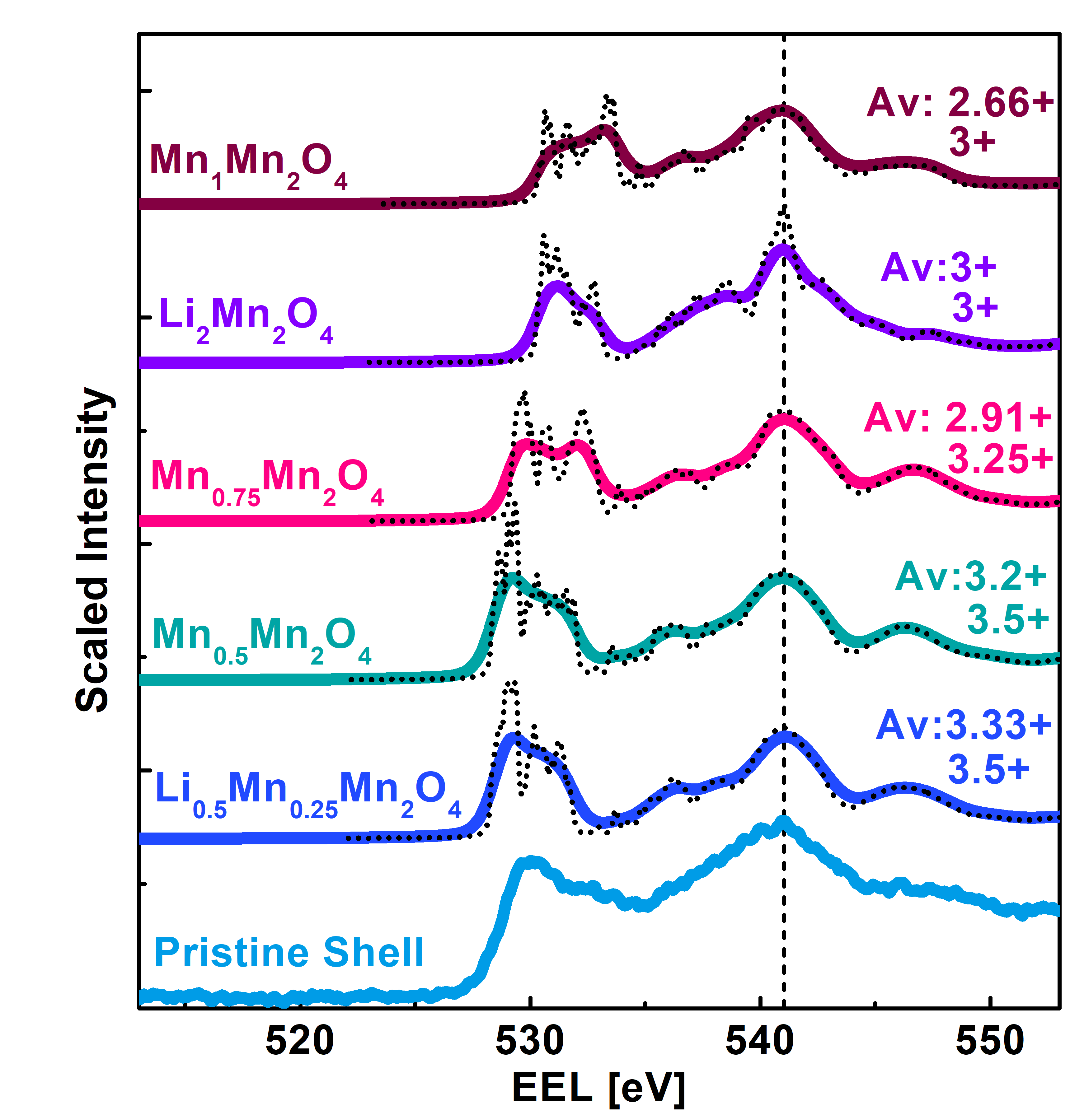}}
	\vspace{-0.43cm}
	\subfloat[]{%
		\includegraphics[width=0.9\columnwidth]{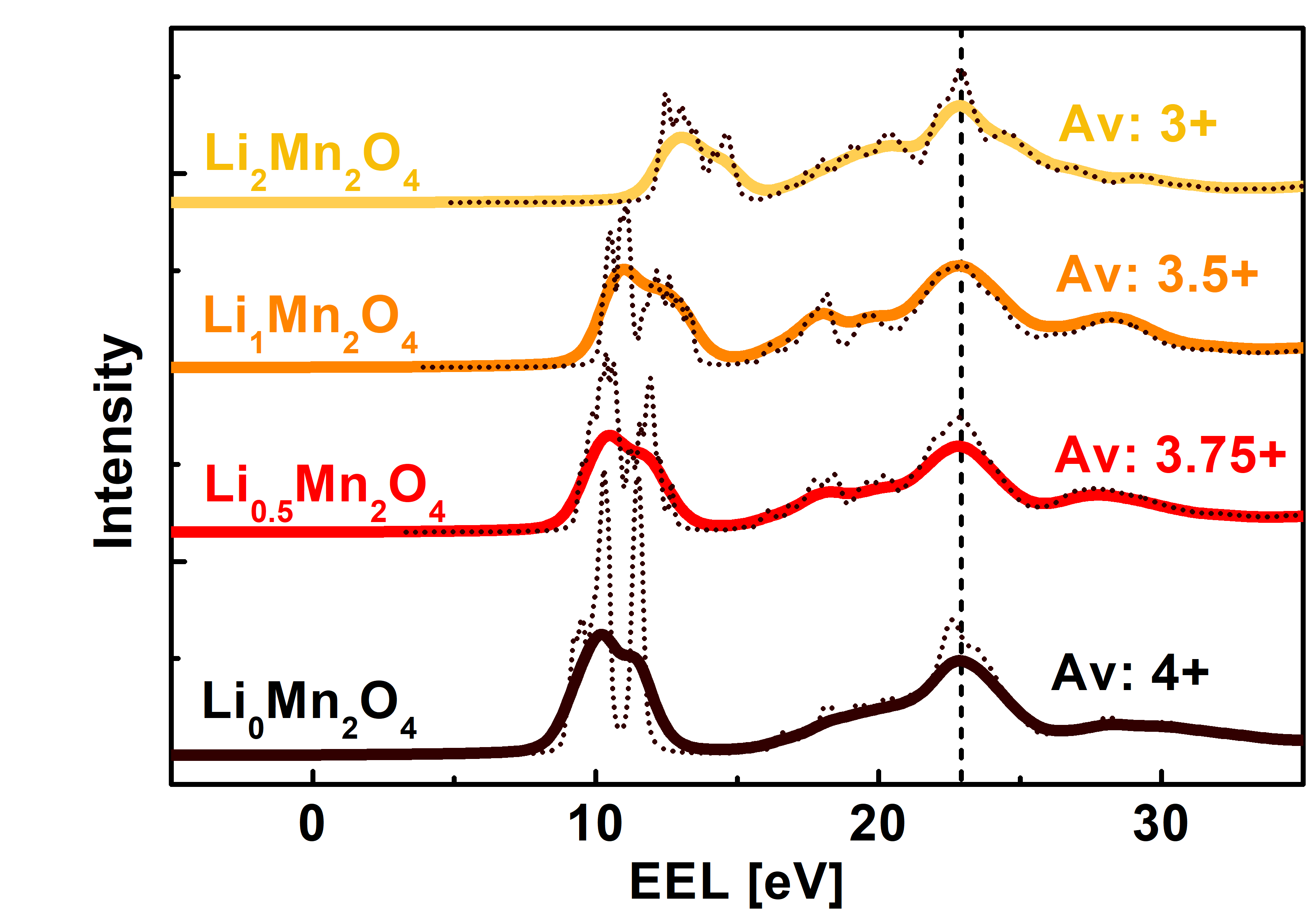}}
	\vspace{-0.25cm}
	\caption{Effect of Mn (a) and Li occupancy (a,b) of tetrahedral sites on the EEL O $K$ edge. The energy axis of the calculated spectra is shifted such that the O $4sp$ peaks align with the experimental shell spectra shown below (dashed vertical lines). Dotted curves show the scattering cross sections resulting from the simulations; solid lines show the spectra after convoluting the simulations with the experimental energy uncertainty. The nominal average Mn valences are indicated for each spectrum. For Mn$^{2+}$ on tetrahedral sites the valence of octahedral Mn is also included.}
	\label{fig:Mn-Defects}
\end{figure}%
Clear trends can be observed with decreasing octahedral Mn valence: the O pre-peak onset shifts to higher energies; the O pre-peak amplitude decreases relative to the O $4sp$ peak; and, the intensity of the high energy shoulder on the convoluted O pre-peak increases relative to the pre-peak intensity. Using the pre-peak onset energy and its relative intensity to the main peak as criteria, we find that the best agreement to the experimental shell spectrum in the O$K$ region is obtained for an octahedral Mn valence of around $3.25+$. Changing the Mn and Li occupancies on the tetrahedral sites while keeping the octahedral Mn valence constant seems to mostly effect the position and intensity of the high energy shoulder on the O pre-peak.  For instance, comparing the simulated spectra for \ce{MnMn2O4} and \ce{Li2Mn2O4} indicates that both Mn and Li are needed on the tetrahedral sites to reproduce the experimental EEL spectrum. We therefore consider the composition [Li$_{0.5}$Mn$_{0.5}$]$_\text{T}$Mn$_2$O$_4$, with full 8a occupancy and 3.25+ octahedral Mn valence, as a good candidate structure for the shell region. The average valence of this composition is 3.0+, quite close to the measured value of 2.9+. Additional spectra with varying Mn compositions are plotted in SI\dag\  Fig.1.\\

The effect of different Li contents on the EEL spectra can be seen in Fig.\ref{fig:Mn-Defects}b for the case of Li$_x$Mn$_2$O$_{4}$ with $x$ between 0 and 2. Literature lattice parameters have been used to set up the simulations, including the change to tetragonal symmetry for $x\geq 1$ \cite{Ohzuku1989}. Increasing the Li content leads to the following evolution of the O $K$ edge: the main O $K$ peak clearly shifts to higher energies; the pre-peak intensity decreases with respect to the O $4sp$ feature and is broadened. Increasing the Li content does move the simulated EEL-spectra closer to the experimentally observed spectrum from the shell region of the nanoparticles, but even the maximum concentration of Li$_2$Mn$_2$O$_{4}$ was not able to fully match the experiments. Furthermore, we know from the Li $K$ edge, that the Li content in the shell region is less than in the core region, making a Li excess very unlikely as an explanation for the shell structure. In contrast, the experimental core region spectra were best fit with a Li composition between $x$ of $1/2$ and $1$ under the assumption that only the Li content varies. Interpolating the ratio of the pre-peak to main peak intensity from the simulations shows that the experimental core spectrum can be reasonably reproduced with the composition Li$_{0.86}$Mn$_2$O$_{4}$. This composition has a nominal valence of $3.57+$ which is quite a bit smaller than the valence calculated directly from the spectrum, which is between $3.82(6)+$ and $3.93(6)+$, depending on the method used. The measured lattice constant of the nanopowder is with $a=\SI{8.234(2)}{\angstrom}$, which somewhat smaller than reported values \cite{Xia2001, Thackeray1997, Hunter1981, Ohzuku1989} and will be discussed in the context of oxidation later in the paper.\\

The effect of oxygen vacancies on the EEL spectra has also been investigated by removing O atoms from the \ce{LiMn2O4} structure, reaching a concentration of LiMn$_2$O$_{3.5}$. The unit cell structure was then relaxed for fixed lattice constant using DFT before the energy levels were calculated. A summary figure of the effect of O vacancies on the EEL-spectra can be found in Fig.2 of the SI\dag. Increasing the O vacancy concentration results in the following changes in the O region of the EEL-spectra: a slight pre-peak shift to higher energies; a slight broadening of the pre-peak; and, the relative intensity of the pre-peak to $4sp$ peak does not change significantly. All in all, introducing O vacancies does not provide an improvement in the agreement between the simulated spectra and the measured O $K$ spectra from the shell region.
\subsection{OER - electrochemical measurements}
\begin{figure} [h]
	\centering
	\subfloat[]{%
		\includegraphics[width=0.6\columnwidth]{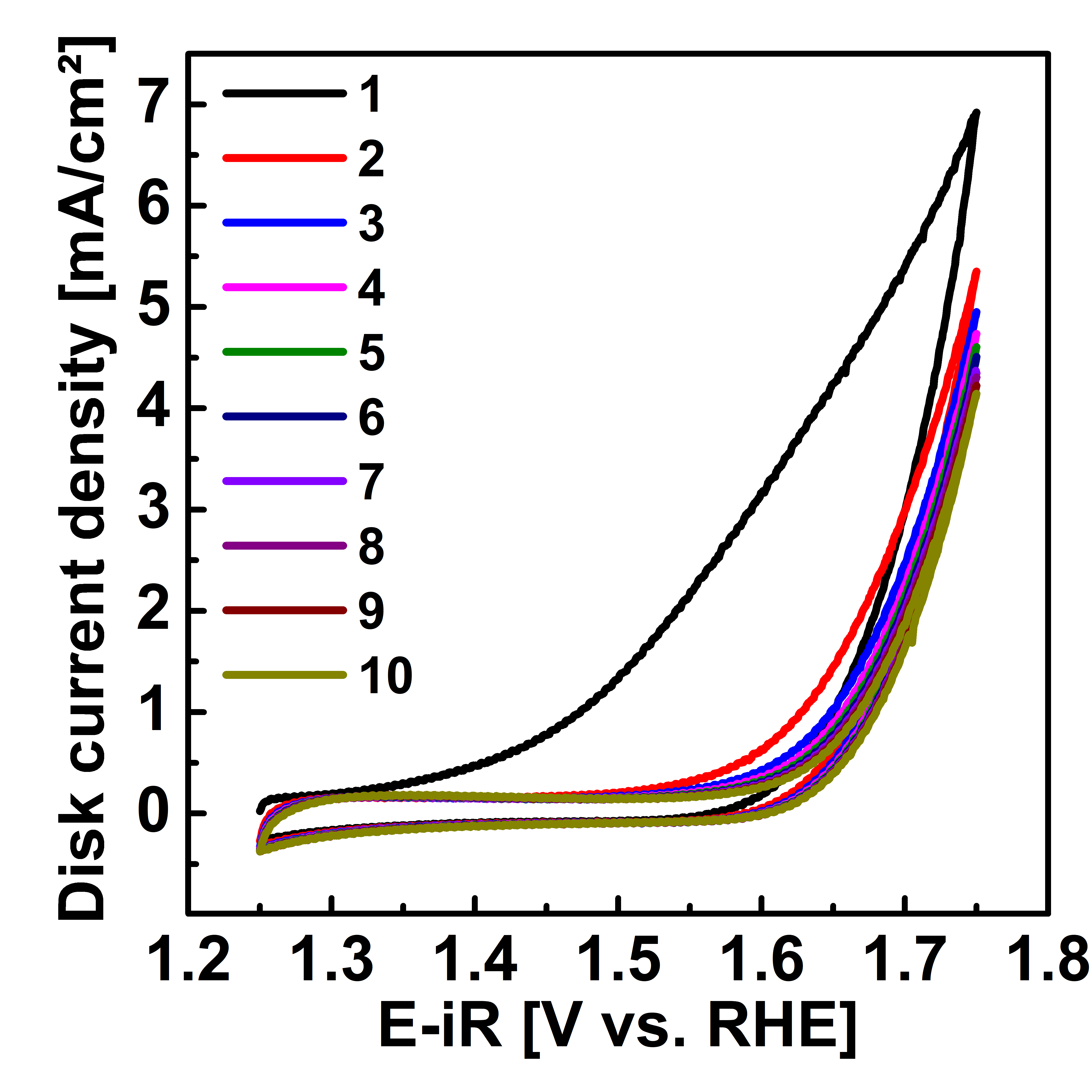}}
	\vspace{-0.43cm}
	\subfloat[]{%
		\includegraphics[width=0.6\columnwidth]{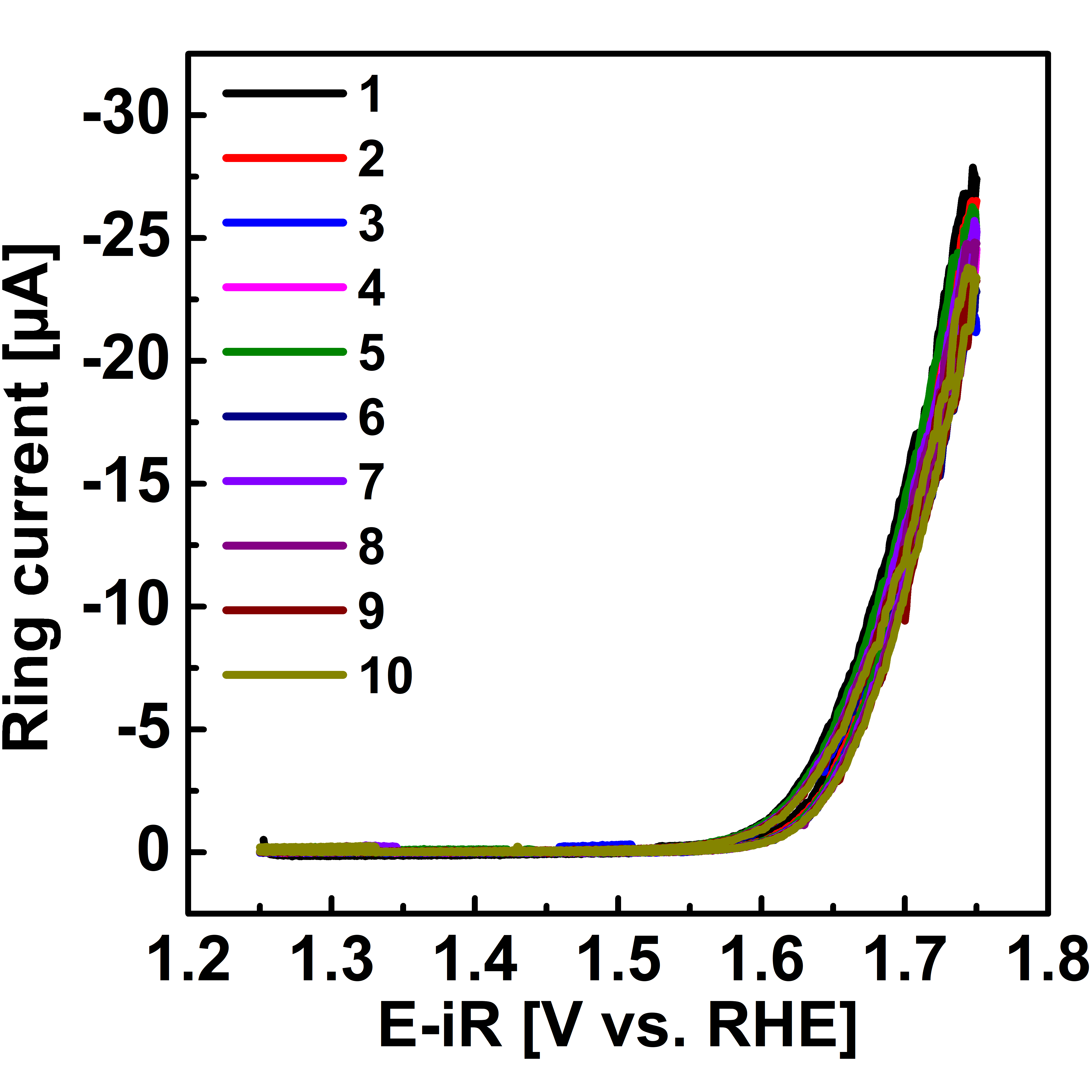}}
	\vspace{-0.2cm}
	\caption{\ce{LiMn2O4} RRDE electrocatalysis of the OER. (a) Disk current normalized to the geometric disk area for the first ten cycles of OER in NaOH (pH13) with an electrode ink containing the pristine powder. (b) The corresponding ring current for oxygen detection at 0.4~V  vs. RHE.}
	\label{fig:DiscCurrent}
\end{figure}

In order to correlate the near-surface structure with OER activity, oxygen electrocatalysis experiments were conducted using the nanoparticles as the active material. Their near-surface structure was subsequently investigated with TEM imaging and STEM-EELS. Ten cyclic voltammetry (CV) cycles were performed at the disk in a RRDE set-up with the ring current set for oxygen detection  (Fig.\ref{fig:DiscCurrent}). A large disk current density is measured in the first three positive-going CV scan and then settles into steady state behavior, with the geometrical current density clearly rising strongly above $E=\SI{1.6}{\V}$ in later cycles, which can be fit with an exponential function, thus indicating kinetically-limited reactivity in these cycles (Fig.\ref{fig:DiscCurrent}a). The corresponding ring current qualitatively reveals the evolution of oxygen above $E=\SI{1.6}{\V}$, and shows steady state behavior during the ten cycles (Fig.\ref{fig:DiscCurrent}b). We conclude that the catalytic evolution of oxygen  probed by the ring is not affected by the drastic changes in the disk current.\\

This behavior is consistent with our previous work under the same electrochemical conditions \cite{Baumung2019a, Kohler2017}, where we identified manganese loss as the cause of the large disk currents in the first few cycles, while the oxygen evolution remained constant. The current during the tenth cycle is mainly caused by double layer capacitance and OER at voltages greater \SI{1.6}{\volt} vs. RHE. We thus subtracted the tenth cycle cycle from the early cycles to isolate the currents due to manganese redox reactions (SI\dag\  Fig.10). A charge of \SI{3.5}{\milli\coulomb} passed before any oxygen is evolved from the sample and a charge of \SI{16.3}{\milli\coulomb} is passed during the first three cycles, which caused significant changes to the shell of the \ce{LiMn2O4} particles.\\

\subsection{Post mortem EELS of cycled particles}
\begin{figure}[h]
	\centering
	\subfloat[]{\includegraphics[width=0.44\columnwidth]{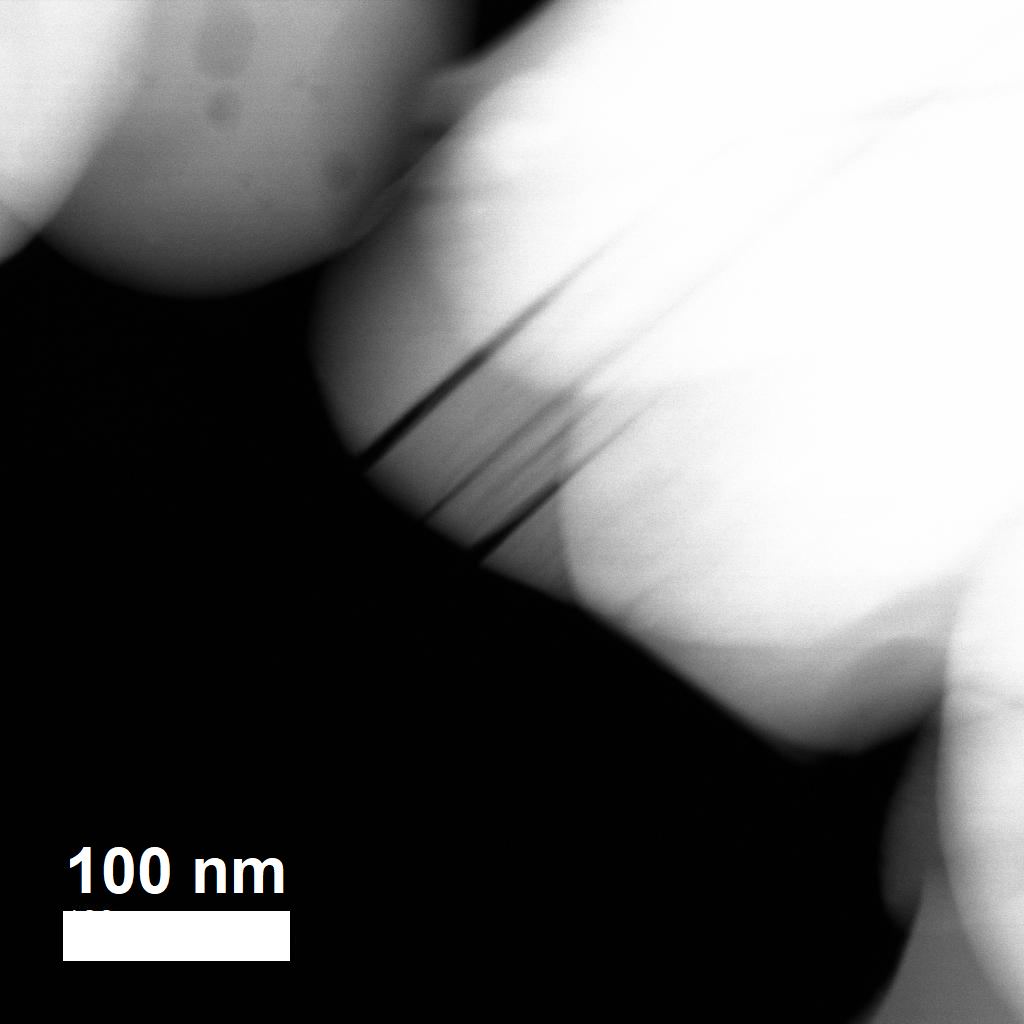}}
	\hspace{0.01\columnwidth}
	\subfloat[]{%
		\includegraphics[width=0.44\columnwidth]{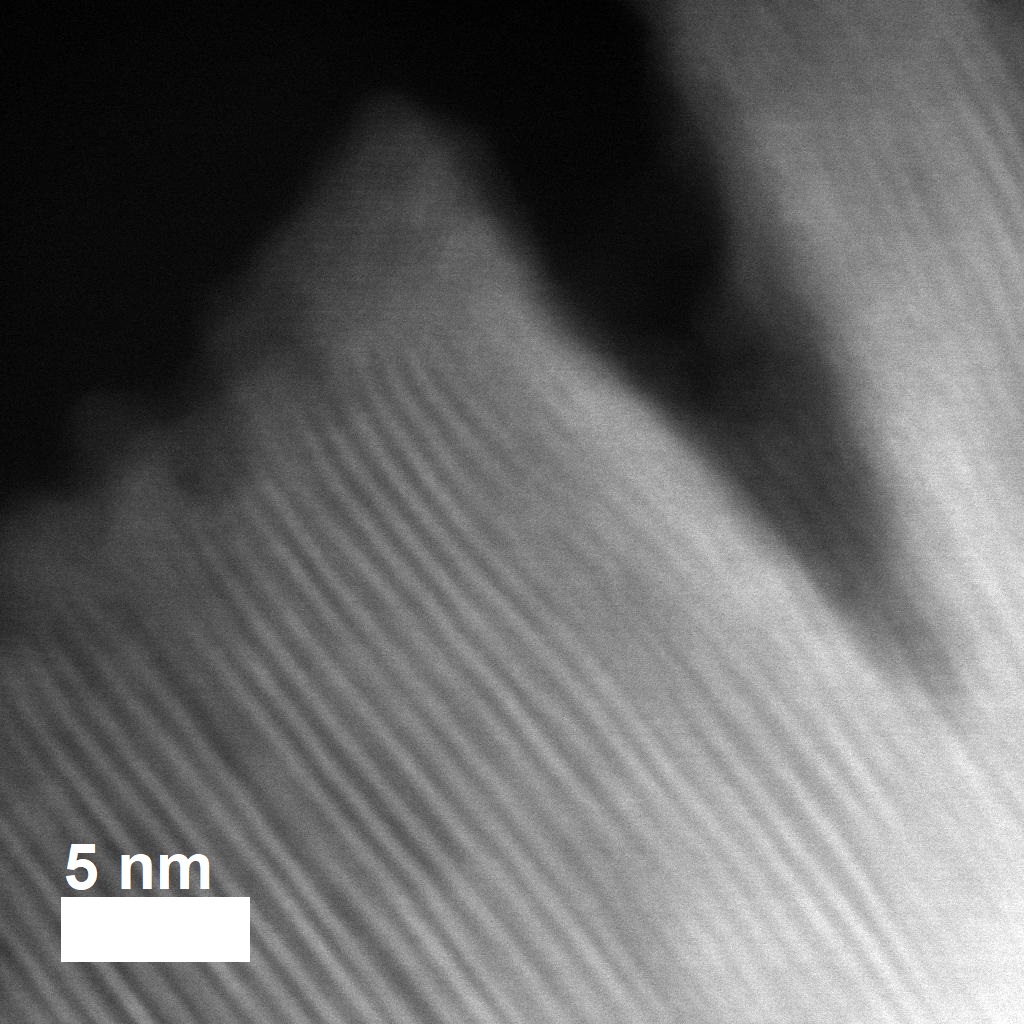}}
	\vspace{-0.4cm}
	\subfloat[]{%
		\includegraphics[width=0.98\columnwidth]{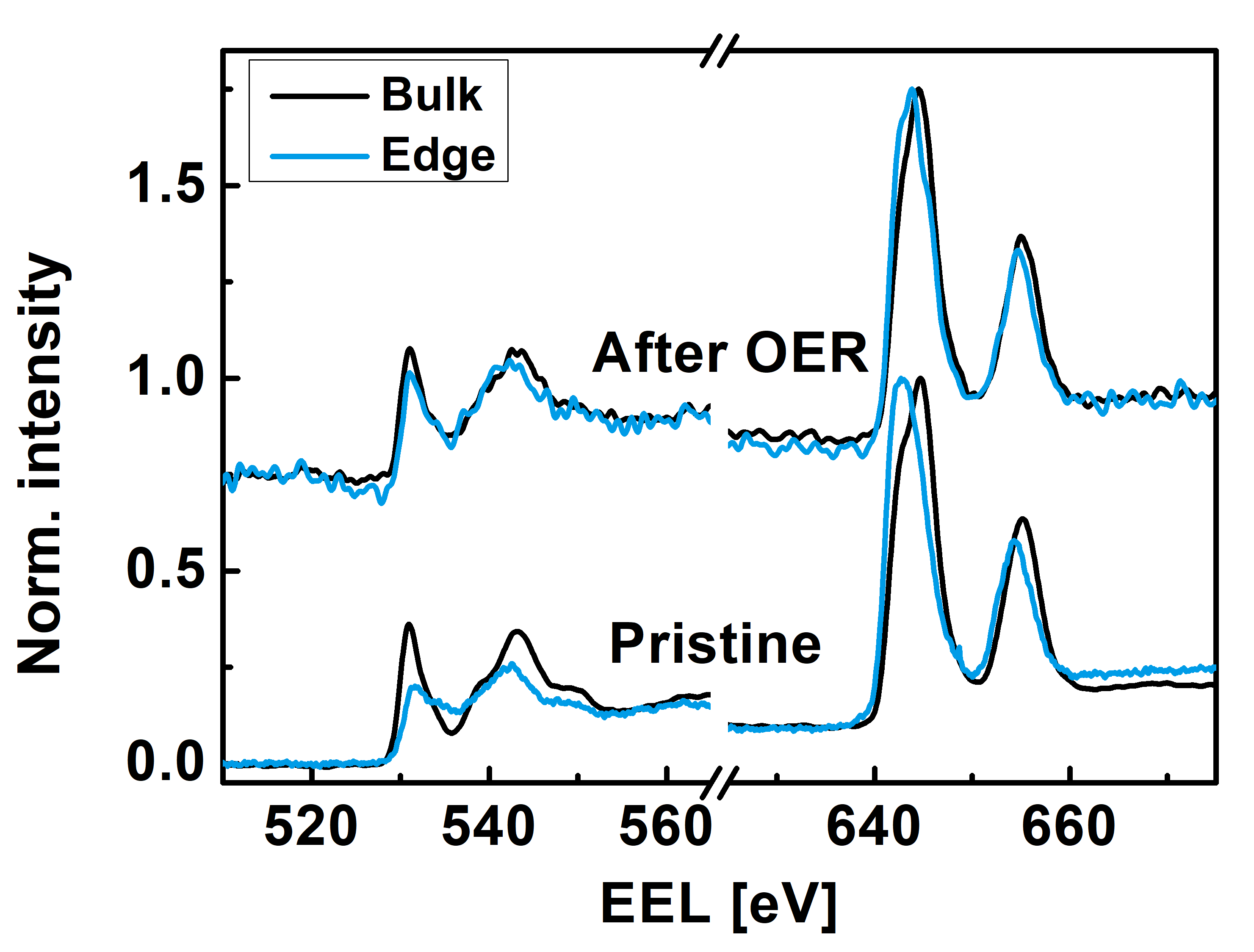}}
	\vspace{-0.25cm}
	\caption{\ce{LiMn2O4} particles after catalysing the OER. (a) and (b) HAADF-STEM (High Angular Annual Dark Field Scanning Transmission Electron Microscopy) images of particle surfaces showing cracks at different length scales and roughened surfaces. (c) STEM-EEL spectra of representative core and shell regions of the OER cycled material compared with the pristine case.}
	\label{fig:Vor-nach-OER}
\end{figure}

TEM images and STEM-EEL spectra of cycled particles are shown in Fig.\ref{fig:Vor-nach-OER}. In contrast to the pristine particles, 6 of the 8 OER-cycled particles studied in the TEM with STEM-EELS revealed significant changes in the shell EEL spectra relative to the pristine case, while the core spectra remained unchanged in all cases (Fig.\ref{fig:Vor-nach-OER}c). In addition, 2 of these particles displayed cracks of various lengths. Both the O $K$- and Mn $L$ edges in the OER cycled shell regions now look almost like the core region of the nanoparticles. EEL spectra from near the cracks look the same as from other surface regions. Compared to the pristine particle surfaces, the EEL spectra of the surface regions of the OER-cycled particles show: an increase in the O pre-peak relative to the $4sp$ peak intensity; an increase in total intensity of the O edge compared to Mn $L_3$ peak; a shift in energy of both the Mn $L_3$ and $L_2$ maxima; and, an increase in the Mn $L_2$ peak relative to $L_3$ peak intensity. Based on the O $K$ pre-peak to Mn $L_3$ peak distance, the average Mn valence in the shell region is found to be $3.50(11)+$, an increase of more than $0.5+$ relative to the pristine state, while the core valence of $3.78(6)+$ is is slightly decreased by voltage cycling. Fig.8 in SI\dag shows that the EEL spectra of core regions are almost identical to pristine state and within the deviations obderved between different particles. A difference in the Li content between the shell and core regions of the cycled particles, as determined from the Li $K$ edge contained in the low loss spectra (Fig.9 in SI\dag), could not be detected within the noise.\\%

A majority of the particles that were investigated with the TEM after CV show oxidized surface regions relative to the pristine state. However, two of the 8 investigated particles were unchanged by OER-cycling. We attribute this to the widely observed heterogeneous activity of composite electrodes. The catalyst nanoparticles were mixed with carbon black particles and drop-cast onto a glassy carbon electrode. Some of the nanoparticles will have poor electrical connections or may not have contact to the electrolyte, leading to a situation where not all particles contribute to the catalytic process \cite{Suntivich2010}.\\

\begin{figure}[h]
	\centering
	\subfloat[]{\includegraphics[width=0.35\columnwidth]{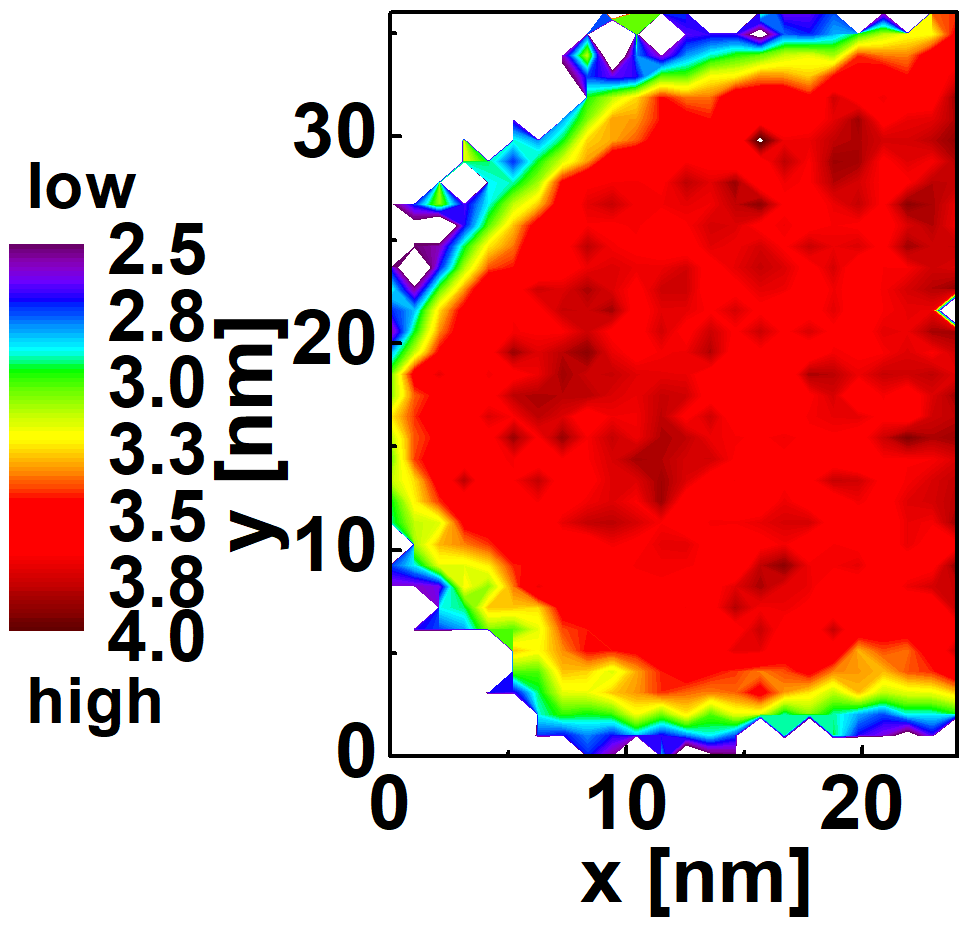}}
	\subfloat[]{\includegraphics[width=0.33\columnwidth]{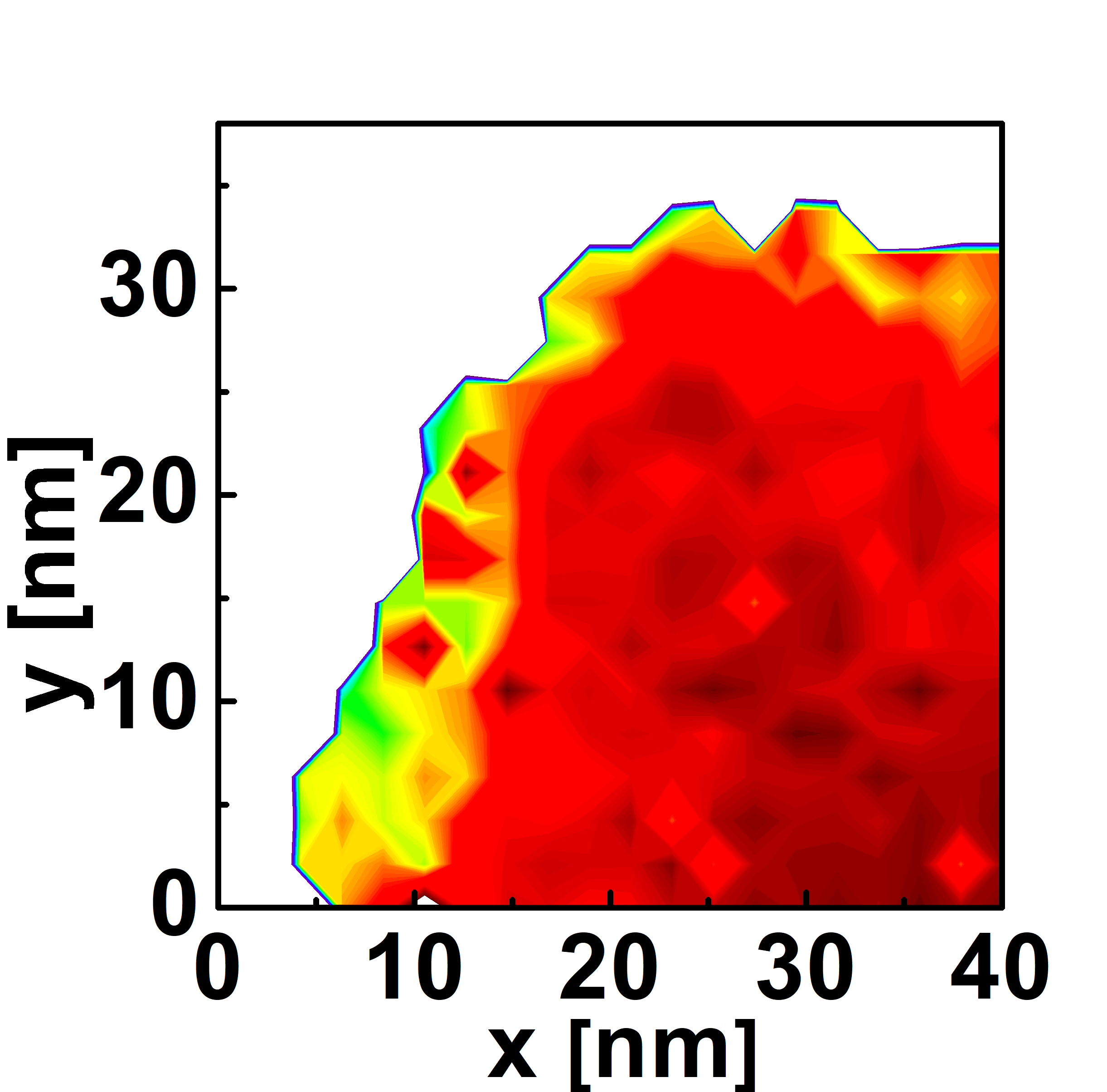}}
	\subfloat[]{\includegraphics[width=0.33\columnwidth]{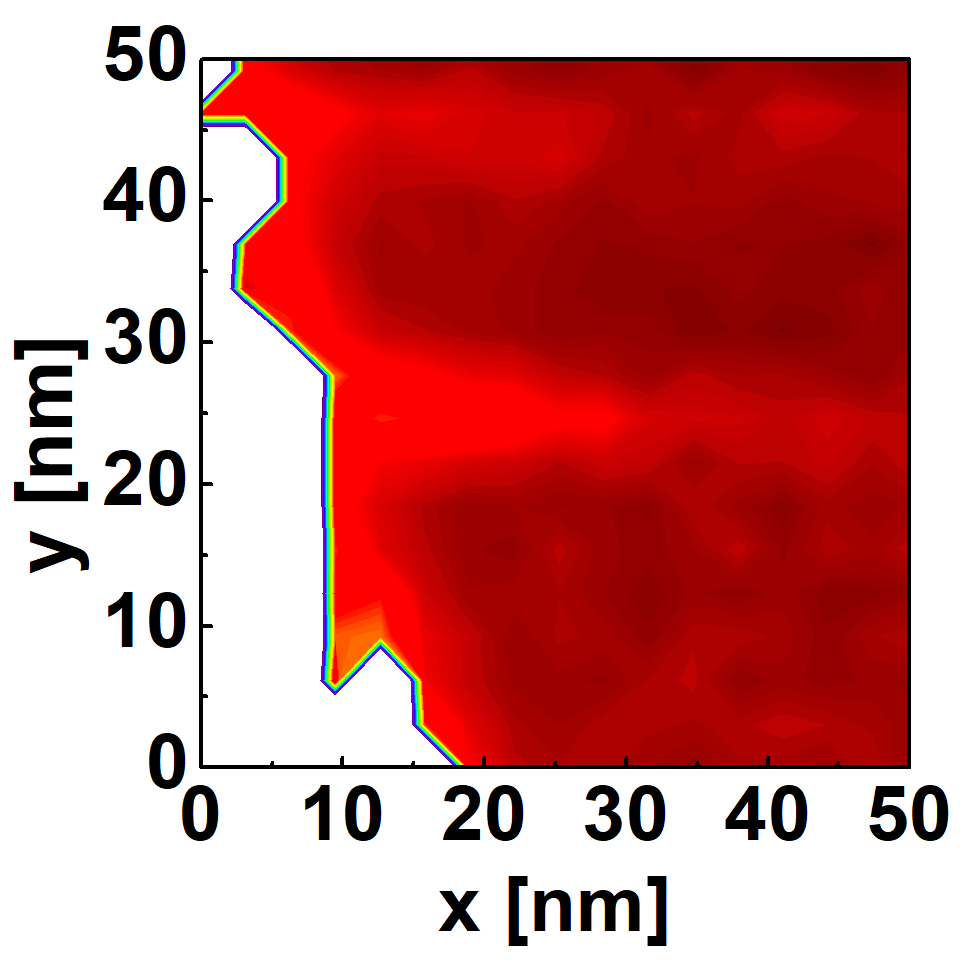}}
	\caption{Mn valence maps of the surface regions of the \ce{LiMn2O4} particles. (a) Map of the pristine state, (b) after 5 minutes in the NaOH pH13 electrolyte, and (c) after 10 cycles in the same electrolyte. The originally Mn reduced surface layer of the pristine particles is slightly further reduced by the electrolyte treatment and strongly oxidized by the voltage cycling.}
	\label{fig:Valencemap-comp}
\end{figure}

To ensure that the observed change in surface Mn valence is actually caused by an electrochemical reaction and is not simply the result of reaction with the electrolyte (\ce{NaOH}, pH13), EEL spectra were also collected from particles held in the electrolyte for 5 minutes without cycling (Fig.4 of the SI\dag), which is comparable with the time that the particles are exposed to the electrolyte prior to CV. The changes due to the electrolyte as well as due to electrochemical cycling are summarized in Mn valence maps (Fig.\ref{fig:Valencemap-comp}) and compared to the pristine state (Fig.\ref{fig:Valencemap-comp}). The valence maps clearly illustrate that reaction with the electrolyte alone does not cause oxidation of the reduced surface layer of the pristine particles (Fig.\ref{fig:Valencemap-comp}a); in fact, the surface layers appear to be slightly more reduced by the electrolyte treatment (Fig.4 of ESI\dag\ and Fig.\ref{fig:Valencemap-comp}b). Therefore, we conclude that oxidation of the originally reduced surface layer, to almost reach the Mn valence value of the \ce{LiMn2O4} in the core region (Fig.\ref{fig:Valencemap-comp}c), is caused by an electrochemical reaction during voltage cycling.\\

An increase in the Mn oxidation state as a result of voltage cycling is also supported by XAS measurements at the Mn $K$ edge. These measurements probe an entire particle ensemble, so that they sample both the core and shell regions. A clear shift of the Mn $K$ edge (normalized absorption 0.5) to higher energies is found as a result of the voltage cycling (Fig.5 of SI\dag), indicating an increase in the average Mn oxidation state of the particles \cite{Gilbert2003}. 
\section{Discussion}
STEM-EELS investigations of the nominal \ce{LiMn2O4} pristine nanoparticles reveal that they have a core shell structure composed of an oxidized core and a reduced shell. By comparing the O $K$ regions of experimental and simulated spectra, we can rule out that the reduced shell region is caused by oxygen vacancies or excess Li, and is instead due to tetrahedral defects. A careful comparison indicates the approximate composition [Li$_{0.5}$Mn$_{0.5}$]$_{\text{T}}$Mn$_2$O$_4$, which is in good agreement with the Mn valence obtained from the O $K$ Mn $L_3$ energy difference. The associated DFT calculations of the DOS (plotted in SI\dag\ Fig. 6 and 7) explain the observed changes in the EEL spectra very well. Tetrahedrally coordinated Mn$^{2+}$ introduces additional unoccupied e$_g$ and t$_{2g}$ states at the high energy limit of the O pre-peak region thereby explaining the observed pre-peak broadening in experiment and spectrum simulations.\\

Surface layers in \ce{LiMn2O4} with similar thicknesses, containing less oxidised Mn, have been reported and discussed in terms of oxygen vacancies \cite{Huang2011} as well as \ce{Mn3O4} formation \cite{Tang2014,Tang2014b,Amos2016,Gao2019,Liu2019}. \ce{Mn3O4} shares the same close-packed oxygen lattice, but in contrast to "perfect" \ce{LiMn2O4}, Mn resides on both octahedral and tetrahedral sites. This leads to large tetragonal distortions \cite{Jarosch1987} and lattice mismatches of \SI{1}{\%} ($a,b$) and \SI{14.6}{\%} ($c$) with respect to the cubic spinel. To our knowledge, no direct observation of the \ce{Mn3O4} structure (for example via XRD) has been reported in the literature, except after several cycles of electrochemical de-/lithiation \cite{Liu2019}. Approximating the LiMn$_2$O$_4$ particles used in this study as spheres with diameters of \SI{44}{\nm} and with a \SI{3}{\nm} thick surface layer, the volume fraction of the shell region is around $1/3$ of the total volume. However, the XRD measurements (Fig.\ref{fig:XRDandoverview}c) clearly limit the Mn$_3$O$_4$ fractions to less than 1$\%$, thereby decisively ruling out the possibility that the reduced surface layer observed here has the Mn$_3$O$_4$ crystal structure. This is supported by the HR-TEM investigations which show that the particles are single crystalline and are not strained at their surfaces as well as by comparison of the measured shell EEL spectra to \ce{Mn3O4} spectra \cite{Kurata1993, Laffont2010}.\\

The evidence of tetrahedral Mn in the near-surface region of the nanoparticles is in good agreement with a number of literature studies that detect Mn contrast on tetrahedral sites in a HR-STEM study along the [110] direction of LiMn$_2$O$_4$ nanoparticles and film surfaces  \cite{Tang2014, Tang2014b, Amos2016, Liu2019, Gao2019}. It also may explain why previous studies have attributed the reduced surface layer to the presence of Mn$_3$O$_4$, which has Mn$^{2+}$ instead of Li$^+$ on the tetrahedral sites. However, the stable phase of Mn$_3$O$_4$ is tetragonal due to Jahn-Teller distortions, which our XRD measurements have been able to rule out.\\

In contrast, the comparison with simulation suggests that the core is slightly oxidized, either through missing tetrahedral Li or by replacement of octahedral Mn with Li  \cite{Strobel2004}. A combination of both effects - transfer of Li from the tetrahedral to octahedral sites - could account for the Mn valence of $3.8+$ calculated for the core, although, the valence numbers are not believed to be as accurate as the comparison between experimental and simulated spectra. This is highlighted by the fact that the  by two independent methods evaluated valence numbers differ. Furthermore, replacement of octahedral Mn by Li in Li rich spinels has been shown to lead to a slight decrease in lattice parameter  \cite{Kopec2008} in agreement with the lattice constant measured here.\\%

During the first few electrochemical cycles in the potential window including the OER, a large disk current was recorded, while the OER activity remained constant (Fig.\ref{fig:DiscCurrent}). Subsequently, STEM-EELS of the voltage cycled particles reveals Mn oxidation near the particle surfaces compared to the pristine state.  Removal of most of the tetrahedral Mn in the pristine shell with composition [Li$_{0.5}$Mn$_{0.5}$]$_{\text{T}}$Mn$_2$O$_4$  would result in an increase in the average Mn valence (by 0.57 at 80\% removal) to a value close to the measured one of $3.5+$, which is close to the nominal valence in \ce{LiMn2O4}. At the same time an increased disc current below OER onset was observed which is explained by Mn dissolution in our previous publication \cite{Baumung2019a}.
The complete dissolution of tetrahedral Mn results in a mass difference of $\SI{2.5}{\mu\g}$ assuming that the shell accounts for $1/3$ of the volume as discussed above. Using Faraday's law of electrolysis, a charge of \SI{17.1}{\milli\coulomb} is expected for complete conversion of the shell due to the transfer of five electrons, i.e. the reaction \mbox{Mn$^{2+}$ + 8 OH$^{-} \rightarrow$ Mn$^{7+}$O$_{4}$ + 4H$_{2}$O + 5$e^{-}$}. The formation of \ce{MnO4} under the investigated conditions is also predicted by thermodynamic calculations, i.e. the E-pH diagram of Mn-O-H \cite{Marcel1}.
The calculated charge of \SI{17.1}{\milli\coulomb} agrees well with our experimental result of \SI{16.3}{\milli\coulomb} in the first three cycles. Since the reaction has not completed after the first three cycles (Fig.\ref{fig:DiscCurrent}(a)) the experimental value is expected to be slightly smaller. 
Mn$^{4+}$ dissolution can be excluded as it cannot account for the observed oxidation of the shell, while Mn$^{3+}$ dissolution is not consistent with the measured additional currents. Since the charge that passed the sample before oxygen evolved for the first time (\SI{3.5}{\milli\coulomb}) is considerably smaller than the necessary charge of \SI{17.1}{\milli\coulomb} for complete shell conversion, the complete dissolution of tetrahedral Mn$^{2+}$ prior to the first OER can also be excluded.\\

Thus, the EEL spectra within \SI{3}{\nm} of the LiMn$_2$O$_4$ particle surfaces clearly reveal oxidation during voltage cycling. Surprisingly, the change in electronic structure has no effect on their OER activity. In combination with DFT-based simulations of the EEL spectra, we conclude that the oxidation occurs as a result of tetrahedral Mn dissolution during voltage cycling. This picture is directly supported by RRDE Mn detection \cite{Baumung2019a}. The average valence in the near-surface region changes from $2.9+$ to $3.5+$, while the valence of the octahedral Mn valence, which is generally believed to be the active site for the rate-limiting steps in the oxygen evolution and reduction reactions  \cite{Wei2017, Cady2015, Hong2016, Huynh2015, Chan2018a}, changes from $3.25+$ to about $3.5+$, without affecting the OER catalytic activity. Thus, we come to the surprising discovery that changing the octahedral Mn valence by removing tetrahedral Mn has no effect on the OER activity.\\

Our observation that the OER activity is unaffected by a change in the octahedral Mn valence runs contrary to the literature on descriptors as well as to previous literature studies \cite{Wei2017} and our own studies on the effect of Li content on the OER activity of Li$_x$Mn$_2$O$_4$. In agreement with the literature, increasing the octahedral Mn valence above $3.5+$ by removing Li leads to a decrease in activity or an increase in overpotential respectively \cite{Baumung2019, Wei2017}. Similarly, an increase in octahedral Mn valence from $3.25+$ to $3.5+$ is also expected to decrease activity significantly \cite{Wei2017}, but is observed here to have no effect. Given the wealth of literature supporting a decrease in activity with decreasing e$_g$ occupancy between one and zero (corresponding to octahedral Mn valence between $3.0+$ and $4.0+$) \cite{Wei2017,Cady2015,Man2011a, Hong2016, Huynh2015, Chan2018a}, we must conclude that either (i) the octahedral Mn sites at the particle surfaces are unchanged even as the octahedral Mn valence within  \SI{1}{\nm} of the surface changes considerably, or (ii) the changes caused to the octahedral surface Mn by removal of tetrahedral Mn from the near-surface region are not relevant to OER activity despite the changes in valence and e$_g$ occupancy. Thus, the current state of observations on the catalytic activity of lithium manganese oxide spinel can be summarized as follows: tuning octahedral Mn valence using the tetrahedral Li content leads to the widely observed change in activity, while tuning it using the tetrahedral Mn content leads to no change in activity.\\

We cannot rule out the first possibility that identical active sites formed during the OER are unaffected by tetrahedral Mn$^{2+}$. At some level it remains surprising that any parameter describing the bulk state can be an effective descriptor of OER activity at the interface between the particle and the electrolyte during catalysis. The active catalytic state differs in almost every respect from the bulk: it is a surface interacting with a strongly perturbing basic aqueous environment, and it is not in equilibrium. Nonetheless, if this is the explanation for our observation, then it is a clear illustration of the difficulties of using bulk properties (or even those measured within \SI{1}{\nm} of the surface) to describe catalytic behavior.\\%

The other possibility that our results raise is whether there is some other parameter that controls activity, which correlates with Mn valence or e$_g$ occupancy when changing the tetrahedral Li content but not when changing the tetrahedral Mn content. In our case, the shell region did not form the tetragonal phase, which is expected to occur due to increased Jahn-Teller distortions if the Mn valence drops below $3.5+$. The tetragonal phase formation in the shell may have been suppressed by the strains imposed through the epitaxial relation with the core. These strains will also affect other aspects of the structure such as Mn-O bond lengths and thereby possibly suppress the expected change in activity. It seems conceivable that in a strongly correlated system such as Li$_x$Mn$_2$O$_4$, complex correlations between details of the electronic and geometric structure parameters could lead to a case where changing Mn versus Li tetrahedral occupancy could have very different effects. In fact, the DFT calculations underlying the EELS simulations show that the two electrons from tetrahedral Mn are more closely bound to the octahedral Mn than the single electron from tetrahedral Li. A careful comparison with other descriptors that have been suggested in the literature\cite{Hong2016}, such as Mn-O bond lengths and angles and the extent of covalency, may help to identify what features of the bulk electronic and geometric structure are most important.
\section{Conclusion}
\balance
This combined experimental-theoretical study shows that Li$_x$Mn$_2$O$_4$ nanoparticles have a ca. \SI{3}{\nm} thick reduced surface layer containing tetrahedral Mn. During voltage cycling, the tetrahedral Mn is leached out, increasing the average Mn valence in the shell region. At the same time, the OER activity remains constant, revealing that the catalytic activity of Li$_x$Mn$_2$O$_4$ is unaffected by the near-surface Mn valence state when it is tuned through the tetrahedral Mn occupancy. This contradicts widely accepted bulk state descriptors for catalytic activity as well as the observed effect of tetrahedral Li occupancy on catalytic activity. We conclude that either the octahedral Mn valence does not affect surface catalytic activity or that e$_g$ occupancy is not a dominant causal parameter controlling the activity. Either way, the results gained here present important challenges to the use of bulk state descriptors and emphasize the need for surface-sensitive operando experiments and complementary theoretical studies of electronic structure.\\
\section*{Conflicts of interest}
There are no conflicts to declare.
\section*{Acknowledgements}
We would like to thank Thomas Brede, Heidrun Sowa and Helmut Klein for support with X-ray diffraction measurements and interpretation and Lifei Xi for performing XAS experiments at the KMC-2 beamline of BESSY II. Niklas Weber is greatfully acknowledged for his support with graphical illustration. The use of equipment in the "Collaborative Laboratory and User Facility for Electron Microscopy" (CLUE) www.clue.physik.uni-goettingen.de is gratefully acknowledged. The presented work is  funded by the Deutsche Forschungsgemeinschaft (DFG, German Research Foundation) - 217133147/SFB 1073, project C05, C03. We also gratefully acknowledge the computing time provided by the Paderborn Center for Parallel Computing (PC$^2$). J.B.\ is grateful for a DFG Heisenberg professorship BE3264/11-2 (Project No.\ 329898176).

\newpage
\bibliography{library, libmarco}
\bibliographystyle{apsrev4-2} 

\end{document}


\title{Supporting Information -- A Criticial View on e$_g$ Occupancy as a Descriptor for Oxygen Evolution Catalytic Activity in \ce{LiMn2O4} Nanoparticles}

\author{Florian Sch\"onewald}
\email{f.schoenewald@uni-goettingen.de}
\affiliation{Universit\"at G\"ottingen, Institut f\"ur Materialphysik, Friedrich-Hund-Platz 1, 37077 G\"ottingen, Germany.}
\author{Marco Eckhoff}
\affiliation{Universit\"at G\"ottingen, Institut f\"ur Physikalische Chemie, Theoretische Chemie, Tammannstra{\ss}e 6, 37077 G\"ottingen, Germany.}
\author{Max Baumung}
\affiliation{Universit\"at G\"ottingen, Institut f\"ur Materialphysik, Friedrich-Hund-Platz 1, 37077 G\"ottingen, Germany.}
\author{Marcel Risch}
\affiliation{Universit\"at G\"ottingen, Institut f\"ur Materialphysik, Friedrich-Hund-Platz 1, 37077 G\"ottingen, Germany.}
\affiliation{Helmholtz-Zentrum Berlin f\"ur Materialien und Energie, Hahn-Meitner-Platz 1, 14109 Berlin, Germany.}
\author{Peter E. Bl\"ochl}
\affiliation{Technische Universit\"at Clausthal, Institut f\"ur Theoretische Physik, Leibnizstra{\ss}e 10, 38678 Clausthal-Zellerfeld, Germany.}
\affiliation{Universit\"at G\"ottingen, Institut f\"ur Theoretische Physik, Friedrich-Hund-Platz 1, 37077 G\"ottingen, Germany.}
\author{J\"org Behler}
\affiliation{Universit\"at G\"ottingen, Institut f\"ur Physikalische Chemie, Theoretische Chemie, Tammannstra{\ss}e 6, 37077 G\"ottingen, Germany.}
\affiliation{The International Center for Advanced Studies of Energy Conversion (ICASEC), University of Göttingen, 37077 Göttingen, Germany.}
\author{Cynthia A. Volkert}
\email{cynthia.volkert@phys.uni-goettingen.de}
\affiliation{Universit\"at G\"ottingen, Institut f\"ur Materialphysik, Friedrich-Hund-Platz 1, 37077 G\"ottingen, Germany.}
\affiliation{The International Center for Advanced Studies of Energy Conversion (ICASEC), University of Göttingen, 37077 Göttingen, Germany.}

\date{\today}

\maketitle
\section{Electrochemical setup}
Electrochemical experiments were performed with an OrigaFlex system consisting of three OGF500 module potentiostats (Origalys SAS) using a three-electrode configuration with a saturated calomel electrode (SCE) (ALS Japan Co Ltd., RE-2B) and a platinum counter electrode. The RRDE-setup consisted of a RRDE-3A rotator (ALS Japan Co Ltd.) and a custom made electrochemical cell made of PTFE. The RRDE-electrode consists of a glassy carbon disk electrode with a diameter of \SI{4}{\mm} and a concentric platinum electrode with \SI{5}{\mm} inner and \SI{7}{\mm} outer diameter. Both working electrodes were separately cleaned, polished to a mirror finish and assembled afterwards to prevent a cross contamination. The SCE reference electrode was calibrated to a RHE (HydroFlex Gaskatel).\\

For the experiment the LiMn$_2$O$_4$ covered disk electrode was conditioned at 1.25~V vs. RHE for three minutes and afterwards a CV between 1.25-1.75~V vs. RHE for ten cycles at a scan rate of 10~mV/s. The Rotation speed of the RRDE was set to 1600~rpm. The ring electrode was set to the detection potential of oxygen\cite{Baumung2019a}. For compensating the voltage for the Ohmic drop, electrochemical impedance spectroscopy was done after the catalytic experiment in a range from \SI{100}{\kilo \hertz} to \SI{1}{\hertz}. The electrochemical experiments were done at room temperature. 

\section{Influence of Li content, O vacancies and Mn on tetrahedral sites on EEL spectra}
\begin{figure}[h!]
\includegraphics[width=0.9\columnwidth]{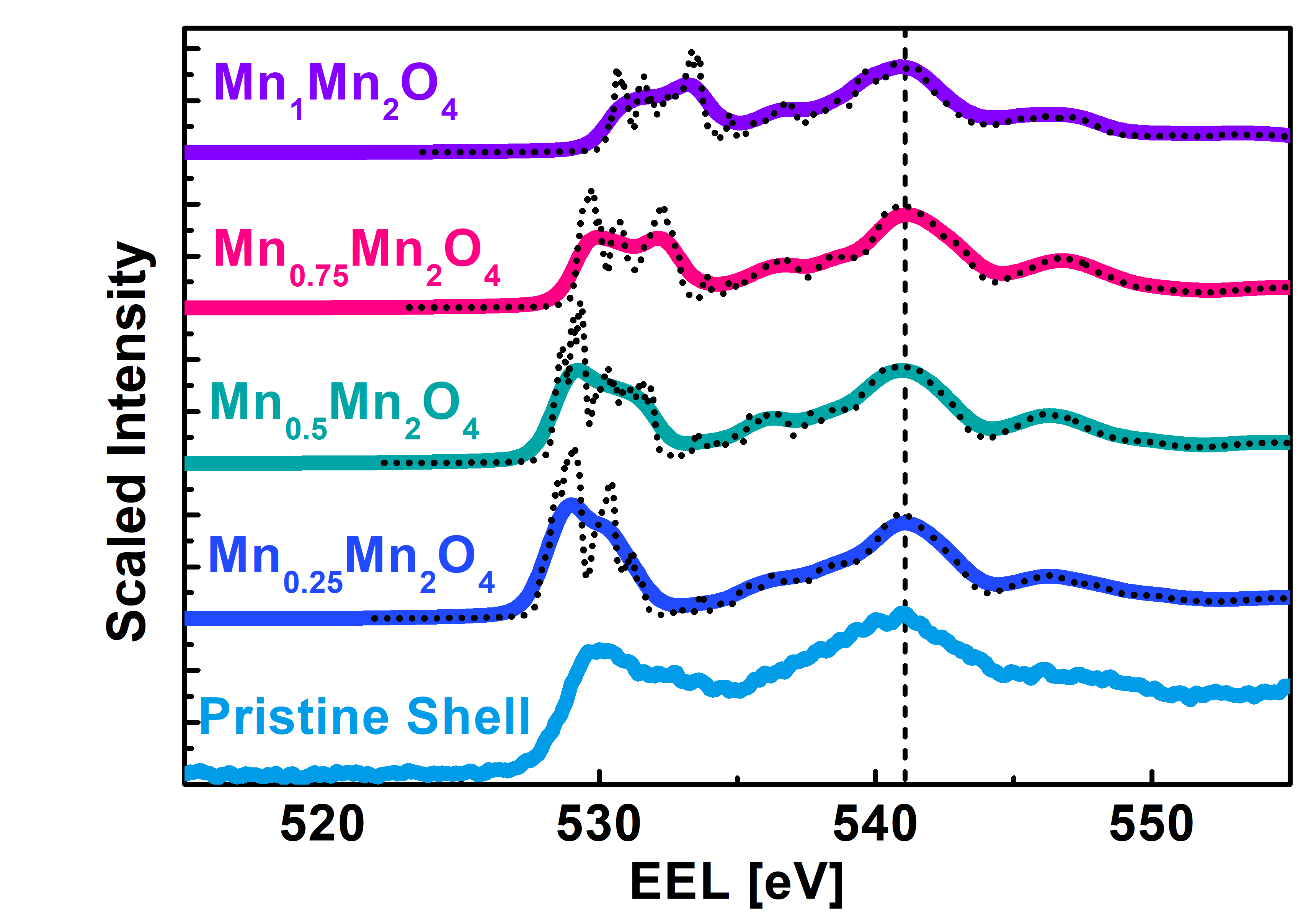}
\vspace*{-0.4cm}
\caption{Lines show broadened, simulated EEL spectra for different Mn occupations $y$ on tetrahedral sites of Mn$_y$Mn$_2$O$_4$ and the experimentally observed shell spectrum. Dotted lines show the non broadened cross sections. Positions on the energy axis were shifted so that the second maxima overlap with the one of experimental spectrum.}
\label{fig:Mn-Defects}
\end{figure}

\begin{figure}[h!]
\includegraphics[width=0.9\columnwidth]{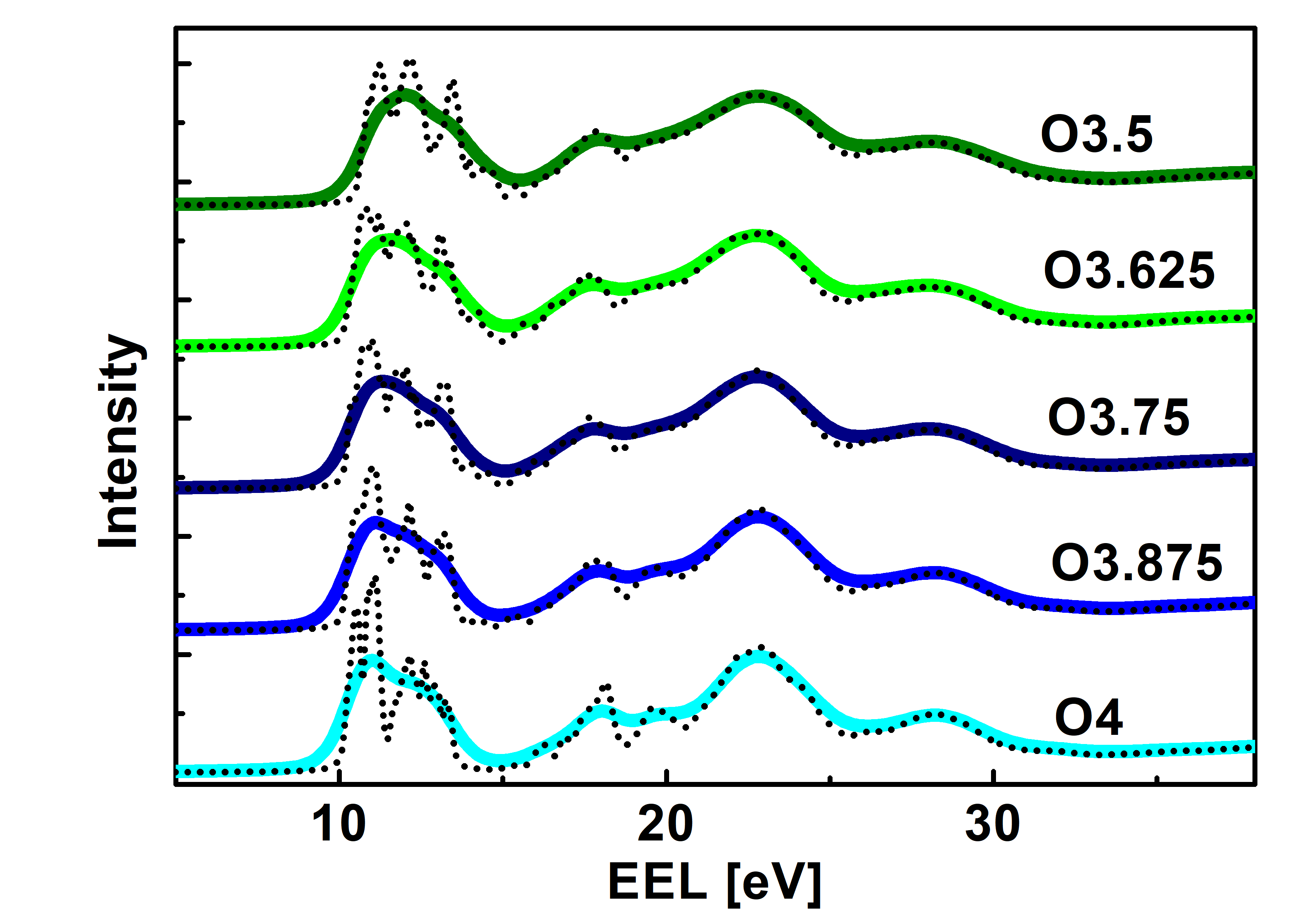}
\vspace*{-0.4cm}
\caption{Lines show simulated EEL spectra broadened with an experimental ZLP for different amounts of oxygen vacancies. Dotted lines show the non broadened cross sections. All simulated structures are based on the cubic spinel O-Mn lattice with the experimentally determined lattice constant. Positions on the energy axis were shifted so that the second maxima overlap. The Oxygen stoichiometry $y$ in Li$_1$Mn$_2$O$_y$ is plotted next to the curves. Positions on the energy axis are arbitrary.}
\label{fig:O-Defects}
\end{figure}

With the help of DFT simulations the influence of Mn defects on tetrahedral sites of the Li empty cubic spinel structure was systematically investigated and is plotted in \ref{fig:Mn-Defects}. The lattice constant was fixed to $a=8.234\,$\AA in agreement with the experimentally measured constant of the particles. Determined electron scattering cross sections from calculations were convoluted with an experimental zero-loss spectrum recorded without sample for better comparison to experimental data.
In the same manner fig. \ref{fig:O-Defects} shows changes to EEL-spectra after O vacancies have been introduced in the Li$_1$Mn$_2$O$_4$ structure. Atom positions could relax, while the lattice constant was hold to the experimental determined.
\newpage
\section{EEL Low Loss Spectrum Pristine}
\begin{figure}[h]
\includegraphics[width=0.9\columnwidth]{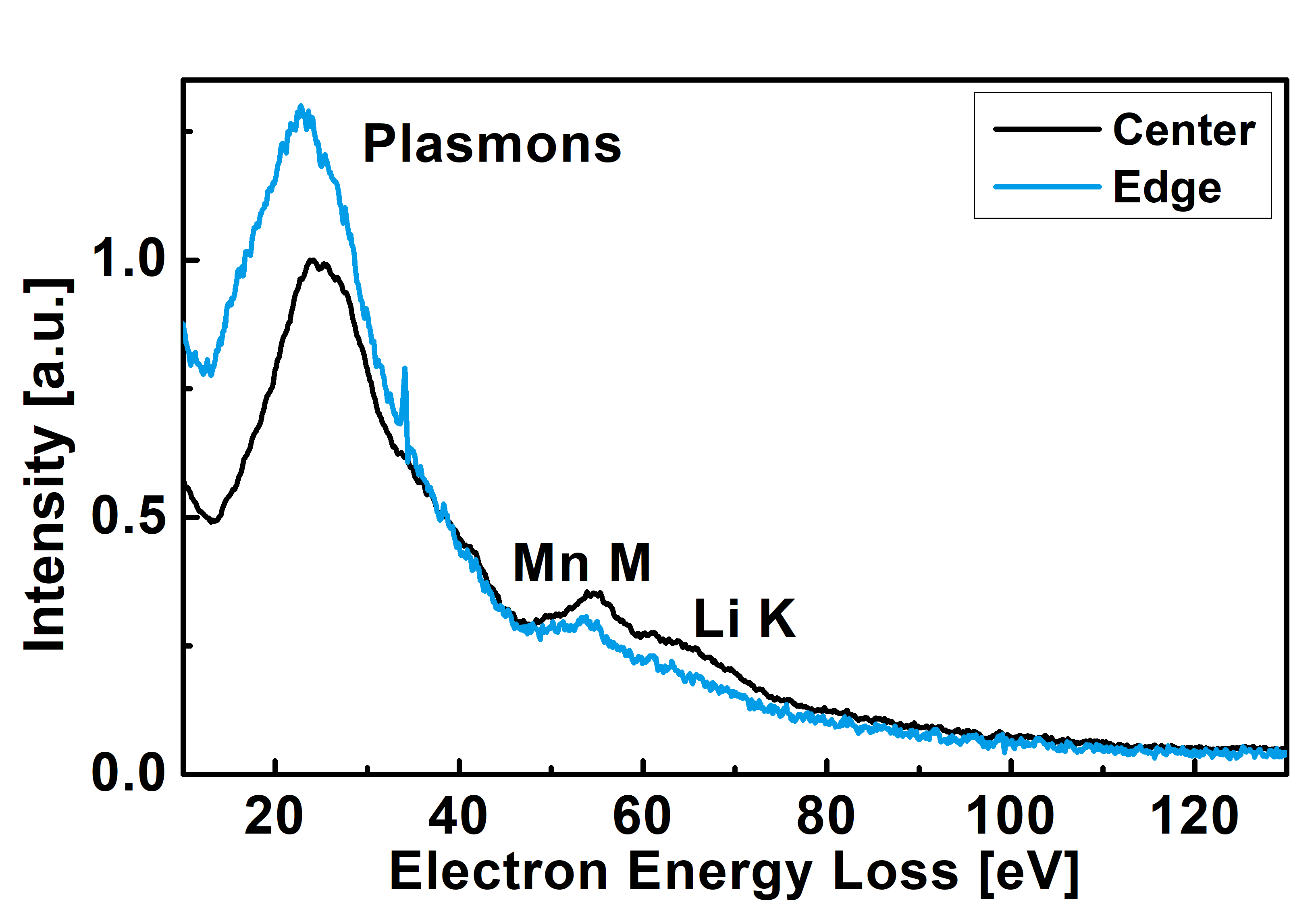}
\vspace*{-0.4cm}
\caption{Low Loss spectra of the core and bulk are plotted of core and shell regions of a pristine nanoparticle. The Mn $M$ and Li $K$ edges are indicated. Intensities are scaled for better comparison}
\label{fig:low-loss}
\end{figure}

The Li $K$ edge ist contained in simultaneously recorded low-loss spectra as a shoulder or small peak next to the Mn $M$ edge. A spectra where surface and bulk regions are compared can be found in \ref{fig:low-loss}. At the Edge, the Li peak is not more pronounced compared to the Mn $M$ edge, meaning that Li excess is not taken as possibility to account for the reduced surface Mn state on the basis of this data.
\section{Influence of electrolyte} 
\begin{figure}[h]
\includegraphics[width=0.9\columnwidth]{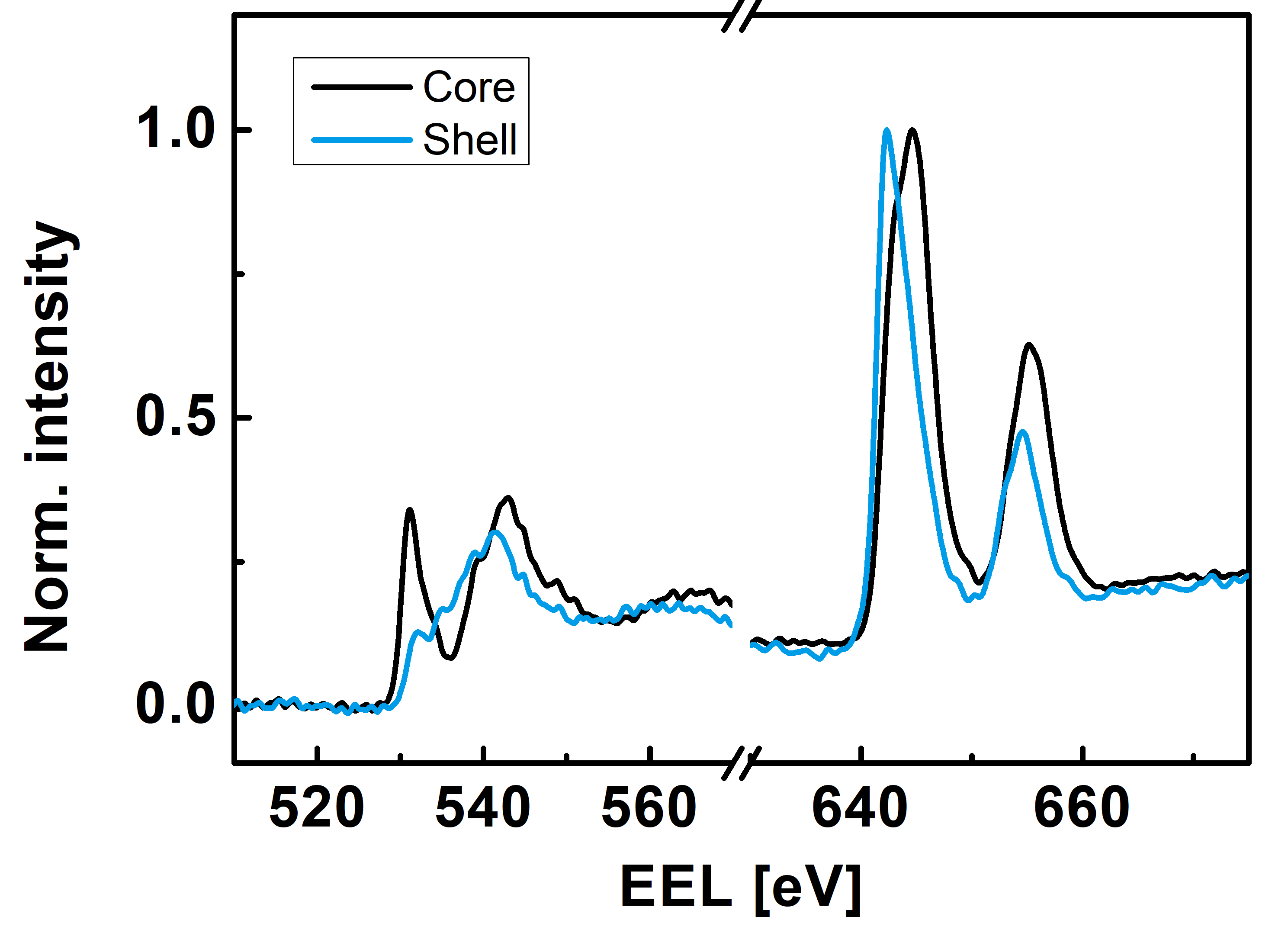}
\vspace*{-0.4cm}
\caption{EEL core and shell spectra of the oxygen $K$ and manganese $L$ edge of nanoparticles dispersed in 0.1 mol NaOH for 5 minutes. NaOH does not cause surface oxidation.}
\label{fig:naoh}
\end{figure}

The influence of the used electrolyte was investigated as a control experiment by dispersing the pristine particles for $5\,$min in 0.1$\,$mol NaOH. The resulting core-shell EEL spectra are plotted in fig. \ref{fig:naoh}
\section{XAS before and after OER}
\begin{figure}[h]
\includegraphics[width=0.8\columnwidth]{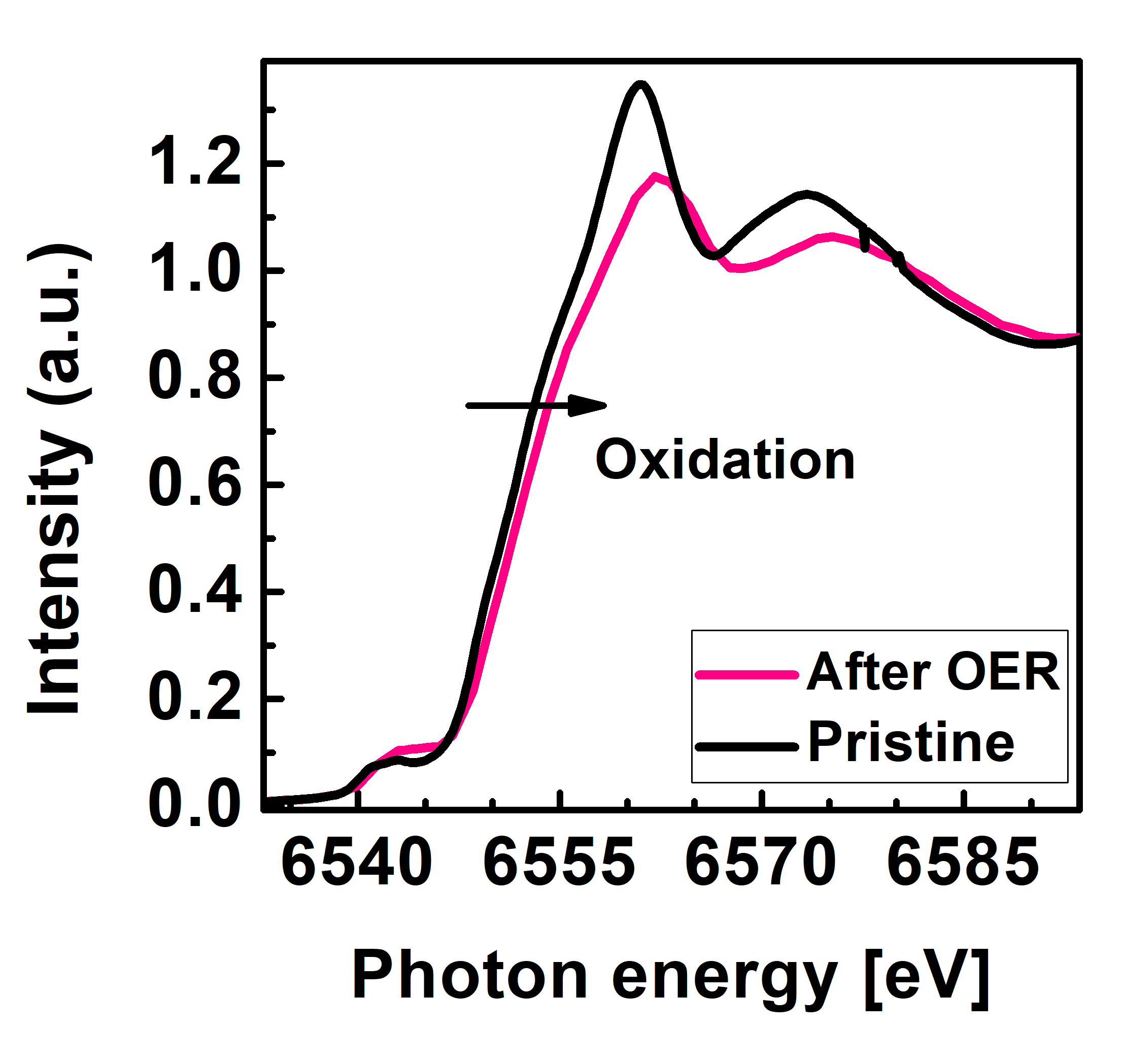}
\vspace*{-0.4cm}
\caption{X-ray-absorption (XAS) spectroscopy at the Mn $K$ edge probing the volume average of pristine nanoparticles and cycled nanoparticles after 10 cycles.}
\label{fig:XAS}
\end{figure}

XAS (X-ray absorption spectroscopy) is a transmissive and in our case non local method to investigate the Mn $K$ edge of the sample, whose energy position is sensitive to the Mn oxidation state \cite{Okumura2014}. A comparison of the pristine particles edge position with the one from cycled particles is shown in fig. \ref{fig:XAS}. After cycling, the edge shifts to higher energies indicating an overall increased Mn
oxidation state. 
For additional XAS measurements the catalytic ink was drop casted on electrode made of graphite foil. After the experiment the electrode was washed off with Milli-Q water, dried and transferred to the synchrotron.  This electrode was cycled with the same parameters. The XAS erperiment was performed as reported in our previous publication. \cite{Baumung2019}  
\section{Density of States of selected simulated structures}
\begin{figure}[h]
	\centering
	\subfloat{%
		\includegraphics[width=1.05\columnwidth]{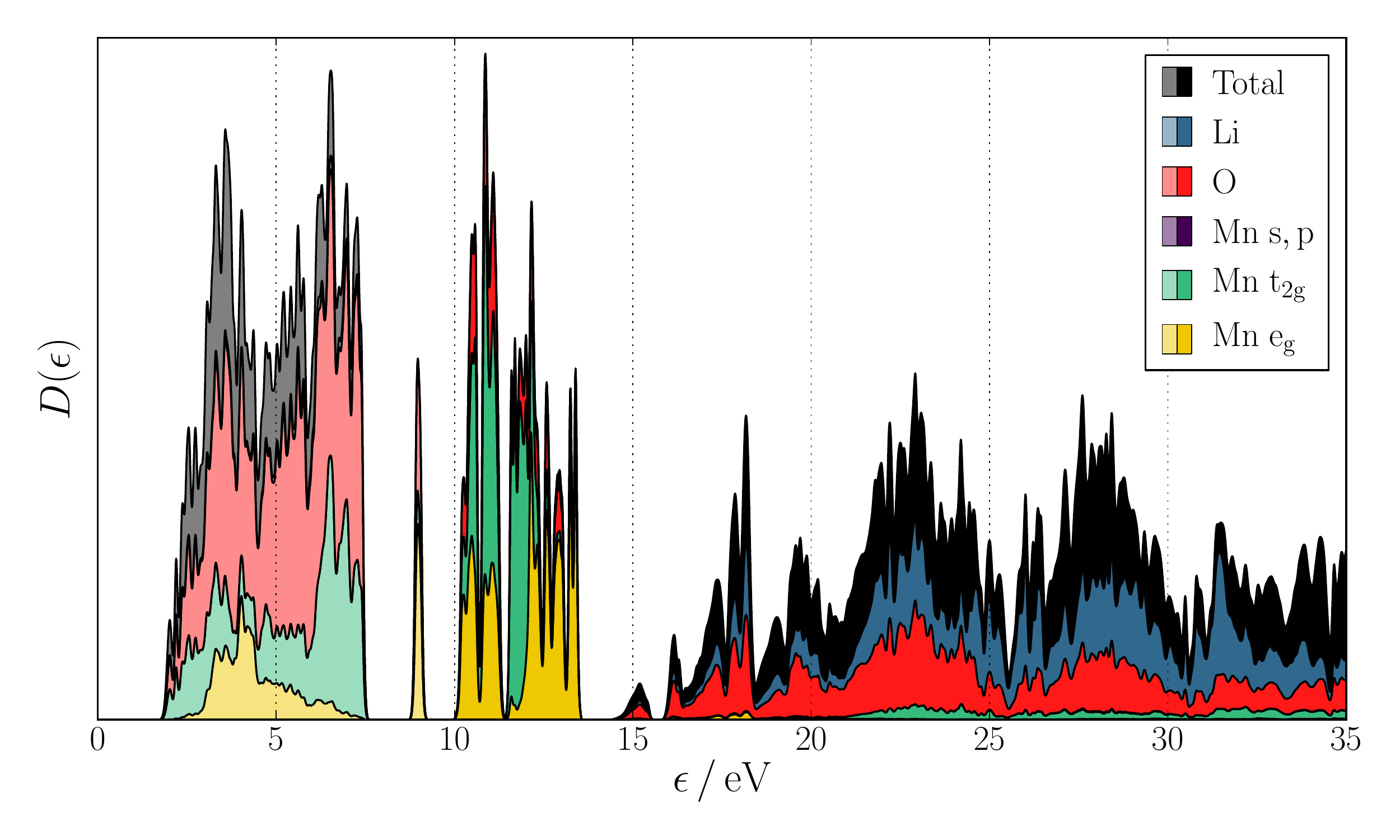}}
	\hspace{0.05\textwidth}
	\vspace*{-0.4cm}
	\subfloat{%
		\includegraphics[width=1.05\columnwidth]{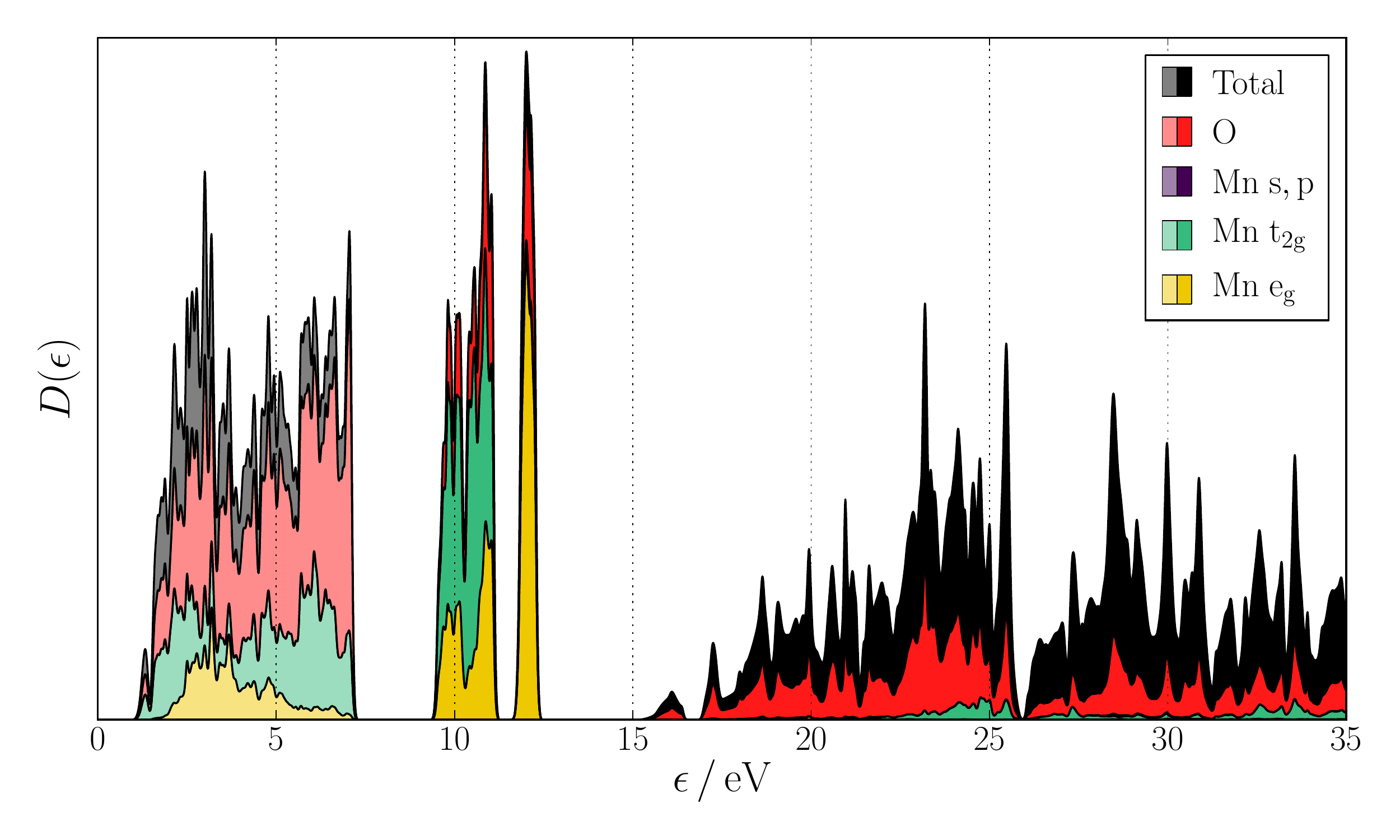}}
	\vspace*{-0.1cm}
	\caption{Calculated DOS (density of states) of Li$_1$Mn$_2$O$_4$ (top) and $\lambda-$Mn$_2$O$_4$ (bottom) showing the influence of Li occupation on tetrahedral sites.}
	\label{fig:DOS1}
\end{figure}

\begin{figure}[h]
	\centering
	\subfloat{%
		\includegraphics[width=1.05\columnwidth]{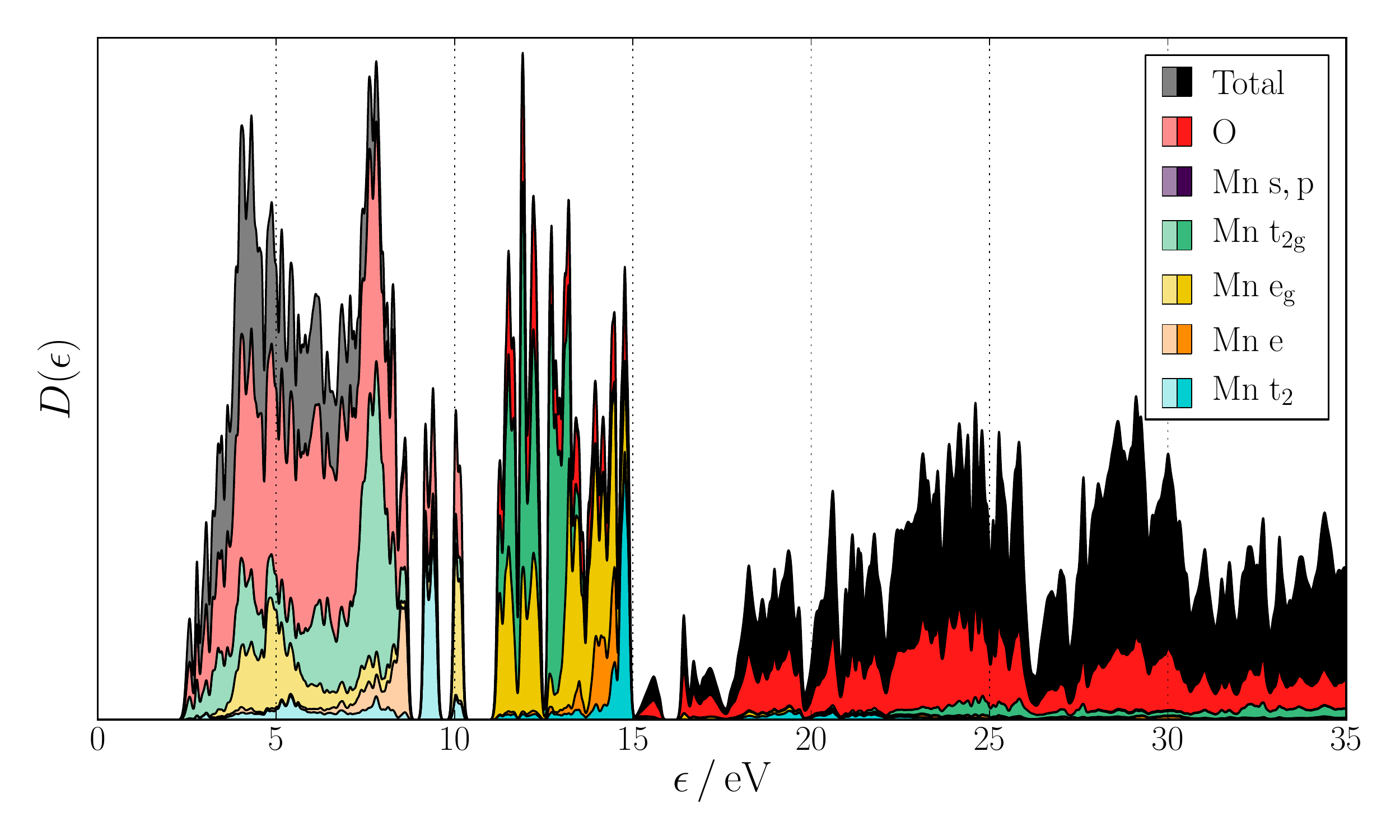}}
	\hspace{0.05\textwidth}
	\vspace*{-0.4cm}
	\subfloat{%
		\includegraphics[width=1.05\columnwidth]{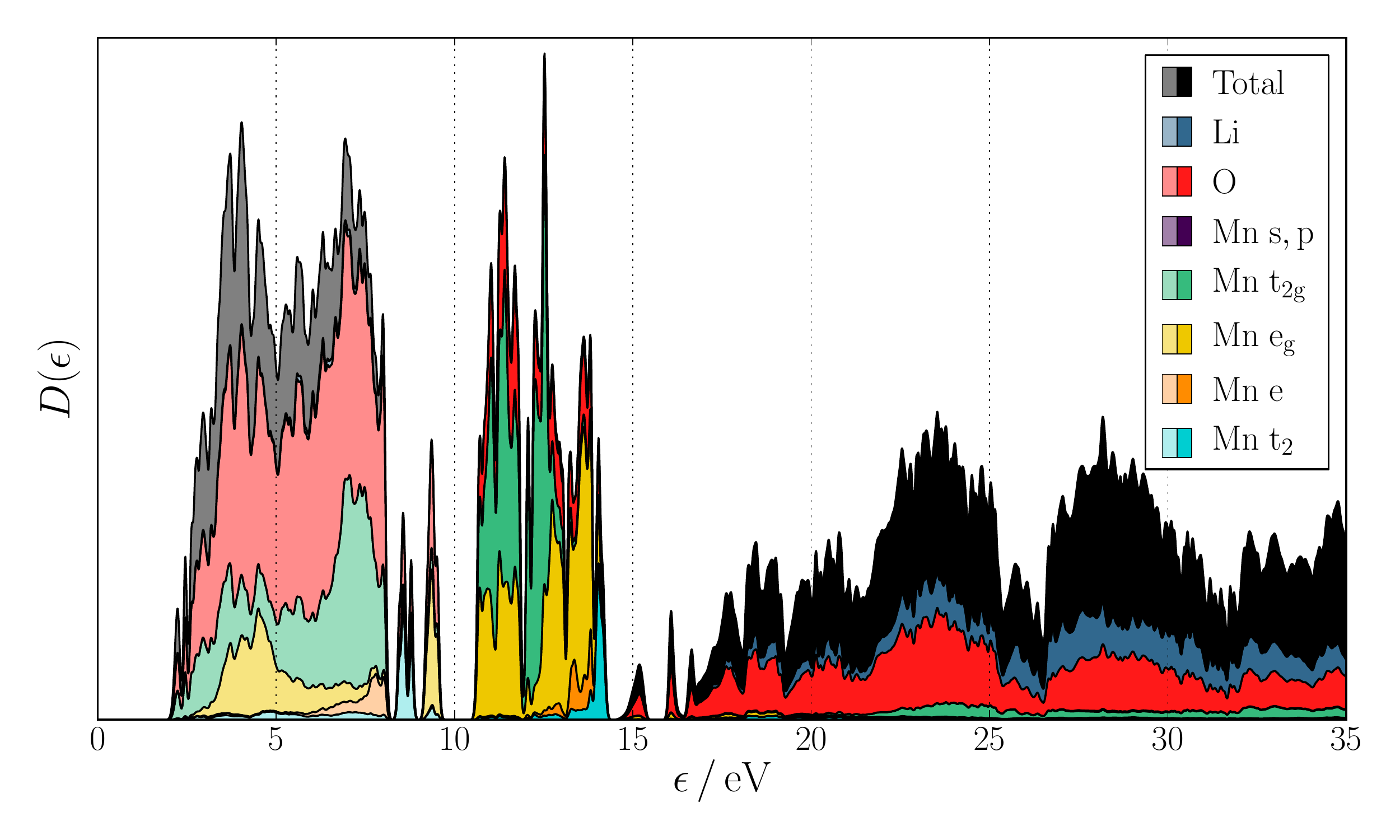}}
	\vspace*{-0.1cm}
	\caption{Calculated DOS (density of states) of $[$Mn$_{0.5}]_T$Mn$_2$O$_4$ (top) and $[$Li$_{0.5}$Mn$_{0.25}]_T$Mn$_2$O$_4$ (bottom) showing the influence of Li occupation on tetrahedral sites.}
	\label{fig:DOS2}
\end{figure}

Looking at the density of states (DOS) in the simulated structures can improve the understanding of the differences if Li or Mn is introduced on tetrahedral sites. Figure \ref{fig:DOS1} compares the the DOS of the "empty" spinel with no Li residing on tetrahedral sites to the case Li$_1$Mn$_2$O$_4$, where the maximum amount of Li occupies the $8a$ tetrahedra in the cubic spinel structure. Pale colours indicate occupied states, while brighter colours indicate unoccupied states relevant for EEL spectra calculations. 
The influence of replacing Li by Mn on tetrahedral sites and having a combination of both species present can be seen in fig. \ref{fig:DOS2}. The plotted Mn states can be differentiated between tetrahedral and octahedral origin. Tetrahedral states are plotted in orange and turquoise and are located at different energy positions compared to octahedral occupied and unoccupied states. Since the tetrahedral unoccupied states are located closely to the gap between about $14\,$eV and $15\,$eV in the graphs, it is expected to cause a more pronounced high energy shoulder at the pre peak as well as a less pronounced local minima before the observed second peak, in agreement with the trends observed in experimental shell spectra. 

\section{Core EEL spectra in pristine state and after 10 cycles}
\begin{figure}[h]
\includegraphics[width=0.9\columnwidth]{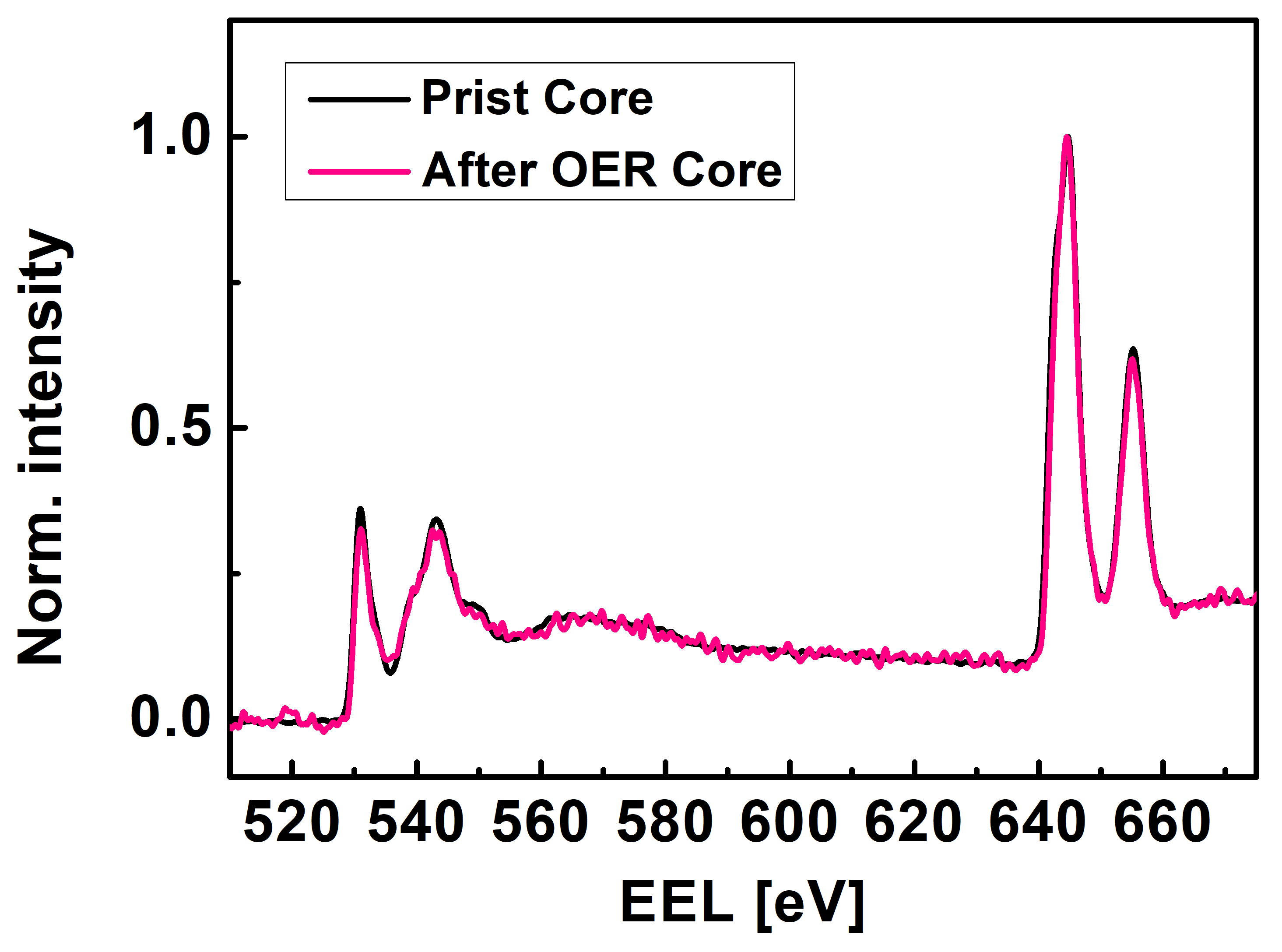}
\vspace*{-0.1cm}
\caption{EEL spectra from the center/core of a pristine particle and of a cycled particle after 10 cycles of OER.}
\label{fig:OER-Bulk}
\end{figure}

The surfaces show strong changes after 10 cycles of OER compared to pristine state. Spectra from the center of the particles (which also contain some signal from the surface due to transmission setup) however do not present significant changes (see fig.\ref{fig:OER-Bulk}).
\section{EEL-Low-Loss Spectra after OER}
\begin{figure}[h]
	\centering
	\subfloat{%
		\includegraphics[width=0.9\columnwidth]{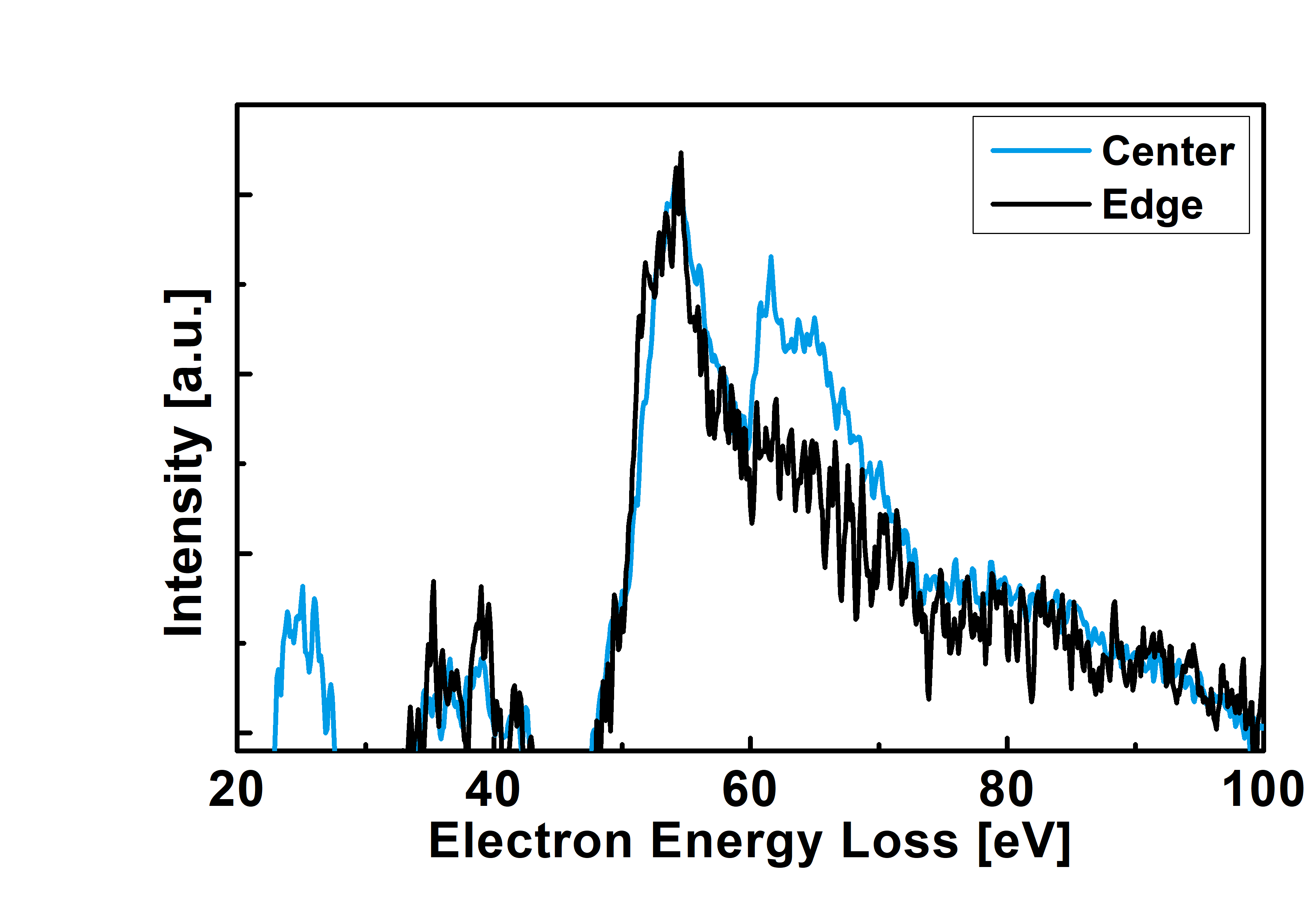}}
	\vspace{-0.5cm}
	\hspace{0.05\textwidth}
	
	\subfloat{%
		\includegraphics[width=0.9\columnwidth]{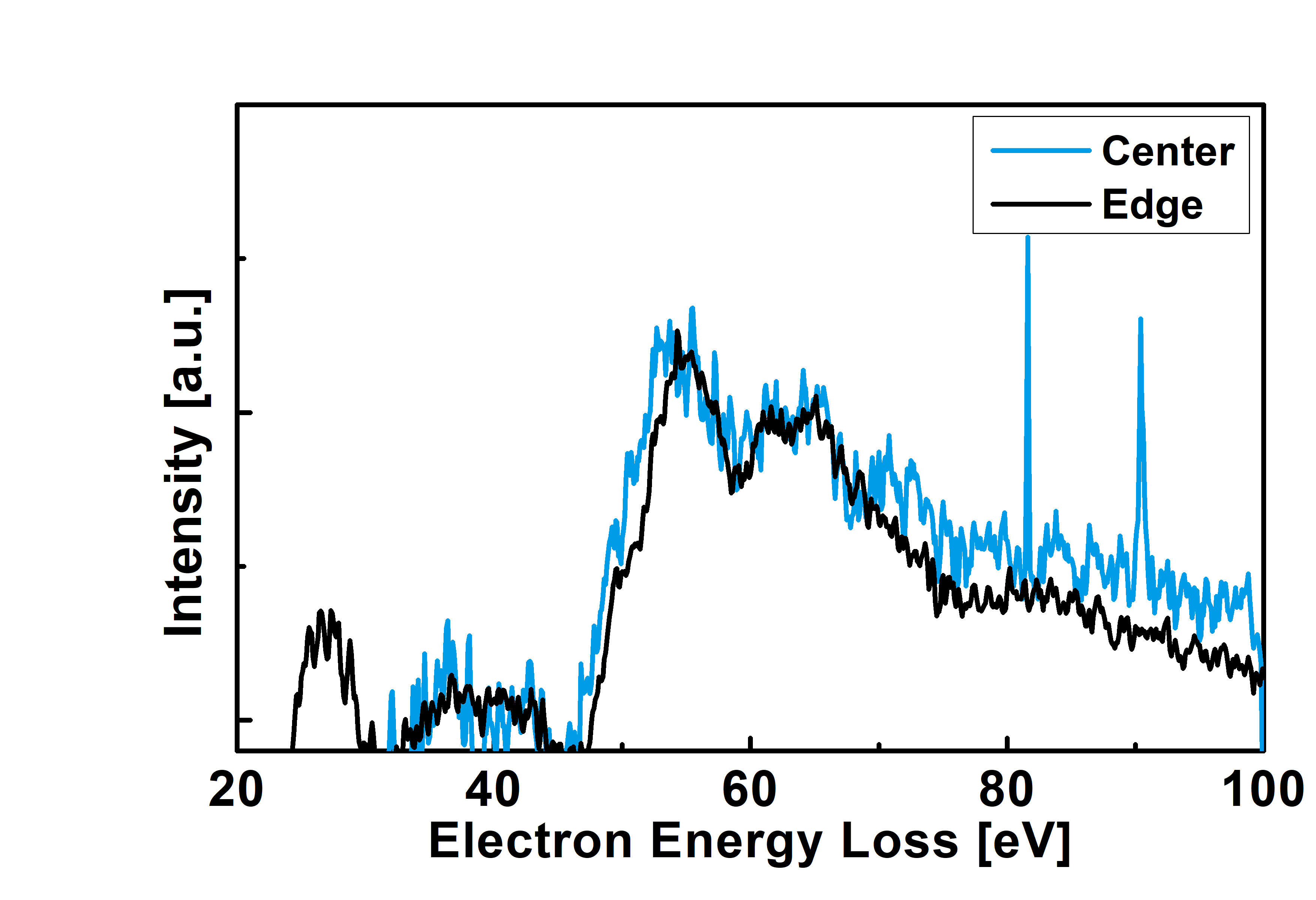}}
	\vspace*{-0.1cm}
	\caption{EEL (Electron Energy Loss) Low Loss spectra of the Mn $M$ and Li $K$ edge of particle surfaces and bulk after cycling of particles exhibiting an oxidiced surface relative to pristine state (Fig.\ref{fig:low-loss})}
	\label{fig:LL-OER}
\end{figure}

Low Loss spectra after OER are plotted in fig.\ref{fig:LL-OER} of cycled particles showing an oxidised surface compared to the pristine case after cycling. In one case, an intensity decrease at the Li $K$ edge position can be observed, while in other cases the spectra do not reveal significant differences above noise level. The effect visible in fig.\ref{fig:LL-OER}(a) might thus be introduced by data treatment (e.g. background subtraction).
\section{Estimation of Mn oxidation current}
\begin{figure}[!h]
	\centering
	\subfloat{%
		\includegraphics[width=0.9\columnwidth]{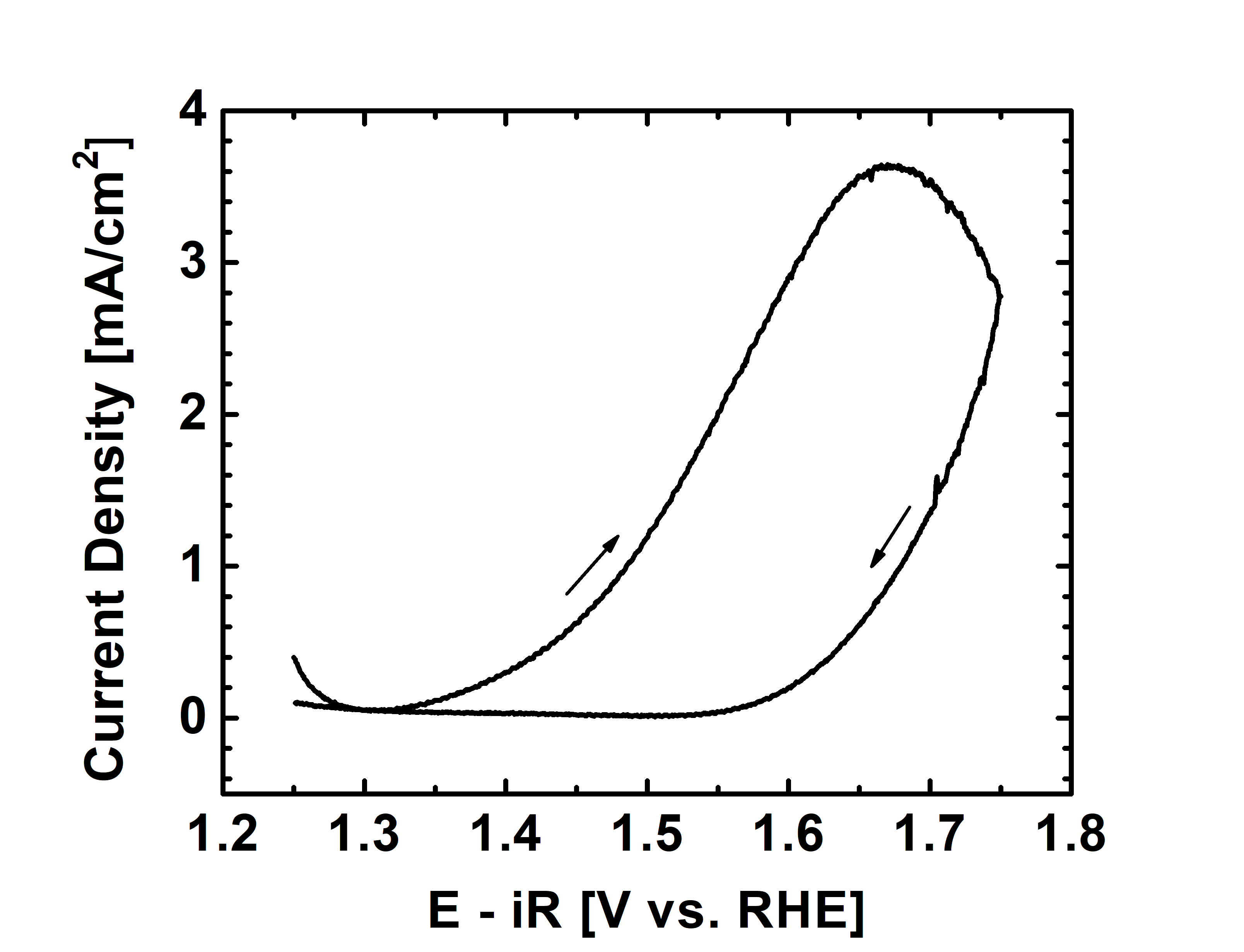}}
	\hspace{0.05\textwidth}
	\subfloat{%
		\includegraphics[width=0.9\columnwidth]{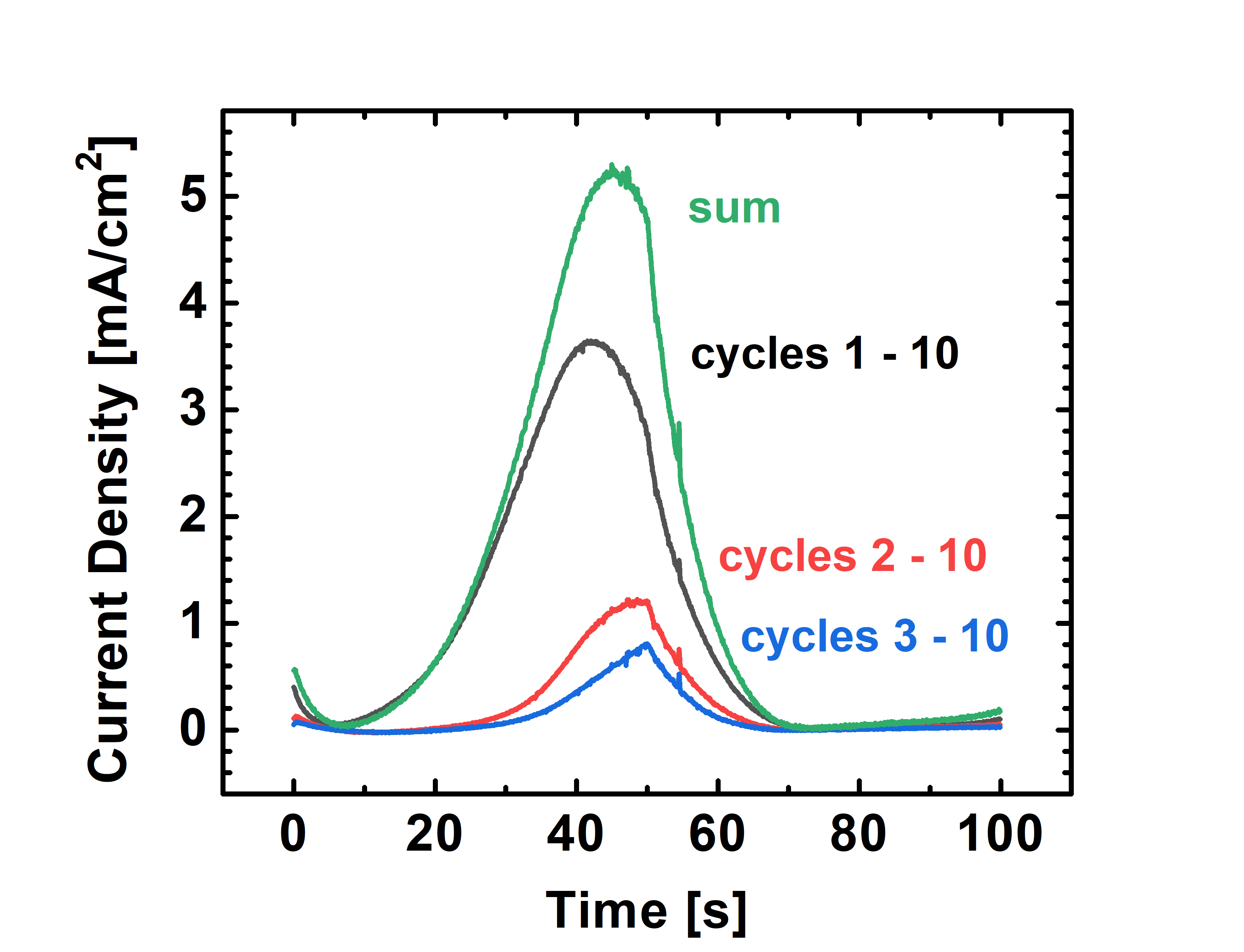}}
	\vspace*{-0.1cm}
	\caption{(a) Exemplary difference between the first and 10th cycle on a voltage axis. (b) The current density of the first, second and third cycle after subtraction of the 10th cycle and their sum as function of the time from the start of the cycle.
}
	\label{fig:Mn-current}
\end{figure}

The current additionally flown to OER activity was determined by subtracting the $I-V$ curve of the tenth cycle from the previous cycles. The results are plotted in Fig.\ref{fig:Mn-current}.
\section{PBE0r Optimized Structures and Atomic Spin Values}

Initial structural data were taken from measured x-ray diffraction structures of $\uplambda$-Mn$_2$O$_4$ at room temperature,\cite{Takahashi2003} Li$_{0.5}$Mn$_2$O$_4$ at 293\,K,\cite{Bianchini2015} LiMn$_2$O$_4$ at 330\,K,\cite{Akimoto2004} Li$_2$Mn$_2$O$_4$ at room temperature,\cite{Mosbah1983} and Mn$_3$O$_4$ at room temperature.\cite{Jarosch1987} The PBE0r functional was used to optimize the atomic positions inside the unit cell whose lattice vectors were fixed and to derive the atomic spins of the atoms from the spin density of the optimized structures. The lattice vectors $\mathbf{a}$, $\mathbf{b}$, and $\mathbf{c}$, the Cartesian coordinates of the atoms $x$, $y$, and $z$, and the atomic spin $S$ of all structures used in this study are given in table \ref{tab_mn2o4} to \ref{tab_limn2o3.5}.
\footnotesize
\begin{table}[H]
  \centering
  \caption{Structure and spin of $\uplambda$-Mn$_2$O$_4$.}
    \begin{tabular}{lrrrr}
    \hline
    Lattice & $x$\,/\,\AA & $y$\,/\,\AA & $z$\,/\,\AA &  \\
    \hline
    $\mathbf{a}$ & 8.0384 & 0.0000 & 0.0000 &  \\
    $\mathbf{b}$ & 0.0000 & 8.0384 & 0.0000 &  \\
    $\mathbf{c}$ & 0.0000 & 0.0000 & 8.0384 &  \\
    \hline
    Element & $x$\,/\,\AA & $y$\,/\,\AA & $z$\,/\,\AA & $S$\,/\,$\hbar$ \\
    \hline
    O & 2.1094 & 2.1096 & 2.1089 & 0.00 \\
    O & 5.9295 & 5.9288 & 5.9290 & 0.00 \\
    O & 3.9194 & 7.9384 & 6.1281 & 0.00 \\
    O & 4.1185 & 0.1000 & 1.9098 & 0.00 \\
    O & 7.9386 & 6.1288 & 3.9199 & 0.00 \\
    O & 0.0993 & 1.9096 & 4.1190 & 0.00 \\
    O & 6.1286 & 3.9192 & 7.9391 & 0.00 \\
    O & 1.9103 & 4.1192 & 0.0998 & 0.00 \\
    O & 0.0993 & 4.1192 & 1.9098 & 0.00 \\
    O & 7.9386 & 3.9192 & 6.1281 & 0.00 \\
    O & 4.1185 & 1.9096 & 0.0998 & 0.00 \\
    O & 3.9194 & 6.1288 & 7.9391 & 0.00 \\
    O & 1.9103 & 0.1000 & 4.1190 & 0.00 \\
    O & 6.1286 & 7.9384 & 3.9199 & 0.00 \\
    O & 2.1094 & 6.1288 & 6.1281 & 0.00 \\
    O & 5.9295 & 1.9096 & 1.9098 & 0.00 \\
    O & 3.9194 & 3.9192 & 2.1089 & 0.00 \\
    O & 4.1185 & 4.1192 & 5.9290 & 0.00 \\
    O & 7.9386 & 2.1096 & 7.9391 & 0.00 \\
    O & 0.0993 & 5.9288 & 0.0998 & 0.00 \\
    O & 0.0993 & 0.1000 & 5.9290 & 0.00 \\
    O & 7.9386 & 7.9384 & 2.1089 & 0.00 \\
    O & 4.1185 & 5.9288 & 4.1190 & 0.00 \\
    O & 3.9194 & 2.1096 & 3.9199 & 0.00 \\
    O & 6.1286 & 2.1096 & 6.1281 & 0.00 \\
    O & 1.9103 & 5.9288 & 1.9098 & 0.00 \\
    O & 2.1094 & 3.9192 & 3.9199 & 0.00 \\
    O & 5.9295 & 4.1192 & 4.1190 & 0.00 \\
    O & 5.9295 & 0.1000 & 0.0998 & 0.00 \\
    O & 2.1094 & 7.9384 & 7.9391 & 0.00 \\
    O & 6.1286 & 6.1288 & 2.1089 & 0.00 \\
    O & 1.9103 & 1.9096 & 5.9290 & 0.00 \\
    Mn & 4.0198 & 4.0192 & 4.0186 & 1.59 \\
    Mn & 2.0090 & 6.0288 & $-$0.0006 & $-$1.59 \\
    Mn & 6.0282 & 2.0096 & $-$0.0006 & 1.59 \\
    Mn & 6.0282 & 0.0000 & 2.0102 & 1.59 \\
    Mn & 2.0090 & 0.0000 & 6.0294 & 1.59 \\
    Mn & 0.0006 & 2.0096 & 6.0294 & $-$1.59 \\
    Mn & 0.0006 & 6.0288 & 2.0102 & 1.59 \\
    Mn & 4.0198 & 0.0000 & $-$0.0006 & $-$1.59 \\
    Mn & 2.0090 & 2.0096 & 4.0186 & 1.59 \\
    Mn & 6.0282 & 6.0288 & 4.0186 & $-$1.59 \\
    Mn & 6.0282 & 4.0192 & 6.0294 & $-$1.59 \\
    Mn & 2.0090 & 4.0192 & 2.0102 & $-$1.59 \\
    Mn & 0.0006 & 4.0192 & $-$0.0006 & 1.59 \\
    Mn & 4.0198 & 2.0096 & 2.0102 & $-$1.59 \\
    Mn & 4.0198 & 6.0288 & 6.0294 & 1.59 \\
    Mn & 0.0006 & 0.0000 & 4.0186 & $-$1.59 \\
    \hline
    \end{tabular}
  \label{tab_mn2o4}
\end{table}
\begin{table}[H]
  \centering
  \caption{Structure and spin of Li$_{0.5}$Mn$_2$O$_4$.}
    \begin{tabular}{lrrrr}
    \hline
    Lattice & $x$\,/\,\AA & $y$\,/\,\AA & $z$\,/\,\AA &  \\
    \hline
    $\mathbf{a}$ & 8.1529 & 0.0000 & 0.0000 &  \\
    $\mathbf{b}$ & 0.0000 & 8.1529 & 0.0000 &  \\
    $\mathbf{c}$ & 0.0000 & 0.0000 & 8.1529 &  \\
    \hline
    Element & $x$\,/\,\AA & $y$\,/\,\AA & $z$\,/\,\AA & $S$\,/\,$\hbar$ \\
    \hline
    Li & 0.0217 & $-$0.0020 & 0.0086 & 0.00 \\
    Li & 4.0547 & 8.1549 & 4.0847 & 0.00 \\
    Li & 4.0982 & 4.0783 & 8.1444 & 0.00 \\
    Li & 8.1312 & 4.0744 & 4.0679 & 0.00 \\
    O & 3.1326 & 3.1333 & 3.1249 & 0.00 \\
    O & 1.1890 & 7.0264 & 1.0224 & $-$0.01 \\
    O & 7.0510 & 7.0620 & 7.0482 & 0.00 \\
    O & 0.8412 & 1.0386 & 3.2162 & 0.00 \\
    O & 0.9439 & 5.0195 & 7.2013 & 0.00 \\
    O & 7.2090 & 0.9431 & 5.0280 & 0.00 \\
    O & 5.0203 & 7.2097 & 0.9515 & 0.00 \\
    O & 2.8871 & 1.1267 & 5.0974 & 0.01 \\
    O & 5.2654 & 5.2031 & 7.1308 & $-$0.01 \\
    O & 6.9638 & 2.9499 & 3.0543 & 0.01 \\
    O & 1.1260 & 1.1654 & 7.0513 & $-$0.01 \\
    O & 2.9505 & 6.9873 & 2.9749 & 0.01 \\
    O & 5.2024 & 2.9110 & 1.1016 & $-$0.01 \\
    O & 7.0269 & 5.2419 & 5.1780 & 0.01 \\
    O & 7.0318 & 1.0176 & 1.1835 & 0.02 \\
    O & 2.9556 & 3.0577 & 6.9695 & 0.02 \\
    O & 1.1211 & 5.0939 & 2.8929 & $-$0.02 \\
    O & 5.1975 & 7.1352 & 5.2599 & $-$0.02 \\
    O & 5.1784 & 1.0910 & 2.9719 & 0.00 \\
    O & 2.9745 & 5.1674 & 1.1045 & 0.00 \\
    O & 1.1019 & 2.9856 & 5.1812 & 0.00 \\
    O & 3.2348 & 7.1158 & 7.2926 & 0.00 \\
    O & 4.9178 & 3.0381 & 4.9366 & 0.00 \\
    O & 7.3116 & 5.1149 & 0.8602 & 0.00 \\
    O & 3.2145 & 0.8425 & 1.0274 & 0.01 \\
    O & 0.8623 & 7.3102 & 5.1023 & $-$0.01 \\
    O & 7.2906 & 3.2338 & 7.1268 & 0.01 \\
    O & 4.9387 & 4.9191 & 3.0502 & $-$0.01 \\
    O & 0.9140 & 3.1917 & 0.8626 & $-$0.02 \\
    O & 4.9904 & 0.8848 & 7.2904 & $-$0.02 \\
    O & 7.2389 & 7.2681 & 3.2139 & 0.02 \\
    O & 3.1625 & 4.9611 & 4.9391 & 0.02 \\
    Mn & 5.0838 & 5.0853 & 5.0748 & $-$2.02 \\
    Mn & 1.0296 & 3.1130 & 7.1074 & $-$1.56 \\
    Mn & 7.1455 & 3.0676 & 0.9984 & 2.02 \\
    Mn & 1.0074 & 7.1441 & 3.0781 & $-$2.02 \\
    Mn & 3.0691 & 1.0089 & 7.1546 & 2.02 \\
    Mn & 3.0468 & 5.0397 & 3.0309 & 1.56 \\
    Mn & 5.1060 & 0.9632 & 1.0457 & $-$1.56 \\
    Mn & 7.1233 & 7.1896 & 5.1219 & 1.56 \\
    Mn & 7.1292 & 1.0346 & 3.1049 & 1.57 \\
    Mn & 5.1001 & 7.1186 & 7.1813 & $-$1.57 \\
    Mn & 3.0528 & 3.0418 & 5.0479 & 1.57 \\
    Mn & 1.0237 & 5.1110 & 0.9716 & $-$1.57 \\
    Mn & 3.0942 & 7.1109 & 1.0260 & 1.58 \\
    Mn & 7.1706 & 5.1186 & 7.1271 & 1.58 \\
    Mn & 5.0588 & 3.0343 & 3.0506 & $-$1.58 \\
    Mn & 0.9824 & 1.0422 & 5.1021 & $-$1.58 \\
    \hline
    \end{tabular}
  \label{tab_li0.5mn2o4}
\end{table}
\begin{table}[H]
  \centering
  \caption{Structure and spin of LiMn$_2$O$_4$.}
    \begin{tabular}{lrrrr}
    \hline
    Lattice & $x$\,/\,\AA & $y$\,/\,\AA & $z$\,/\,\AA &  \\
    \hline
    $\mathbf{a}$ & 8.2455 & 0.0000 & 0.0000 &  \\
    $\mathbf{b}$ & 0.0000 & 8.2455 & 0.0000 &  \\
    $\mathbf{c}$ & 0.0000 & 0.0000 & 8.2455 &  \\
    \hline
    Element & $x$\,/\,\AA & $y$\,/\,\AA & $z$\,/\,\AA & $S$\,/\,$\hbar$ \\
    \hline
    Li & 1.0309 & 1.0567 & 1.0326 & 0.00 \\
    Li & 7.1720 & 3.0924 & 3.0919 & 0.00 \\
    Li & 5.1532 & 1.0047 & 5.1554 & 0.00 \\
    Li & 7.2577 & 7.2145 & 7.2146 & 0.00 \\
    Li & 5.1534 & 5.1804 & 1.0293 & 0.00 \\
    Li & 3.0461 & 7.2149 & 3.0916 & 0.00 \\
    Li & 1.0307 & 5.1265 & 5.1520 & 0.00 \\
    Li & 3.1381 & 3.0919 & 7.2144 & 0.00 \\
    O & 2.1759 & 2.1498 & 2.1903 & 0.01 \\
    O & 6.0287 & 4.3008 & 4.1523 & 0.02 \\
    O & 4.0082 & 8.1572 & 6.3130 & $-$0.01 \\
    O & 0.1554 & 6.0060 & 0.0293 & $-$0.02 \\
    O & 2.2391 & 8.2288 & 8.2211 & 0.01 \\
    O & 0.1277 & 4.0455 & 1.9155 & $-$0.03 \\
    O & 3.9450 & 2.0781 & 4.0984 & $-$0.01 \\
    O & 6.0564 & 6.2616 & 6.0383 & 0.03 \\
    O & 6.3632 & 4.1134 & 8.2157 & 0.02 \\
    O & 4.2492 & $-$0.0708 & 1.9176 & $-$0.02 \\
    O & 8.0664 & 6.1935 & 4.0932 & $-$0.02 \\
    O & 6.0289 & 1.8828 & 2.0312 & 0.02 \\
    O & 8.0667 & 8.2355 & 2.0919 & $-$0.01 \\
    O & 4.2492 & 6.2541 & 4.2664 & $-$0.02 \\
    O & 1.9349 & 4.0528 & 0.1437 & 0.02 \\
    O & 2.1761 & 4.0346 & 3.9939 & 0.01 \\
    O & 8.1328 & 2.1495 & 8.1155 & $-$0.01 \\
    O & 6.0562 & $-$0.0785 & 0.1459 & 0.03 \\
    O & 8.1325 & 4.0348 & 6.3142 & $-$0.01 \\
    O & 0.1279 & 2.1398 & 4.2686 & $-$0.03 \\
    O & 2.2395 & 6.2004 & 6.2078 & 0.01 \\
    O & 0.1553 & 0.1786 & 6.1539 & $-$0.02 \\
    O & 3.9446 & 4.1065 & 2.0850 & $-$0.01 \\
    O & 4.0080 & 6.2723 & 8.1167 & $-$0.01 \\
    O & 6.2968 & 8.1574 & 3.9928 & 0.01 \\
    O & 4.2781 & 4.3038 & 6.1641 & $-$0.02 \\
    O & 4.2778 & 1.8798 & 0.0206 & $-$0.02 \\
    O & 1.9063 & 0.1816 & 4.1434 & 0.02 \\
    O & 6.3629 & 2.0712 & 6.2146 & 0.01 \\
    O & 1.9060 & 6.0031 & 2.0413 & 0.02 \\
    O & 1.9349 & 2.1322 & 6.0403 & 0.02 \\
    O & 6.2972 & 6.2720 & 2.1915 & 0.01 \\
    Mn & 4.1217 & 4.1645 & 4.0945 & $-$2.02 \\
    Mn & 4.1186 & 6.2250 & 6.1654 & $-$1.56 \\
    Mn & 2.0624 & 6.1424 & $-$0.0283 & 2.02 \\
    Mn & 2.0655 & 4.0819 & 2.0426 & 1.56 \\
    Mn & $-$0.0032 & 2.1018 & 6.1659 & $-$1.56 \\
    Mn & 6.1845 & 2.0173 & $-$0.0315 & 2.02 \\
    Mn & 6.1873 & $-$0.0405 & 2.0432 & 1.56 \\
    Mn & 4.1188 & $-$0.0415 & 0.0187 & $-$1.56 \\
    Mn & 6.1876 & 6.2239 & 4.1411 & 1.56 \\
    Mn & 6.1844 & 4.1667 & 6.2159 & 2.02 \\
    Mn & $-$0.0003 & 6.1401 & 2.0931 & $-$2.02 \\
    Mn & 4.1215 & 2.0196 & 2.0893 & $-$2.02 \\
    Mn & 2.0627 & 0.0418 & 6.2121 & 2.02 \\
    Mn & 2.0653 & 2.1029 & 4.1414 & 1.56 \\
    Mn & $-$0.0034 & 4.0830 & 0.0182 & $-$1.56 \\
    Mn & $-$0.0004 & 0.0441 & 4.0913 & $-$2.02 \\
    \hline
    \end{tabular}
  \label{tab_limn2o4}
\end{table}
\begin{table}[H]
  \centering
  \caption{Structure and spin of Li$_2$Mn$_2$O$_4$.}
    \begin{footnotesize}
    \begin{tabular}{lrrrr}
    \hline
    Lattice & $x$\,/\,\AA & $y$\,/\,\AA & $z$\,/\,\AA &  \\
    \hline
    $\mathbf{a}$ & 11.3008 & 0.0000 & 0.0000 &  \\
    $\mathbf{b}$ & 0.0000 & 5.6504 & 0.0000 &  \\
    $\mathbf{c}$ & 0.0000 & 0.0000 & 9.2420 &  \\
    \hline
    Element & $x$\,/\,\AA & $y$\,/\,\AA & $z$\,/\,\AA & $S$\,/\,$\hbar$ \\
    \hline
    Li & 0.0000 & 0.0000 & 4.6210 & $-$0.01 \\
    Li & 1.4126 & 4.2378 & 6.9315 & $-$0.01 \\
    Li & 2.8252 & 0.0000 & 0.0000 & $-$0.01 \\
    Li & 1.4126 & 1.4126 & 2.3105 & $-$0.01 \\
    Li & 4.2378 & 1.4126 & 2.3105 & 0.01 \\
    Li & 4.2378 & 4.2378 & 6.9315 & 0.01 \\
    Li & 2.8252 & 2.8252 & 0.0000 & 0.01 \\
    Li & 0.0000 & 2.8252 & 4.6210 & 0.01 \\
    Li & 5.6504 & 0.0000 & 4.6210 & $-$0.01 \\
    Li & 7.0630 & 4.2378 & 6.9315 & $-$0.01 \\
    Li & 8.4756 & 0.0000 & 0.0000 & $-$0.01 \\
    Li & 7.0630 & 1.4126 & 2.3105 & $-$0.01 \\
    Li & 9.8882 & 1.4126 & 2.3105 & 0.01 \\
    Li & 9.8882 & 4.2378 & 6.9315 & 0.01 \\
    Li & 8.4756 & 2.8252 & 0.0000 & 0.01 \\
    Li & 5.6504 & 2.8252 & 4.6210 & 0.01 \\
    O & 0.0000 & 2.9204 & 2.3292 & 0.03 \\
    O & 4.1426 & 4.2378 & 4.6397 & 0.03 \\
    O & 2.8252 & 2.7300 & 6.9502 & 0.03 \\
    O & 4.3330 & 1.4126 & 0.0187 & 0.03 \\
    O & 0.0000 & 2.7300 & 6.9128 & 0.03 \\
    O & 4.3330 & 4.2378 & 9.2233 & 0.03 \\
    O & 2.8252 & 2.9204 & 2.2918 & 0.03 \\
    O & 4.1426 & 1.4126 & 4.6023 & 0.03 \\
    O & 1.5078 & 1.4126 & 4.6023 & $-$0.03 \\
    O & 1.3174 & 4.2378 & 9.2233 & $-$0.03 \\
    O & 1.3174 & 1.4126 & 0.0187 & $-$0.03 \\
    O & 1.5078 & 4.2378 & 4.6397 & $-$0.03 \\
    O & 2.8252 & 0.0952 & 6.9502 & $-$0.03 \\
    O & 0.0000 & 5.5552 & 2.3292 & $-$0.03 \\
    O & 2.8252 & 5.5552 & 2.2918 & $-$0.03 \\
    O & 0.0000 & 0.0952 & 6.9128 & $-$0.03 \\
    O & 5.6504 & 2.9204 & 2.3292 & 0.03 \\
    O & 9.7930 & 4.2378 & 4.6397 & 0.03 \\
    O & 8.4756 & 2.7300 & 6.9502 & 0.03 \\
    O & 9.9834 & 1.4126 & 0.0187 & 0.03 \\
    O & 5.6504 & 2.7300 & 6.9128 & 0.03 \\
    O & 9.9834 & 4.2378 & 9.2233 & 0.03 \\
    O & 8.4756 & 2.9204 & 2.2918 & 0.03 \\
    O & 9.7930 & 1.4126 & 4.6023 & 0.03 \\
    O & 7.1582 & 1.4126 & 4.6023 & $-$0.03 \\
    O & 6.9678 & 4.2378 & 9.2233 & $-$0.03 \\
    O & 6.9678 & 1.4126 & 0.0187 & $-$0.03 \\
    O & 7.1582 & 4.2378 & 4.6397 & $-$0.03 \\
    O & 8.4756 & 0.0952 & 6.9502 & $-$0.03 \\
    O & 5.6504 & 5.5552 & 2.3292 & $-$0.03 \\
    O & 8.4756 & 5.5552 & 2.2918 & $-$0.03 \\
    O & 5.6504 & 0.0952 & 6.9128 & $-$0.03 \\
    Mn & $-$0.0001 & 0.0000 & 0.0000 & $-$2.01 \\
    Mn & 1.4126 & 4.2377 & 2.3105 & $-$2.01 \\
    Mn & 2.8253 & 0.0000 & 4.6210 & $-$2.01 \\
    Mn & 1.4126 & 1.4127 & 6.9315 & $-$2.01 \\
    Mn & 4.2378 & 1.4125 & 6.9315 & 2.01 \\
    Mn & 4.2378 & 4.2379 & 2.3105 & 2.01 \\
    Mn & 2.8251 & 2.8252 & 4.6210 & 2.01 \\
    Mn & 0.0001 & 2.8252 & 0.0000 & 2.01 \\
    Mn & 5.6503 & 0.0000 & 0.0000 & $-$2.01 \\
    Mn & 7.0630 & 4.2377 & 2.3105 & $-$2.01 \\
    Mn & 8.4757 & 0.0000 & 4.6210 & $-$2.01 \\
    Mn & 7.0630 & 1.4127 & 6.9315 & $-$2.01 \\
    Mn & 9.8882 & 1.4125 & 6.9315 & 2.01 \\
    Mn & 9.8882 & 4.2379 & 2.3105 & 2.01 \\
    Mn & 8.4755 & 2.8252 & 4.6210 & 2.01 \\
    Mn & 5.6505 & 2.8252 & 0.0000 & 2.01 \\
    \hline
    \end{tabular}
    \end{footnotesize}
  \label{tab_li2mn2o4}
\end{table}
\begin{table}[H]
  \centering
  \caption{Structure and spin of Mn$_3$O$_4$.}
    \begin{tabular}{lrrrr}
    \hline
    Lattice & $x$\,/\,\AA & $y$\,/\,\AA & $z$\,/\,\AA &  \\
    \hline
    $\mathbf{a}$ & 11.5300 & 0.0000 & 0.0000 &  \\
    $\mathbf{b}$ & 0.0000 & 5.7650 & 0.0000 &  \\
    $\mathbf{c}$ & 0.0000 & 0.0000 & 9.4420 &  \\
    \hline
    Element & $x$\,/\,\AA & $y$\,/\,\AA & $z$\,/\,\AA & $S$\,/\,$\hbar$ \\
    \hline
    O & $-$0.0021 & 2.7378 & 2.4393 & $-$0.02 \\
    O & 4.4723 & 4.3257 & 4.7996 & $-$0.04 \\
    O & 2.8846 & 3.0272 & 7.1603 & $-$0.02 \\
    O & 4.1752 & 1.4393 & 0.0786 & $-$0.04 \\
    O & 0.0019 & 3.0310 & 7.0029 & $-$0.04 \\
    O & 4.1790 & 4.3216 & 9.3632 & $-$0.02 \\
    O & 2.8806 & 2.7340 & 2.2819 & $-$0.04 \\
    O & 4.4685 & 1.4434 & 4.6422 & $-$0.02 \\
    O & 1.2927 & 1.4432 & 4.6424 & 0.04 \\
    O & 1.5898 & 4.3218 & 9.3635 & 0.04 \\
    O & 1.5860 & 1.4391 & 0.0788 & 0.02 \\
    O & 1.2965 & 4.3259 & 4.7998 & 0.02 \\
    O & 2.8844 & 5.6165 & 7.1601 & 0.04 \\
    O & $-$0.0019 & 0.1485 & 2.4391 & 0.04 \\
    O & 2.8804 & 0.1447 & 2.2817 & 0.02 \\
    O & 0.0021 & 5.6203 & 7.0027 & 0.02 \\
    O & 5.7629 & 2.7378 & 2.4393 & $-$0.02 \\
    O & 10.2373 & 4.3257 & 4.7996 & $-$0.04 \\
    O & 8.6496 & 3.0272 & 7.1603 & $-$0.02 \\
    O & 9.9402 & 1.4393 & 0.0786 & $-$0.04 \\
    O & 5.7669 & 3.0310 & 7.0029 & $-$0.04 \\
    O & 9.9440 & 4.3216 & 9.3632 & $-$0.02 \\
    O & 8.6456 & 2.7340 & 2.2819 & $-$0.04 \\
    O & 10.2335 & 1.4434 & 4.6422 & $-$0.02 \\
    O & 7.0577 & 1.4432 & 4.6424 & 0.04 \\
    O & 7.3548 & 4.3218 & 9.3635 & 0.04 \\
    O & 7.3510 & 1.4391 & 0.0788 & 0.02 \\
    O & 7.0615 & 4.3259 & 4.7998 & 0.02 \\
    O & 8.6494 & 5.6165 & 7.1601 & 0.04 \\
    O & 5.7631 & 0.1485 & 2.4391 & 0.04 \\
    O & 8.6454 & 0.1447 & 2.2817 & 0.02 \\
    O & 5.7671 & 5.6203 & 7.0027 & 0.02 \\
    Mn & 0.0009 & 1.4404 & 8.2618 & $-$2.40 \\
    Mn & $-$0.0009 & 4.3229 & 1.1803 & 2.40 \\
    Mn & 2.8816 & 4.3247 & 3.5408 & $-$2.40 \\
    Mn & 2.8834 & 1.4422 & 5.9013 & 2.40 \\
    Mn & 0.0011 & 2.8799 & 4.7208 & $-$2.03 \\
    Mn & 4.3212 & 4.3249 & 7.0817 & $-$2.03 \\
    Mn & 2.8814 & 2.8851 & $-$0.0002 & $-$2.03 \\
    Mn & 4.3263 & 1.4401 & 2.3607 & $-$2.03 \\
    Mn & 1.4438 & 1.4424 & 2.3603 & 2.03 \\
    Mn & 1.4387 & 4.3226 & 7.0813 & 2.03 \\
    Mn & 2.8836 & 0.0026 & 0.0002 & 2.03 \\
    Mn & $-$0.0011 & $-$0.0026 & 4.7212 & 2.03 \\
    Mn & 5.7659 & 1.4404 & 8.2618 & $-$2.40 \\
    Mn & 5.7641 & 4.3229 & 1.1803 & 2.40 \\
    Mn & 8.6466 & 4.3247 & 3.5408 & $-$2.40 \\
    Mn & 8.6484 & 1.4422 & 5.9013 & 2.40 \\
    Mn & 5.7661 & 2.8799 & 4.7208 & $-$2.03 \\
    Mn & 10.0862 & 4.3249 & 7.0817 & $-$2.03 \\
    Mn & 8.6464 & 2.8851 & $-$0.0002 & $-$2.03 \\
    Mn & 10.0913 & 1.4401 & 2.3607 & $-$2.03 \\
    Mn & 7.2088 & 1.4424 & 2.3603 & 2.03 \\
    Mn & 7.2037 & 4.3226 & 7.0813 & 2.03 \\
    Mn & 8.6486 & 0.0026 & 0.0002 & 2.03 \\
    Mn & 5.7639 & $-$0.0026 & 4.7212 & 2.03 \\
    \hline
    \end{tabular}
  \label{tab_mn3o4}
\end{table}
\begin{table}[H]
  \centering
  \caption{Structure and spin of Li$_{0.5}$Mn$_{0.25}$Mn$_2$O$_4$.}
    \begin{tabular}{lrrrr}
    \hline
    Lattice & $x$\,/\,\AA & $y$\,/\,\AA & $z$\,/\,\AA &  \\
    \hline
    $\mathbf{a}$ & 8.2455 & 0.0000 & 0.0000 &  \\
    $\mathbf{b}$ & 0.0000 & 8.2455 & 0.0000 &  \\
    $\mathbf{c}$ & 0.0000 & 0.0000 & 8.2455 &  \\
    \hline
    Element & $x$\,/\,\AA & $y$\,/\,\AA & $z$\,/\,\AA & $S$\,/\,$\hbar$ \\
    \hline
    Li & 5.1539 & 1.0505 & 5.1535 & 0.00 \\
    Li & 7.2551 & 7.2210 & 7.1889 & 0.00 \\
    Li & 3.0491 & 7.2205 & 3.1244 & 0.00 \\
    Li & 1.0290 & 5.1552 & 5.1522 & 0.00 \\
    O & 2.1645 & 2.2122 & 2.1971 & 0.01 \\
    O & 6.0547 & 4.3844 & 4.2338 & 0.00 \\
    O & 4.0220 & 8.2123 & 6.3381 & 0.00 \\
    O & 0.1546 & 6.0216 & $-$0.0835 & $-$0.04 \\
    O & 2.2480 & 8.0701 & 8.3052 & 0.04 \\
    O & 0.1261 & 4.1718 & 1.8812 & 0.00 \\
    O & 3.9660 & 2.0213 & 4.0200 & $-$0.03 \\
    O & 6.0342 & 6.3517 & 6.0651 & 0.04 \\
    O & 6.3732 & 4.1369 & 8.0897 & 0.01 \\
    O & 4.2595 & $-$0.1812 & 1.9282 & $-$0.04 \\
    O & 8.0771 & 6.3501 & 4.2223 & $-$0.02 \\
    O & 6.0879 & 1.8177 & 1.8984 & $-$0.01 \\
    O & 8.0640 & 8.0769 & 1.9979 & $-$0.01 \\
    O & 4.2752 & 6.3429 & 4.2315 & $-$0.05 \\
    O & 1.9543 & 4.1702 & 0.1842 & $-$0.01 \\
    O & 2.1281 & 4.0052 & 4.0433 & 0.01 \\
    O & 8.1426 & 2.2135 & 8.1076 & 0.01 \\
    O & 6.0330 & $-$0.1792 & 0.1310 & 0.04 \\
    O & 8.1829 & 4.0079 & 6.2667 & 0.00 \\
    O & 0.1204 & 1.9990 & 4.2821 & 0.00 \\
    O & 2.2311 & 6.3444 & 6.0797 & 0.02 \\
    O & 0.1586 & 0.1794 & 6.2882 & $-$0.04 \\
    O & 3.9312 & 4.1423 & 2.2221 & $-$0.04 \\
    O & 3.9949 & 6.2446 & 8.0566 & $-$0.01 \\
    O & 6.2738 & 8.2142 & 3.9632 & 0.00 \\
    O & 4.2611 & 4.3844 & 6.0797 & $-$0.01 \\
    O & 4.2045 & 1.8229 & 0.1648 & 0.01 \\
    O & 1.8968 & 0.1821 & 4.0160 & 0.05 \\
    O & 6.3386 & 2.0233 & 6.2877 & 0.03 \\
    O & 1.9133 & 6.0314 & 2.1489 & 0.04 \\
    O & 1.9372 & 1.9943 & 6.0346 & 0.01 \\
    O & 6.3164 & 6.2488 & 2.2460 & $-$0.01 \\
    Mn & 1.0322 & 1.0586 & 1.0282 & 2.39 \\
    Mn & 5.1555 & 5.1377 & 1.0272 & $-$2.39 \\
    Mn & 4.1177 & 4.1474 & 4.1837 & $-$2.03 \\
    Mn & 4.1163 & 6.2985 & 6.0890 & $-$1.57 \\
    Mn & 2.0462 & 6.0925 & $-$0.0831 & 2.02 \\
    Mn & 2.0307 & 4.1616 & 2.1160 & 1.55 \\
    Mn & 0.0037 & 2.0594 & 6.1701 & $-$1.55 \\
    Mn & 6.1437 & 1.9931 & $-$0.0343 & 2.03 \\
    Mn & 6.1606 & $-$0.0750 & 2.0184 & 1.59 \\
    Mn & 4.1265 & $-$0.0650 & 0.0399 & $-$1.56 \\
    Mn & 6.1915 & 6.2958 & 4.2080 & 1.55 \\
    Mn & 6.1959 & 4.1566 & 6.1350 & 2.01 \\
    Mn & 0.0297 & 6.1061 & 2.1479 & $-$2.01 \\
    Mn & 4.1489 & 1.9770 & 2.0959 & $-$2.02 \\
    Mn & 2.0692 & 0.0474 & 6.1675 & 2.02 \\
    Mn & 2.0563 & 2.0598 & 4.1471 & 1.58 \\
    Mn & 0.0526 & 4.1513 & $-$0.0476 & $-$1.58 \\
    Mn & $-$0.0191 & 0.0559 & 4.1259 & $-$2.01 \\
    \hline
    \end{tabular}
  \label{tab_li0.5mn0.25mn2o4}
\end{table}
\begin{table}[H]
  \centering
  \caption{Structure and spin of Mn$_{0.25}$Mn$_2$O$_4$.}
    \begin{tabular}{lrrrr}
    \hline
    Lattice & $x$\,/\,\AA & $y$\,/\,\AA & $z$\,/\,\AA &  \\
    \hline
    $\mathbf{a}$ & 8.2455 & 0.0000 & 0.0000 &  \\
    $\mathbf{b}$ & 0.0000 & 8.2455 & 0.0000 &  \\
    $\mathbf{c}$ & 0.0000 & 0.0000 & 8.2455 &  \\
    \hline
    Element & $x$\,/\,\AA & $y$\,/\,\AA & $z$\,/\,\AA & $S$\,/\,$\hbar$ \\
    \hline
    O & 2.2128 & 2.1470 & 2.2399 & 0.01 \\
    O & 6.0545 & 4.2935 & 4.2461 & 0.00 \\
    O & 4.0109 & 8.1438 & 6.2826 & $-$0.01 \\
    O & 0.1413 & 5.9838 & 0.1074 & $-$0.01 \\
    O & 2.2225 & 8.2644 & 8.0801 & 0.03 \\
    O & 0.1516 & 4.0082 & 1.8830 & $-$0.02 \\
    O & 3.9832 & 2.2577 & 4.0953 & 0.00 \\
    O & 6.0518 & 6.1077 & 6.0690 & 0.00 \\
    O & 6.3437 & 4.1438 & 8.0780 & 0.01 \\
    O & 4.2572 & $-$0.1082 & 1.8885 & $-$0.02 \\
    O & 8.1066 & 6.3808 & 4.0920 & 0.00 \\
    O & 6.0419 & 1.8611 & 1.9535 & 0.01 \\
    O & 8.0859 & 8.2662 & 2.2296 & $-$0.01 \\
    O & 4.2571 & 6.1028 & 4.2354 & $-$0.01 \\
    O & 1.9266 & 4.0146 & 0.1726 & 0.02 \\
    O & 2.1732 & 4.0207 & 4.0240 & 0.01 \\
    O & 8.1015 & 2.1512 & 8.0626 & 0.01 \\
    O & 6.0327 & $-$0.1151 & 0.1787 & 0.02 \\
    O & 8.1361 & 4.0234 & 6.2794 & 0.00 \\
    O & 0.1321 & 1.9850 & 4.2382 & 0.00 \\
    O & 2.2009 & 6.3812 & 6.2127 & 0.00 \\
    O & 0.1303 & 0.1712 & 6.0613 & 0.00 \\
    O & 3.9616 & 4.1416 & 2.2271 & $-$0.03 \\
    O & 3.9718 & 6.2704 & 8.0668 & $-$0.01 \\
    O & 6.2933 & 8.1458 & 4.0277 & 0.00 \\
    O & 4.2557 & 4.2959 & 6.0696 & $-$0.01 \\
    O & 4.2556 & 1.8666 & 0.0979 & $-$0.01 \\
    O & 1.9288 & 0.1732 & 4.2371 & 0.01 \\
    O & 6.3234 & 2.2584 & 6.2157 & 0.00 \\
    O & 1.9283 & 5.9887 & 1.9613 & 0.01 \\
    O & 1.9272 & 1.9805 & 6.0714 & 0.01 \\
    O & 6.3284 & 6.2738 & 2.2447 & $-$0.01 \\
    Mn & 1.0310 & 1.0639 & 1.0330 & 2.39 \\
    Mn & 5.1523 & 5.1870 & 1.0285 & $-$2.39 \\
    Mn & 4.0979 & 4.1590 & 4.1675 & $-$1.63 \\
    Mn & 4.0999 & 6.2366 & 6.1130 & $-$1.60 \\
    Mn & 2.0240 & 6.1066 & $-$0.0532 & 2.02 \\
    Mn & 2.0240 & 4.1096 & 2.0715 & 1.57 \\
    Mn & $-$0.0215 & 2.1113 & 6.1169 & $-$1.58 \\
    Mn & 6.1412 & 1.9940 & $-$0.0612 & 2.03 \\
    Mn & 6.1286 & $-$0.0221 & 2.0747 & 1.58 \\
    Mn & 4.1602 & $-$0.0124 & $-$0.0104 & $-$1.57 \\
    Mn & 6.2059 & 6.2338 & 4.1904 & 1.58 \\
    Mn & 6.2090 & 4.1651 & 6.1503 & 1.60 \\
    Mn & 0.0431 & 6.1167 & 2.1225 & $-$2.03 \\
    Mn & 4.1601 & 1.9840 & 2.1140 & $-$2.02 \\
    Mn & 2.0868 & 0.0370 & 6.1392 & 1.63 \\
    Mn & 2.0844 & 2.1137 & 4.1939 & 1.60 \\
    Mn & 0.0553 & 4.1010 & $-$0.0130 & $-$1.58 \\
    Mn & $-$0.0250 & 0.0422 & 4.1570 & $-$1.60 \\
    \hline
    \end{tabular}
  \label{tab_mn0.25mn2o4}
\end{table}
\begin{table}[H]
  \centering
  \caption{Structure and spin of Mn$_{0.5}$Mn$_2$O$_4$.}
    \begin{tabular}{lrrrr}
    \hline
    Lattice & $x$\,/\,\AA & $y$\,/\,\AA & $z$\,/\,\AA &  \\
    \hline
    $\mathbf{a}$ & 8.2455 & 0.0000 & 0.0000 &  \\
    $\mathbf{b}$ & 0.0000 & 8.2455 & 0.0000 &  \\
    $\mathbf{c}$ & 0.0000 & 0.0000 & 8.2455 &  \\
    \hline
    Element & $x$\,/\,\AA & $y$\,/\,\AA & $z$\,/\,\AA & $S$\,/\,$\hbar$ \\
    \hline
    O & 2.1964 & 2.1665 & 2.2112 & $-$0.01 \\
    O & 6.0149 & 4.3212 & 4.1633 & 0.02 \\
    O & 3.9877 & 8.1404 & 6.3340 & 0.01 \\
    O & 0.1692 & 5.9857 & 0.0403 & $-$0.02 \\
    O & 2.2373 & 8.1980 & 8.1966 & 0.01 \\
    O & 0.1465 & 4.0478 & 1.9052 & $-$0.03 \\
    O & 3.9468 & 2.1089 & 4.0738 & $-$0.01 \\
    O & 6.0376 & 6.2593 & 6.0280 & 0.03 \\
    O & 6.3620 & 4.0813 & 8.1920 & 0.01 \\
    O & 4.2677 & $-$0.0676 & 1.9070 & $-$0.02 \\
    O & 8.0676 & 6.2256 & 4.0694 & $-$0.01 \\
    O & 6.0183 & 1.8586 & 2.0096 & 0.02 \\
    O & 8.0551 & 8.2040 & 2.1176 & $-$0.03 \\
    O & 4.2705 & 6.2563 & 4.2753 & $-$0.02 \\
    O & 1.9137 & 4.0506 & 0.1526 & 0.02 \\
    O & 2.1996 & 4.0237 & 3.9801 & 0.01 \\
    O & 8.1100 & 2.1603 & 8.1027 & $-$0.01 \\
    O & 6.0347 & $-$0.0818 & 0.1547 & 0.02 \\
    O & 8.1120 & 4.0182 & 6.3346 & 0.01 \\
    O & 0.1494 & 2.1431 & 4.2774 & $-$0.02 \\
    O & 2.2502 & 6.2310 & 6.2340 & 0.03 \\
    O & 0.1659 & 0.2028 & 6.1324 & $-$0.02 \\
    O & 3.9339 & 4.0759 & 2.1113 & $-$0.03 \\
    O & 3.9846 & 6.2832 & 8.1029 & $-$0.01 \\
    O & 6.3196 & 8.1466 & 3.9800 & 0.01 \\
    O & 4.2887 & 4.3279 & 6.1432 & $-$0.02 \\
    O & 4.2913 & 1.8594 & 0.0310 & $-$0.02 \\
    O & 1.8929 & 0.2020 & 4.1538 & 0.02 \\
    O & 6.3746 & 2.1027 & 6.2404 & 0.03 \\
    O & 1.8954 & 5.9790 & 2.0204 & 0.02 \\
    O & 1.9165 & 2.1290 & 6.0297 & 0.02 \\
    O & 6.3176 & 6.2886 & 2.2118 & $-$0.01 \\
    Mn & 1.0319 & 1.0587 & 1.0322 & $-$2.38 \\
    Mn & 5.1522 & 1.0026 & 5.1550 & 2.38 \\
    Mn & 5.1544 & 5.1823 & 1.0298 & $-$2.38 \\
    Mn & 1.0297 & 5.1246 & 5.1525 & 2.38 \\
    Mn & 4.1405 & 4.1791 & 4.1176 & $-$2.01 \\
    Mn & 4.1477 & 6.2052 & 6.1385 & $-$1.52 \\
    Mn & 2.0437 & 6.1278 & $-$0.0052 & 2.01 \\
    Mn & 2.0365 & 4.1017 & 2.0158 & 1.52 \\
    Mn & 0.0260 & 2.0820 & 6.1391 & $-$1.52 \\
    Mn & 6.1664 & 2.0031 & $-$0.0080 & 2.01 \\
    Mn & 6.1581 & $-$0.0206 & 2.0163 & 1.52 \\
    Mn & 4.1552 & $-$0.0182 & 0.0400 & $-$1.57 \\
    Mn & 6.1509 & 6.2010 & 4.1625 & 1.57 \\
    Mn & 6.1626 & 4.1856 & 6.1989 & 2.02 \\
    Mn & 0.0215 & 6.1213 & 2.0761 & $-$2.02 \\
    Mn & 4.1436 & 2.0006 & 2.0726 & $-$2.02 \\
    Mn & 2.0406 & 0.0608 & 6.1953 & 2.02 \\
    Mn & 2.0289 & 2.0796 & 4.1628 & 1.57 \\
    Mn & 0.0332 & 4.1060 & 0.0397 & $-$1.57 \\
    Mn & 0.0177 & 0.0583 & 4.1147 & $-$2.01 \\
    \hline
    \end{tabular}
  \label{tab_mn0.5mn2o4}
\end{table}
\begin{table}[H]
  \centering
  \caption{Structure and spin of Mn$_{0.75}$Mn$_2$O$_4$.}
    \begin{tabular}{lrrrr}
    \hline
    Lattice & $x$\,/\,\AA & $y$\,/\,\AA & $z$\,/\,\AA &  \\
    \hline
    $\mathbf{a}$ & 8.2455 & 0.0000 & 0.0000 &  \\
    $\mathbf{b}$ & 0.0000 & 8.2455 & 0.0000 &  \\
    $\mathbf{c}$ & 0.0000 & 0.0000 & 8.2455 &  \\
    \hline
    Element & $x$\,/\,\AA & $y$\,/\,\AA & $z$\,/\,\AA & $S$\,/\,$\hbar$ \\
    \hline
    O & 2.2012 & 2.1710 & 2.0631 & 0.02 \\
    O & 5.9914 & 4.2851 & 4.1218 & 0.02 \\
    O & 3.9629 & 8.2961 & 6.3370 & $-$0.02 \\
    O & 0.1300 & 6.0900 & 0.1483 & $-$0.02 \\
    O & 2.2456 & 8.2830 & 8.1209 & 0.02 \\
    O & 0.2401 & 4.0322 & 2.0603 & $-$0.06 \\
    O & 3.9652 & 2.2295 & 4.1561 & $-$0.02 \\
    O & 6.0679 & 6.1694 & 6.0035 & 0.03 \\
    O & 6.3471 & 3.9588 & 8.2763 & 0.02 \\
    O & 4.2292 & 0.0252 & 1.8819 & $-$0.02 \\
    O & 8.0603 & 6.1509 & 4.0053 & $-$0.02 \\
    O & 5.9832 & 2.0053 & 2.0824 & 0.04 \\
    O & 8.0753 & 8.2634 & 2.2363 & $-$0.06 \\
    O & 4.2415 & 6.1280 & 4.1811 & 0.01 \\
    O & 1.8894 & 4.0788 & 0.1362 & 0.05 \\
    O & 2.1943 & 4.0677 & 3.9586 & 0.01 \\
    O & 8.1101 & 2.1055 & 8.0815 & $-$0.02 \\
    O & 6.0644 & 0.0441 & 0.0719 & $-$0.01 \\
    O & 8.1064 & 4.0136 & 6.1955 & $-$0.02 \\
    O & 0.1699 & 2.1090 & 4.2602 & $-$0.05 \\
    O & 2.2354 & 6.1666 & 6.3514 & 0.06 \\
    O & 0.0832 & 0.1839 & 6.0914 & 0.00 \\
    O & 3.9426 & 3.9836 & 1.9815 & $-$0.04 \\
    O & 4.0191 & 6.3637 & 8.1448 & $-$0.02 \\
    O & 6.2770 & 8.0666 & 4.0102 & 0.02 \\
    O & 4.3261 & 4.1851 & 6.2140 & $-$0.04 \\
    O & 4.3223 & 1.8972 & $-$0.0127 & $-$0.02 \\
    O & 1.9414 & 0.1093 & 4.2704 & 0.01 \\
    O & 6.3651 & 2.1970 & 6.1072 & 0.04 \\
    O & 1.9774 & 5.9969 & 1.9778 & 0.00 \\
    O & 1.8187 & 2.1516 & 6.1910 & 0.06 \\
    O & 6.3496 & 6.1245 & 2.2141 & 0.02 \\
    Mn & 7.1680 & 3.0992 & 3.1268 & $-$2.34 \\
    Mn & 3.1380 & 3.0854 & 7.2514 & 2.34 \\
    Mn & 1.0081 & 0.9951 & 1.0236 & $-$2.37 \\
    Mn & 5.1573 & 0.9916 & 5.1219 & 2.37 \\
    Mn & 5.1497 & 5.1924 & 0.9980 & $-$2.37 \\
    Mn & 1.0529 & 5.1847 & 5.1451 & 2.37 \\
    Mn & 4.0919 & 4.2007 & 4.1072 & $-$2.00 \\
    Mn & 4.1468 & 6.2323 & 6.1339 & $-$1.98 \\
    Mn & 2.0452 & 6.2150 & 0.0598 & 2.00 \\
    Mn & 2.0695 & 4.0878 & 2.0297 & 1.48 \\
    Mn & $-$0.0087 & 2.0940 & 6.1544 & $-$1.48 \\
    Mn & 6.2187 & 1.9775 & $-$0.0187 & 2.00 \\
    Mn & 6.1557 & $-$0.0496 & 2.0142 & 1.99 \\
    Mn & 4.1736 & 0.0222 & $-$0.0043 & $-$1.52 \\
    Mn & 6.1307 & 6.1605 & 4.1161 & 1.52 \\
    Mn & 6.1969 & 4.1979 & 6.2258 & 2.04 \\
    Mn & 0.0514 & 6.1545 & 2.0618 & $-$2.01 \\
    Mn & 4.1109 & 1.9909 & 2.0965 & $-$2.04 \\
    Mn & 2.0137 & 0.0307 & 6.1804 & 2.01 \\
    Mn & 2.0692 & 2.0640 & 4.1376 & 2.01 \\
    Mn & $-$0.0058 & 4.1226 & 0.0154 & $-$2.01 \\
    Mn & 0.0154 & $-$0.0256 & 4.1794 & $-$2.00 \\
    \hline
    \end{tabular}
  \label{tab_mn0.75mn2o4}
\end{table}
\begin{table}[H]
  \centering
  \caption{Structure and spin of Mn$_3$O$_4$ epitaxial to LiMn$_2$O$_4$.}
    \begin{tabular}{lrrrr}
    \hline
    Lattice & $x$\,/\,\AA & $y$\,/\,\AA & $z$\,/\,\AA &  \\
    \hline
    $\mathbf{a}$ & 11.6609 & 0.0000 & 0.0000 &  \\
    $\mathbf{b}$ & 0.0000 & 5.8304 & 0.0000 &  \\
    $\mathbf{c}$ & 0.0000 & 0.0000 & 9.4420 &  \\
    \hline
    Element & $x$\,/\,\AA & $y$\,/\,\AA & $z$\,/\,\AA & $S$\,/\,$\hbar$ \\
    \hline
    O & $-$0.0015 & 2.7530 & 2.4176 & $-$0.02 \\
    O & 4.5432 & 4.3705 & 4.7744 & $-$0.04 \\
    O & 2.9167 & 3.0775 & 7.1386 & $-$0.02 \\
    O & 4.2024 & 1.4600 & 0.0534 & $-$0.04 \\
    O & $-$0.0024 & 3.0856 & 7.0281 & $-$0.04 \\
    O & 4.2106 & 4.3713 & 9.3849 & $-$0.02 \\
    O & 2.9176 & 2.7448 & 2.3071 & $-$0.04 \\
    O & 4.5351 & 1.4591 & 4.6639 & $-$0.02 \\
    O & 1.2872 & 1.4553 & 4.6676 & 0.04 \\
    O & 1.6280 & 4.3752 & 9.3886 & 0.04 \\
    O & 1.6199 & 1.4561 & 0.0571 & 0.02 \\
    O & 1.2954 & 4.3743 & 4.7781 & 0.02 \\
    O & 2.9129 & 5.6601 & 7.1349 & 0.04 \\
    O & 0.0024 & 0.1704 & 2.4139 & 0.04 \\
    O & 2.9137 & 0.1623 & 2.3034 & 0.02 \\
    O & 0.0015 & 5.6682 & 7.0244 & 0.02 \\
    O & 5.8290 & 2.7530 & 2.4176 & $-$0.02 \\
    O & 10.3737 & 4.3705 & 4.7744 & $-$0.04 \\
    O & 8.7472 & 3.0775 & 7.1386 & $-$0.02 \\
    O & 10.0329 & 1.4600 & 0.0534 & $-$0.04 \\
    O & 5.8281 & 3.0856 & 7.0281 & $-$0.04 \\
    O & 10.0410 & 4.3713 & 9.3849 & $-$0.02 \\
    O & 8.7480 & 2.7448 & 2.3071 & $-$0.04 \\
    O & 10.3655 & 1.4591 & 4.6639 & $-$0.02 \\
    O & 7.1177 & 1.4553 & 4.6676 & 0.04 \\
    O & 7.4585 & 4.3752 & 9.3886 & 0.04 \\
    O & 7.4503 & 1.4561 & 0.0571 & 0.02 \\
    O & 7.1258 & 4.3743 & 4.7781 & 0.02 \\
    O & 8.7433 & 5.6601 & 7.1349 & 0.04 \\
    O & 5.8328 & 0.1704 & 2.4139 & 0.04 \\
    O & 8.7442 & 0.1623 & 2.3034 & 0.02 \\
    O & 5.8320 & 5.6682 & 7.0244 & 0.02 \\
    Mn & 0.0041 & 1.4535 & 8.2618 & $-$2.41 \\
    Mn & $-$0.0041 & 4.3687 & 1.1803 & 2.41 \\
    Mn & 2.9111 & 4.3770 & 3.5408 & $-$2.41 \\
    Mn & 2.9194 & 1.4618 & 5.9013 & 2.41 \\
    Mn & 0.0049 & 2.9063 & 4.7205 & $-$2.04 \\
    Mn & 4.3639 & 4.3778 & 7.0820 & $-$2.04 \\
    Mn & 2.9103 & 2.9242 & $-$0.0005 & $-$2.04 \\
    Mn & 4.3818 & 1.4527 & 2.3610 & $-$2.04 \\
    Mn & 1.4666 & 1.4626 & 2.3600 & 2.04 \\
    Mn & 1.4487 & 4.3679 & 7.0810 & 2.04 \\
    Mn & 2.9202 & 0.0090 & 0.0005 & 2.04 \\
    Mn & $-$0.0049 & $-$0.0090 & 4.7215 & 2.04 \\
    Mn & 5.8346 & 1.4535 & 8.2618 & $-$2.41 \\
    Mn & 5.8263 & 4.3687 & 1.1803 & 2.41 \\
    Mn & 8.7415 & 4.3770 & 3.5408 & $-$2.41 \\
    Mn & 8.7498 & 1.4618 & 5.9013 & 2.41 \\
    Mn & 5.8354 & 2.9063 & 4.7205 & $-$2.04 \\
    Mn & 10.1943 & 4.3778 & 7.0820 & $-$2.04 \\
    Mn & 8.7407 & 2.9242 & $-$0.0005 & $-$2.04 \\
    Mn & 10.2122 & 1.4527 & 2.3610 & $-$2.04 \\
    Mn & 7.2970 & 1.4626 & 2.3600 & 2.04 \\
    Mn & 7.2791 & 4.3679 & 7.0810 & 2.04 \\
    Mn & 8.7506 & 0.0090 & 0.0005 & 2.04 \\
    Mn & 5.8255 & $-$0.0090 & 4.7215 & 2.04 \\
    \hline
    \end{tabular}
  \label{tab_mn3o4_epitaxial}
\end{table}
\begin{table}[H]
  \centering
  \caption{Structure and spin of LiMn$_3$O$_{3.875}$.}
    \begin{tabular}{lrrrr}
    \hline
    Lattice & $x$\,/\,\AA & $y$\,/\,\AA & $z$\,/\,\AA &  \\
    \hline
    $\mathbf{a}$ & 8.2455 & 0.0000 & 0.0000 &  \\
    $\mathbf{b}$ & 0.0000 & 8.2455 & 0.0000 &  \\
    $\mathbf{c}$ & 0.0000 & 0.0000 & 8.2455 &  \\
    \hline
    Element & $x$\,/\,\AA & $y$\,/\,\AA & $z$\,/\,\AA & $S$\,/\,$\hbar$ \\
    \hline
    Li & 1.0048 & 1.0636 & 1.0066 & 0.00 \\
    Li & 7.1913 & 3.0534 & 3.0661 & 0.00 \\
    Li & 5.1616 & 0.9804 & 5.1714 & 0.00 \\
    Li & 7.2286 & 7.2127 & 7.2255 & 0.00 \\
    Li & 5.1725 & 5.2019 & 1.0580 & 0.00 \\
    Li & 2.8232 & 7.4129 & 2.8180 & 0.00 \\
    Li & 1.0577 & 5.1968 & 5.1290 & 0.00 \\
    Li & 3.1126 & 3.0829 & 7.2005 & 0.00 \\
    O & 2.1650 & 2.1677 & 2.1806 & 0.00 \\
    O & 6.0468 & 4.3019 & 4.1850 & 0.00 \\
    O & 4.0065 & 8.1762 & 6.3495 & 0.00 \\
    O & 0.1414 & 6.0163 & 0.0586 & $-$0.01 \\
    O & 2.2194 & 8.1257 & 8.3545 & 0.02 \\
    O & 0.1289 & 4.0256 & 1.8982 & $-$0.03 \\
    O & 3.9637 & 2.0373 & 4.0391 & $-$0.05 \\
    O & 6.0263 & 6.3129 & 5.9873 & 0.04 \\
    O & 6.3537 & 4.1716 & 8.1209 & 0.02 \\
    O & 4.2668 & $-$0.0987 & 1.9670 & $-$0.04 \\
    O & 8.2345 & 6.3492 & 4.0341 & 0.05 \\
    O & 6.0522 & 1.9080 & 1.9485 & 0.00 \\
    O & 8.0324 & 8.2712 & 1.9777 & $-$0.03 \\
    O & 1.9652 & 4.1388 & 0.1451 & 0.00 \\
    O & 2.1487 & 4.0071 & 4.0238 & 0.00 \\
    O & 8.1504 & 2.1788 & 8.1100 & 0.00 \\
    O & 6.0162 & $-$0.1132 & 0.1255 & 0.04 \\
    O & 8.1656 & 4.0212 & 6.2846 & 0.00 \\
    O & 0.1045 & 2.0124 & 4.2628 & 0.00 \\
    O & 2.2263 & 6.3110 & 6.1155 & 0.02 \\
    O & 0.1498 & 0.1641 & 6.2536 & $-$0.04 \\
    O & 3.9582 & 4.1921 & 2.2187 & $-$0.03 \\
    O & 3.9843 & 6.2603 & 8.2608 & $-$0.06 \\
    O & 6.2512 & 8.1665 & 3.9630 & 0.00 \\
    O & 4.2295 & 4.3087 & 6.0453 & $-$0.01 \\
    O & 4.2169 & 1.8850 & 0.0848 & 0.00 \\
    O & 1.9212 & 0.1852 & 4.0538 & 0.04 \\
    O & 6.3484 & 2.0286 & 6.2759 & 0.02 \\
    O & 1.9041 & 5.9852 & 2.1257 & 0.04 \\
    O & 1.9519 & 2.0197 & 6.0476 & 0.00 \\
    O & 6.3011 & 6.2667 & 2.1765 & 0.00 \\
    Mn & 4.1224 & 4.1222 & 4.1229 & $-$1.97 \\
    Mn & 4.1303 & 6.2633 & 6.1758 & $-$1.95 \\
    Mn & 2.0781 & 6.1505 & $-$0.0034 & 2.01 \\
    Mn & 2.0454 & 4.1043 & 2.0867 & 1.59 \\
    Mn & $-$0.0067 & 2.0419 & 6.1991 & $-$1.57 \\
    Mn & 6.1709 & 2.0178 & $-$0.0274 & 2.02 \\
    Mn & 6.1994 & $-$0.0465 & 2.0076 & 1.55 \\
    Mn & 4.1011 & $-$0.0880 & 0.0806 & $-$1.56 \\
    Mn & 6.2088 & 6.2359 & 4.1173 & 1.90 \\
    Mn & 6.2020 & 4.1833 & 6.1781 & 2.03 \\
    Mn & 0.0126 & 6.1760 & 2.0884 & $-$2.02 \\
    Mn & 4.1220 & 2.0326 & 2.0609 & $-$2.03 \\
    Mn & 2.0582 & 0.0213 & 6.2290 & 2.02 \\
    Mn & 2.0698 & 2.0493 & 4.1157 & 1.57 \\
    Mn & 0.0036 & 4.1169 & $-$0.0136 & $-$1.57 \\
    Mn & 0.0001 & 0.0343 & 4.0733 & $-$2.01 \\
    \hline
    \end{tabular}
  \label{tab_limn2o3.875}
\end{table}
\begin{table}[H]
  \centering
  \caption{Structure and spin of LiMn$_3$O$_{3.75}$.}
    \begin{tabular}{lrrrr}
	\hline
    Lattice & $x$\,/\,\AA & $y$\,/\,\AA & $z$\,/\,\AA &  \\
    \hline
    $\mathbf{a}$ & 8.2455 & 0.0000 & 0.0000 &  \\
    $\mathbf{b}$ & 0.0000 & 8.2455 & 0.0000 &  \\
    $\mathbf{c}$ & 0.0000 & 0.0000 & 8.2455 &  \\
    \hline
    Element & $x$\,/\,\AA & $y$\,/\,\AA & $z$\,/\,\AA & $S$\,/\,$\hbar$ \\
    \hline
    Li & 1.0098 & 1.0372 & 0.9934 & 0.00 \\
    Li & 6.8940 & 2.8463 & 3.3774 & 0.00 \\
    Li & 5.1430 & 0.9571 & 5.1851 & 0.00 \\
    Li & 7.2174 & 7.2119 & 7.2384 & 0.00 \\
    Li & 5.1837 & 5.1781 & 1.0558 & 0.00 \\
    Li & 2.8185 & 7.4044 & 2.8142 & 0.00 \\
    Li & 1.0550 & 5.1715 & 5.0983 & 0.00 \\
    Li & 3.1108 & 3.1041 & 7.2105 & 0.00 \\
    O & 2.1223 & 2.1762 & 2.1812 & 0.01 \\
    O & 6.0455 & 4.3085 & 4.1797 & 0.00 \\
    O & 4.0077 & 8.1733 & 6.3512 & 0.00 \\
    O & 0.1278 & 6.0233 & 0.1435 & $-$0.01 \\
    O & 2.2256 & 8.1173 & 8.3553 & 0.02 \\
    O & 3.9611 & 2.0285 & 4.0254 & $-$0.04 \\
    O & 6.0320 & 6.3114 & 5.9873 & 0.04 \\
    O & 6.3525 & 4.1551 & 8.1582 & 0.03 \\
    O & 4.2707 & $-$0.0979 & 1.9687 & $-$0.03 \\
    O & 8.2370 & 6.3254 & 3.9916 & 0.05 \\
    O & 6.0650 & 1.8976 & 1.9559 & 0.00 \\
    O & 8.0420 & 8.2545 & 1.9750 & $-$0.06 \\
    O & 1.9343 & 4.1310 & 0.1697 & 0.00 \\
    O & 2.1437 & 4.0000 & 4.0293 & 0.00 \\
    O & 8.1483 & 2.1940 & 8.1210 & $-$0.01 \\
    O & 6.0161 & $-$0.1101 & 0.1278 & 0.04 \\
    O & 8.1160 & 3.9916 & 6.1249 & $-$0.06 \\
    O & 0.0991 & 2.0039 & 4.2536 & 0.00 \\
    O & 2.2154 & 6.3068 & 6.1182 & 0.02 \\
    O & 0.1509 & 0.1635 & 6.2518 & $-$0.04 \\
    O & 4.0932 & 4.1726 & 2.2415 & 0.01 \\
    O & 3.9765 & 6.2545 & 8.2613 & $-$0.06 \\
    O & 6.2591 & 8.1665 & 3.9625 & 0.00 \\
    O & 4.2433 & 4.3116 & 6.0487 & $-$0.01 \\
    O & 4.2155 & 1.8846 & 0.0980 & 0.00 \\
    O & 1.9150 & 0.1829 & 4.0516 & 0.04 \\
    O & 6.3430 & 2.0207 & 6.2707 & 0.02 \\
    O & 1.8980 & 5.9807 & 2.1370 & 0.04 \\
    O & 1.9490 & 2.0225 & 6.0512 & 0.00 \\
    O & 6.2942 & 6.2544 & 2.1766 & $-$0.01 \\
    Mn & 4.1310 & 4.1267 & 4.1076 & $-$1.97 \\
    Mn & 4.1316 & 6.2549 & 6.1766 & $-$1.95 \\
    Mn & 2.0777 & 6.1483 & 0.0016 & 2.02 \\
    Mn & 2.0742 & 4.1169 & 2.0952 & 1.91 \\
    Mn & $-$0.0094 & 2.0563 & 6.1933 & $-$1.56 \\
    Mn & 6.1765 & 2.0321 & $-$0.0264 & 2.02 \\
    Mn & 6.2007 & $-$0.0574 & 2.0064 & 1.57 \\
    Mn & 4.1046 & $-$0.0874 & 0.0805 & $-$1.56 \\
    Mn & 6.2011 & 6.2278 & 4.1122 & 1.91 \\
    Mn & 6.2141 & 4.1795 & 6.1825 & 2.03 \\
    Mn & 0.0149 & 6.2173 & 2.0909 & $-$1.96 \\
    Mn & 4.1109 & 2.0531 & 2.0565 & $-$2.02 \\
    Mn & 2.0566 & 0.0231 & 6.2261 & 2.03 \\
    Mn & 2.0627 & 2.0558 & 4.1106 & 1.58 \\
    Mn & $-$0.0009 & 4.1229 & $-$0.0276 & $-$1.95 \\
    Mn & 0.0038 & 0.0307 & 4.0792 & $-$2.02 \\
    \hline
    \end{tabular}
  \label{tab_limn2o3.75}
\end{table}
\begin{table}[H]
  \centering
  \caption{Structure and spin of LiMn$_3$O$_{3.625}$.}
    \begin{tabular}{lrrrr}
    \hline
    Lattice & $x$\,/\,\AA & $y$\,/\,\AA & $z$\,/\,\AA &  \\
    \hline
    $\mathbf{a}$ & 8.2455 & 0.0000 & 0.0000 &  \\
    $\mathbf{b}$ & 0.0000 & 8.2455 & 0.0000 &  \\
    $\mathbf{c}$ & 0.0000 & 0.0000 & 8.2455 &  \\
    \hline
    Element & $x$\,/\,\AA & $y$\,/\,\AA & $z$\,/\,\AA & $S$\,/\,$\hbar$ \\
    \hline
    Li & 1.0099 & 1.0443 & 0.9854 & 0.00 \\
    Li & 6.8325 & 2.7442 & 3.4534 & 0.00 \\
    Li & 5.1021 & 0.9367 & 5.1784 & 0.00 \\
    Li & 7.2046 & 7.2173 & 7.2382 & 0.00 \\
    Li & 5.3894 & 4.9999 & 1.2212 & 0.00 \\
    Li & 2.8986 & 7.3754 & 2.8980 & 0.00 \\
    Li & 1.0581 & 5.1510 & 5.1453 & 0.00 \\
    Li & 3.1239 & 3.0722 & 7.2104 & 0.00 \\
    O & 2.1106 & 2.1925 & 2.1580 & 0.00 \\
    O & 6.0358 & 4.2998 & 4.1851 & 0.01 \\
    O & 3.9733 & 8.1295 & 6.3046 & $-$0.01 \\
    O & $-$0.0081 & 6.0153 & 0.1642 & 0.04 \\
    O & 2.2288 & 8.0386 & 8.3419 & 0.03 \\
    O & 3.9475 & 2.0422 & 4.0174 & $-$0.03 \\
    O & 6.0092 & 6.2947 & 6.0196 & 0.04 \\
    O & 6.3432 & 4.1221 & 8.1626 & 0.03 \\
    O & 4.2802 & $-$0.1597 & 1.9368 & $-$0.03 \\
    O & 8.3131 & 6.3289 & 3.9685 & 0.04 \\
    O & 6.0404 & 1.8700 & 2.0036 & 0.01 \\
    O & 8.0449 & 8.2720 & 1.9804 & $-$0.08 \\
    O & 1.9554 & 4.1890 & 0.1355 & 0.00 \\
    O & 2.1424 & 4.0311 & 4.0073 & 0.00 \\
    O & 8.1432 & 2.1968 & 8.1120 & $-$0.01 \\
    O & 6.0333 & $-$0.0964 & 0.1218 & 0.03 \\
    O & 8.0995 & 3.9885 & 6.1062 & $-$0.07 \\
    O & 0.0964 & 1.9729 & 4.2477 & 0.00 \\
    O & 2.1879 & 6.2073 & 6.3404 & 0.01 \\
    O & 0.1590 & 0.1510 & 6.2769 & $-$0.05 \\
    O & 4.1145 & 4.2101 & 2.2652 & 0.02 \\
    O & 6.2660 & 8.1496 & 3.9705 & 0.00 \\
    O & 4.2363 & 4.2974 & 6.1098 & $-$0.01 \\
    O & 4.2557 & 2.0325 & 0.1050 & $-$0.05 \\
    O & 1.9428 & 0.1711 & 4.0865 & 0.03 \\
    O & 6.3429 & 2.0197 & 6.2282 & 0.02 \\
    O & 1.9631 & 6.0399 & 1.9778 & 0.01 \\
    O & 1.9425 & 2.0416 & 6.0306 & 0.01 \\
    O & 6.3349 & 6.2426 & 2.1966 & 0.00 \\
    Mn & 4.1168 & 4.1835 & 4.1494 & $-$1.95 \\
    Mn & 4.1098 & 6.2182 & 6.1815 & $-$1.68 \\
    Mn & 2.0529 & 6.1542 & 0.0117 & 1.96 \\
    Mn & 2.1202 & 4.1369 & 2.0568 & 1.93 \\
    Mn & $-$0.0241 & 2.0488 & 6.1833 & $-$1.56 \\
    Mn & 6.1625 & 2.0434 & $-$0.0173 & 2.02 \\
    Mn & 6.1919 & $-$0.0672 & 2.0052 & 1.57 \\
    Mn & 4.1335 & $-$0.0952 & 0.0345 & $-$1.93 \\
    Mn & 6.2193 & 6.2250 & 4.1407 & 1.96 \\
    Mn & 6.1903 & 4.1661 & 6.1864 & 2.03 \\
    Mn & 0.0411 & 6.2493 & 2.0582 & $-$1.96 \\
    Mn & 4.1038 & 2.0521 & 2.0522 & $-$2.02 \\
    Mn & 2.0470 & 0.0189 & 6.2458 & 2.04 \\
    Mn & 2.0604 & 2.0615 & 4.0999 & 1.59 \\
    Mn & $-$0.0056 & 4.1396 & $-$0.0670 & $-$1.96 \\
    Mn & 0.0216 & 0.0079 & 4.0952 & $-$2.02 \\
    \hline
    \end{tabular}
  \label{tab_limn2o3.625}
\end{table}
\begin{table}[H]
  \centering
  \caption{Structure and spin of LiMn$_3$O$_{3.5}$.}
    \begin{tabular}{lrrrr}
    \hline
    Lattice & $x$\,/\,\AA & $y$\,/\,\AA & $z$\,/\,\AA &  \\
    \hline
    $\mathbf{a}$ & 8.2455 & 0.0000 & 0.0000 &  \\
    $\mathbf{b}$ & 0.0000 & 8.2455 & 0.0000 &  \\
    $\mathbf{c}$ & 0.0000 & 0.0000 & 8.2455 &  \\
    \hline
    Element & $x$\,/\,\AA & $y$\,/\,\AA & $z$\,/\,\AA & $S$\,/\,$\hbar$ \\
    \hline
    Li & 1.0093 & 1.0950 & 1.0610 & 0.00 \\
    Li & 6.1854 & 2.4844 & 3.9048 & 0.01 \\
    Li & 5.1238 & 0.8489 & 5.2139 & 0.00 \\
    Li & 7.1865 & 7.2272 & 7.2069 & 0.00 \\
    Li & 5.4224 & 4.9428 & 1.2552 & 0.00 \\
    Li & 2.8461 & 7.4070 & 2.8288 & 0.00 \\
    Li & 1.0383 & 5.1235 & 5.1423 & 0.00 \\
    Li & 3.1196 & 3.0973 & 7.2177 & 0.00 \\
    O & 2.1384 & 2.2189 & 2.1614 & 0.01 \\
    O & 6.0063 & 4.3211 & 4.1791 & 0.01 \\
    O & 3.9733 & 8.1207 & 6.3278 & $-$0.01 \\
    O & $-$0.0476 & 6.0140 & 0.1499 & 0.04 \\
    O & 2.2214 & 8.0434 & 8.3555 & 0.04 \\
    O & 4.1633 & 2.0822 & 4.0081 & 0.01 \\
    O & 6.0018 & 6.2871 & 6.0293 & 0.03 \\
    O & 6.2971 & 4.0082 & 8.1658 & 0.01 \\
    O & 4.2990 & $-$0.1828 & 1.9283 & $-$0.03 \\
    O & 8.2798 & 6.1562 & 3.9524 & $-$0.02 \\
    O & 6.0673 & 1.8597 & 2.1397 & 0.05 \\
    O & 8.0913 & 8.2974 & 2.1812 & $-$0.05 \\
    O & 1.9356 & 4.2086 & 0.1289 & 0.00 \\
    O & 2.1216 & 4.0415 & 4.0102 & 0.01 \\
    O & 8.1236 & 2.1931 & 8.2608 & $-$0.06 \\
    O & 6.0968 & 0.0647 & 0.1279 & 0.00 \\
    O & 8.0952 & 3.9845 & 6.1202 & $-$0.06 \\
    O & 2.1801 & 6.2380 & 6.3210 & 0.00 \\
    O & 0.1255 & 0.1042 & 6.0240 & $-$0.01 \\
    O & 4.0905 & 4.2664 & 2.2556 & 0.02 \\
    O & 6.2799 & 8.1336 & 3.9893 & 0.00 \\
    O & 4.2208 & 4.2963 & 6.1039 & $-$0.01 \\
    O & 4.2609 & 2.0597 & 0.1280 & $-$0.06 \\
    O & 1.8981 & 0.2062 & 4.0224 & 0.03 \\
    O & 6.3851 & 2.0451 & 5.9332 & 0.04 \\
    O & 1.9275 & 6.0443 & 1.9664 & 0.01 \\
    O & 1.9443 & 2.0037 & 6.0236 & 0.00 \\
    O & 6.3195 & 6.2641 & 2.1698 & 0.01 \\
    Mn & 4.0877 & 4.2161 & 4.1353 & $-$1.94 \\
    Mn & 4.1089 & 6.2084 & 6.1984 & $-$1.67 \\
    Mn & 2.0279 & 6.1670 & 0.0000 & 1.97 \\
    Mn & 2.1028 & 4.1458 & 2.0565 & 1.93 \\
    Mn & $-$0.0067 & 2.0542 & 6.1536 & $-$1.95 \\
    Mn & 6.1853 & 2.0592 & $-$0.0079 & 2.03 \\
    Mn & 6.1839 & $-$0.0291 & 2.0552 & 1.59 \\
    Mn & 4.1288 & $-$0.0494 & 0.0409 & $-$1.94 \\
    Mn & 6.2390 & 6.1960 & 4.1384 & 1.95 \\
    Mn & 6.1798 & 4.1429 & 6.1775 & 2.02 \\
    Mn & 0.0088 & 6.1826 & 2.0684 & $-$1.97 \\
    Mn & 4.1449 & 2.0857 & 2.0833 & $-$2.03 \\
    Mn & 2.0500 & $-$0.0009 & 6.1897 & 2.05 \\
    Mn & 2.1280 & 2.0827 & 4.1114 & 1.93 \\
    Mn & $-$0.0174 & 4.1043 & $-$0.0370 & $-$1.97 \\
    Mn & 0.0199 & $-$0.0288 & 4.0812 & $-$1.96 \\
    \hline
    \end{tabular}
  \label{tab_limn2o3.5}
\end{table}

\bibliography{library,libmarco}